\documentclass[11pt,a4paper]{article}
\usepackage{jheppub}
\usepackage[american]{babel}
\usepackage{bbm}

\newcommand{\dvol}{d\mathrm{vol}}
\newcommand{\vol}{\mathrm{vol}}

\newcommand{\trans}{{\sf T}}

\newcommand{\parfrac}[2]{\frac{\partial #1}{\partial #2}}

\newcommand{\ud}[2]{^{#1}_{\phantom{#1}#2}}

\newcommand{\smath}[1]{\text{\small#1}}

\newcommand{\rep}[1]{\ensuremath{\mathbf{#1}}}

\newcommand{\wb}{\overline}
\newcommand{\wt}{\widetilde}
\newcommand{\wh}{\widehat}
\newcommand{\pdag}{{\phantom{\dag}}}

\newcommand{\eg}{\textit{e.g.}}

\newcommand{\ie}{\textit{i.e.}}

\numberwithin{equation}{section}

\newcommand{\Dslash}{D\!\!\!\!\slash\,}
\newcommand{\nn}{\nonumber}
\newcommand{\mat}[1]{\begin{pmatrix} #1 \end{pmatrix}}
\newcommand{\smat}[1]{\big( \begin{smallmatrix} #1 \end{smallmatrix} \big)}
\newcommand{\be}{\begin{equation}} \newcommand{\ee}{\end{equation}}
\newcommand{\bea}{\begin{equation} \begin{aligned}} \newcommand{\eea}{\end{aligned} \end{equation}}

\newcommand{\cA}{\mathcal{A}}

\newcommand{\cC}{\mathcal{C}}
\newcommand{\cD}{\mathcal{D}}

\newcommand{\cG}{\mathcal{G}}
\newcommand{\cH}{\mathcal{H}}

\newcommand{\cL}{\mathcal{L}}
\newcommand{\cM}{\mathcal{M}}
\newcommand{\cN}{\mathcal{N}}
\newcommand{\cO}{\mathcal{O}}
\newcommand{\cP}{\mathcal{P}}
\newcommand{\cQ}{\mathcal{Q}}

\newcommand{\cV}{\mathcal{V}}

\newcommand{\cZ}{\mathcal{Z}}

\newcommand{\bC}{\mathbb{C}}

\newcommand{\bN}{\mathbb{N}}

\newcommand{\bP}{\mathbb{P}}

\newcommand{\bR}{\mathbb{R}}
\newcommand{\bZ}{\mathbb{Z}}
\newcommand{\fg}{\mathfrak{g}}
\newcommand{\fh}{\mathfrak{h}}
\newcommand{\fm}{\mathfrak{m}}
\newcommand{\fM}{\mathfrak{M}}
\newcommand{\fn}{\mathfrak{n}}

\newcommand{\fR}{\mathfrak{R}}

\newcommand{\ft}{\mathfrak{t}}

\newcommand{\unit}{\mathbbm{1}}
\newcommand{\se}{\mathsf{e}}
\newcommand{\sg}{\mathsf{g}}
\newcommand{\sQ}{\mathsf{Q}}

\def\su{\mathfrak{su}}

\DeclareMathOperator{\Tr}{Tr}

\DeclareMathOperator{\sign}{sign}
\DeclareMathOperator{\rank}{rank}
\DeclareMathOperator{\re}{\mathbb{R}e}
\DeclareMathOperator{\im}{\mathbb{I}m}
\DeclareMathOperator*{\Res}{Res}
\DeclareMathOperator*{\JKres}{JK-Res}

\DeclareMathOperator{\Pexp}{\mathcal{P}exp}
\DeclareMathOperator{\Ad}{Ad}

\title{A topologically twisted index \\ for three-dimensional supersymmetric theories}

\author[a,b]{Francesco Benini}
\author[c,d]{and Alberto Zaffaroni}

\affiliation[a]{Delta Institute for Theoretical Physics, University of Amsterdam, \\
Science Park 904, 1098 XH Amsterdam, the Netherlands}
\affiliation[b]{Blackett Laboratory, Imperial College London, \\
South Kensington Campus, London SW7 2AZ, United Kingdom}
\affiliation[c]{Dipartimento di Fisica, Universit\`a di Milano-Bicocca, \\
I-20126 Milano, Italy}
\affiliation[d]{INFN, sezione di Milano-Bicocca, I-20126 Milano, Italy}

\emailAdd{f.benini@imperial.ac.uk}
\emailAdd{alberto.zaffaroni@mib.infn.it}

\preprint{Imperial/TP/2015/FB/01}

\abstract{We provide a general formula for the partition function of three-dimensional $\cN=2$ gauge theories placed on $S^2 \times S^1$ with a topological twist along $S^2$, which can be interpreted as an index for chiral states of the theories immersed in background magnetic fields. The result is expressed as a sum over magnetic fluxes of the residues of a meromorphic form which is a function of the scalar zero-modes. The partition function depends on a collection of background magnetic fluxes and fugacities for the global symmetries. We illustrate our formula in many examples of 3d Yang-Mills-Chern-Simons theories with matter, including Aharony and Giveon-Kutasov dualities. Finally, our formula generalizes to $\Omega$-backgrounds, as well as two-dimensional theories on $S^2$ and four-dimensional theories on $S^2 \times T^2$. In particular this provides an alternative way to compute genus-zero A-model topological amplitudes and Gromov-Witten invariants.}

\begin{document}

\setcounter{tocdepth}{2}
\maketitle

%
%

\section{Introduction}

In the last few years there has been a huge development in the study of supersymmetric quantum field theories on compact manifolds in various dimensions. Localization techniques \cite{Witten:1991zz} often allow to exactly evaluate the path-integral of the theory on a compact manifold, possibly with the insertion of local and non-local operators that respect some supersymmetry. Most of the examples studied in the last few years are not topologically twisted \cite{Pestun:2007rz} (see \cite{Marino:2011nm} for a nice review and references). In this paper, instead, we consider a very simple example: the partition function of $\cN=2$ supersymmetric gauge theories in three dimensions on $S^2 \times S^1$ with a semi-topological A-twist on $S^2$ \cite{Witten:1991zz}. Despite the conceptual simplicity, we will see that the model presents several interesting features.

We study generic three-dimensional $\cN=2$ gauge theories with an R-symmetry (with the constraint that the R-charges be integers). The topological twist is equivalent to turning on a background for the R-symmetry, which is a quantized magnetic flux on $S^2$.  Similar magnetic fluxes can be turned on  for all global flavor symmetries of the theory. We use localization to compute the partition function, with a method similar to that recently used to evaluate the elliptic genus of two-dimensional gauge theories \cite{Benini:2013nda, Benini:2013xpa} and the Witten index of quantum mechanical sigma models \cite{Hori:2014tda, Hwang:2014uwa, Cordova:2014oxa}. The path-integral localizes on a set of BPS configurations which are specified by some data of the gauge multiplet: a magnetic gauge flux $\fm$ on $S^2$ and a complex mode $u= A_t + i \beta \sigma$ encoding the vacuum expectation value $\sigma$ of the real scalar in the vector multiplet  and the holonomy of the gauge field $A_t$ along $S^1$ ($\beta$ is the radius of $S^1$).
The final partition function is given by a contour integral,
\be
Z_{S^2 \times S^1} = \frac1{|W|} \sum_{\fm \,\in\, \Gamma_\fh} \oint_\cC \; Z_\text{int}(u; \fm) \;,
\ee
of a meromorphic form in the variables $u$, which encodes the classical and one-loop contributions around BPS configurations, summed over all magnetic gauge fluxes.  Supersymmetric localization selects a particular contour of integration $\cC$, and therefore it picks some of the residues of the form $Z_\text{int}(u; \fm)$. The choice of the correct integration contour is one of the challenges of this computation, that we solve using the methods introduced in \cite{Benini:2013nda, Benini:2013xpa}. In many cases, we can formulate the result of integration in terms of a geometrical operation called the \emph{Jeffrey-Kirwan residue} \cite{JeffreyKirwan} (see also \cite{Brion1999, Szenes:2004}). The final result depends on the magnetic fluxes $\fm_f$ and chemical potentials $u_f$ for the flavor symmetries of the theory. It is, in particular, an analytic function of the fugacities $e^{i u_f}$. For the reader's convenience, we summarize the main features of our formula in section \ref{sec: summary}. 

To avoid confusion, we should stress that we are not computing the superconformal index, which is the partition function on $S^2\times S^1$ without  twist. In fact, our partition function is not counting operators and there is no fugacity corresponding to the dimension, or R-charge, of the operators.  We can still interpret our formula as a twisted index by writing the partition function as a trace over the Hilbert space $\cH$ of states on a sphere, in the presence of a magnetic background for the R- and flavor symmetries,
\be
Z_{S^2 \times S^1} = \Tr_\cH \, (-1)^F \, e^{-\beta H} \, e^{i J_f  A_f} \;,
\ee
where $J_f$ are the generators of the flavor symmetries. Because of the supersymmetry algebra $\cQ^2 =H-\sigma_f J_f$ (see for example (\ref{susyalgebra})), only the states with $H = \sigma_f J_f$ contribute, and $Z$ is actually holomorphic in $u_f$. Thus the partition function represents a twisted Witten index getting contributions from chiral states with energy proportional to the charge.

Upon dimensional reduction on $S^2$, we can compare our results with a recent computation of the Witten index of quantum mechanical sigma models via localization \cite{Hori:2014tda, Hwang:2014uwa, Cordova:2014oxa} and we find indeed many similarities. The three-dimensional nature of the original theory manifests itself in the existence of magnetic gauge fluxes on $S^2$, and makes the quantum mechanical interpretation of the result quite complicated.

The twisted partition function can be used as a new tool to investigate non-perturbative aspects of three-dimensional gauge theories. In this paper we consider several examples of Yang-Mills-Chern-Simons theories in 3d: we evaluate their twisted partition functions and compare them with general results about Chern-Simons theories and known three-dimensional dualities. In particular we provide further evidence for Aharony \cite{Aharony:1997gp} and Giveon-Kutasov \cite{Giveon:2008zn} dualities. 

There are various  generalizations of the setup. First, we can replace $S^2$ with a generic Riemann surface $\Sigma$. We do not discuss the higher genus case in details in this paper, but expressions for the one-loop determinants are explicitly given. Second, we can refine the index by the angular momentum on $S^2$: in the path-integral formulation this corresponds to turning on an $\Omega$-background on $S^2$. This more general formulation makes contact with the ``factorization'' of 3d partition functions \cite{Pasquetti:2011fj, Beem:2012mb, Cecotti:2013mba, Benini:2013yva} (as well as with a similar factorization in two \cite{Benini:2012ui, Doroud:2012xw} and four \cite{Yoshida:2014qwa, Peelaers:2014ima} dimensions).

Third, we can study two-dimensional gauge theories on $S^2$, and four-dimensional gauge theories on $S^2 \times T^2$, with A-twist on the sphere. There is a nice geometrical interpretation of what we are computing in various dimensions. In two dimensions, we compute the partition function of the topological A-model on $S^2$. For a gauged linear sigma model (GLSM), this is the same as the partition function of the non-linear sigma model (NLSM) to which it flows in the infrared (IR), when such a flow exists. In particular, the target of the NLSM is a holomorphic submanifold of the K\"ahler quotient realized by the GLSM, and the NLSM partition function is the equivariant Euler characteristic of the moduli space of holomorphic maps to the target. We can easily include local twisted chiral operators at arbitrary positions. Thus our formula provides an alternative new method to evaluate amplitudes in the topological A-model; some simple examples are provided in section \ref{sec: other dim}. We stress that, when applied to non-Abelian GLSMs, our formula does lead to new results.

When lifted to three dimensions, the same GLSMs (with no Chern-Simons terms) realize the quantum mechanics over the moduli space of holomorphic maps, therefore they compute the K-theoretic Euler characteristic. Finally, going to four dimensions one computes the elliptic genus of that moduli space.

We should mention similar results in the literature. The partition function of Chern-Simons-matter theories on generic Seifert manifolds has been evaluated in \cite{Ohta:2012ev} and reduced to a matrix model. The $S^2\times S^1$ case is a special case of Seifert manifold, but it is difficult to compare the results. Precisely in the case of  $S^2\times S^1$ there are additional fermionic zero-modes, which are instrumental in reducing the path-integral to a contour integral of a meromorphic form and in selecting the correct contour. As already mentioned, our formula for three-dimensional theories is formally similar to those obtained for the quantum mechanical Witten index in \cite{Hori:2014tda, Hwang:2014uwa, Cordova:2014oxa}. Analogously, our results for four-dimensional theories are formally similar to those obtained for the elliptic genus in \cite{Gadde:2013dda, Benini:2013nda, Benini:2013xpa}, and expressions for the one-loop determinants on $S^2 \times T^2$ have appeared in \cite{Closset:2013sxa, Nishioka:2014zpa}. Finally, the partition function of Chern-Simons theories with one adjoint on $\Sigma \times S^1$ has recently been computed in \cite{Gukov:2015sna}.

The paper is organized as follows. In section \ref{sec: loc} we derive our formula for the partition function by performing a supersymmetric localization. In section \ref{sec:examples} we provide examples in various Abelian and non-Abelian Yang-Mills-Chern-Simons-matter theories in 3d, and test our formula against known dualities. In section \ref{sec: refinement} we add a refinement for the angular momentum. In section \ref{sec: other dim} we discuss the generalizations of our formula to two and four-dimensions. Finally, in the appendices we give details about the computation of one-loop determinants. For the reader's convenience, we summarize our main finding in the next subsection: the reader not interested in its formal derivation, can read it and then safely jump to section \ref{sec:examples}.

\

\noindent
\textit{Note added:} while we were finishing our work, we became aware that some overlapping results have been obtained by Closset, Cremonesi and Park, and will appear in \cite{Closset:toappear}.

\subsection{The main result}
\label{sec: summary}

For an ${\cal N}=2$  gauge theory on $S^2\times S^1$ with gauge group $G$ of rank $r$ (and Lie algebra $\fg$), the topologically twisted path-integral localizes on a set of BPS configurations specified by a gauge magnetic flux on $S^2$, $\fm = \frac1{2\pi} \int_{S^2} F$, a flat connection%
\footnote{For disconnected groups, we should more properly talk about a holonomy $g = e^{i \oint A_t dt}$ along $S^1$.}
$A_t$ along $S^1$, and the value $\sigma$ of the real scalar in the vector multiplet, all mutually commuting. Up to gauge transformations,  the magnetic fluxes $\fm$ live in the co-root lattice $\Gamma_\fh$ of $G$ while the scalar zero-modes parameterize the connected components $\fM=H \times \fh$ of the BPS manifold, where $H$ is a maximal torus in $G$ and  $\fh$ is the corresponding Cartan subalgebra. Configurations connected by the Weyl group $W$ of $G$ are gauge-equivalent. It is convenient to combine the components of the zero-modes into the holomorphic Cartan combinations $u = A_t + i \beta\sigma$, where $\beta$ is the radius of $S^1$, and define $x = e^{iu}$. Here $u$ represents $r$ modes. For $G = U(1)$,  $u \simeq u+2\pi$ lives on a cylinder while $x\in\bC^*$. For a generic connected group $G$, $u \simeq u+2\pi \zeta$ where $\zeta$ is an element of the co-root lattice. 

The contribution of a chiral multiplet to the one-loop determinant is given by
\be
\label{chiral1-loop}
Z_\text{1-loop}^\text{chiral} = \prod_{\rho \in \fR} \Big( \frac{x^{\rho/2}}{1-x^\rho} \Big)^{\rho(\fm) - q +1}
\ee
where $\fR$ is the representation under the  gauge group $G$, $\rho$ the corresponding weights and $q$ the R-charge of the field. 
The contribution of a vector multiplet   to the one-loop determinant is instead given by
\be
\label{vector1-loop}
Z_\text{1-loop}^\text{gauge} =   \prod_{\alpha \in G} (1-x^\alpha) \; (i\, du)^r
\ee
where $\alpha$ are the roots of $G$. $Z^\text{gauge}_\text{1-loop}$ is naturally interpreted as a middle-dimensional holomorphic form on  $H\times \fh$. The classical action contributes a factor 
\be
\label{classical}
Z_\text{class}^\text{CS} = x^{k \fm} =\prod_{i=1}^r x_i^{k \fm_i}
\ee
where $k$ is the Chern-Simons coupling of $G$ (each Abelian and simple factor has its own coupling). For Abelian factors $G_1$ and $G_2$ in $G$, a mixed Chern-Simons coupling $k_{12} A_{(1)}\wedge F_{(2)}$ is possible and it contributes $x_1^{k_{12} \fm_2} x_2^{k_{12} \fm_1}$. 

The theory can have flavor symmetries: we denote the corresponding holonomies by $y = e^{iv}$ and the  magnetic fluxes by $\fn$. Then the 1-loop determinant is modified as
\be
x^\rho \to x^\rho \, y^{\rho_f} \;,\qquad\qquad \rho(\fm) \to \rho(\fm) + \rho_f(\fn) \;,
\ee
where $\rho_f$ is the weight under the flavor group. A $U(1)$ topological symmetry with holonomy $\xi =e^{i z}$ and flux $\ft$ contributes
\be
\label{topological}
Z_\text{class}^\text{top} = x^\ft \, \xi^\fm \;.
\ee

The contribution of the classical action and the one-loop determinant in each sector of magnetic flux $\fm$,
\be
Z_\text{int}(u; \fm)= Z_\text{class} \, Z_\text{1-loop} \;,
\ee
is a meromorphic $r$-form on $H \times \fh$. $Z_\text{int}(u; \fm)$ has pole singularities along the hyperplanes $e^{i\rho(u) + i \rho_f(v)} = 1$ determined by the chiral fields and goes to zero or infinity at the boundaries  of $H \times \fh$. The path-integral reduces to an $r$-dimensional contour integral of $Z_\text{int}(u; \fm)$,
\be
Z_{S^2 \times S^1} = \frac1{|W|} \; \sum_{\fm \,\in\, \Gamma_\fh} \; \oint_\cC \; Z_\text{int}(u; \fm) \;,
\ee
summed over all magnetic fluxes in the co-root lattice $\Gamma_\fh$. The contour $\cC$ is a specific sum of $r$-dimensional contours going around the hyperplane singularities or living at the boundary of $H \times \fh$.

For a $U(1)$ theory with chiral fields with charges $Q_i$ (and more generally when $r=1$), it is easy to specify the integration contour $\cC$.  The behavior at the two boundaries of $H \times \fh$ is determined by the effective CS coupling 
\be
\label{keff0}
k_\text{eff}(\sigma) = k+ \frac12 \sum_i Q_i^2 \sign( Q_i \sigma) \;.
\ee
The path-integral can be conveniently written as a sum of residues  of the meromorphic form $Z_\text{int}(u; \fm)$ in the $x=e^{i u}$ plane.  The two boundaries of $H \times \fh$  map to two circles around $x=0$ and $x=\infty$. We can use one of two equivalent prescriptions and sum either
\begin{itemize}
\item all residues at the poles created by fields with positive charge, the residue at $x=0$ if $k_\text{eff}(+\infty)<0$ and the residue at $x=\infty$ if $k_\text{eff}(-\infty)>0$; or
\item minus the residues at the poles created by fields with negative charge, minus the residue at $x=0$ if $k_\text{eff}(+\infty)>0$ and minus the residue at $x=\infty$ if $k_\text{eff}(-\infty)<0$.
\end{itemize}
This prescription can be written in a compact form by assigning charges to the boundaries at $x=0$ and $x=\infty$,
\be
\label{chargesboundary}
Q_{x=0} = - k_\text{eff}(+\infty) \;,\qquad\qquad Q_{x=\infty} = k_\text{eff}(-\infty) \;,
\ee
and using the Jeffrey-Kirwan residue defined as \cite{JeffreyKirwan, Brion1999, Szenes:2004}
\be
\JKres_{y=0} \big(Q, \eta \big) \; \frac{dy}y =  \theta( Q\eta) \sign(Q) \;.
\ee
Here $\eta\neq0$ is a parameter. The final formula for a $U(1)$ theory is then
\be
\label{U(1)formula}
Z_{S^2 \times S^1} = \sum_{\fm \in \bZ} \bigg[ \sum_{\; x_* \in\, \fM_\text{sing}} \hspace{-.2em} \JKres_{x=x_*} \big( Q_{x_*}, \eta \big) \; Z_\text{int} (x; \fm) \;+\; \JKres_{x=0,\infty} \big( Q_x, \eta \big) \; Z_\text{int}(x ; \fm) \bigg] \;,
\ee
where $\fM_\text{sing}$ is the set of singular points in $\fM$ where $Z_\text{int}$ has poles, and $Q_{x_*}$ is the charge of the chiral field creating the pole. To perform the computation one has to choose a parameter $\eta\neq 0$, but the result is independent of such a choice. The Jeffrey-Kirwan residue appears in a similar way in the localization of the elliptic genus for 2d theories \cite{Benini:2013nda, Benini:2013xpa} and of the Witten index in 1d \cite{Cordova:2014oxa, Hwang:2014uwa, Hori:2014tda}. 

For gauge groups of rank $r>1$, the choice of contour is more complicated. The path-integral is still  given by a sum of Jeffrey-Kirwan residues at a finite number of points in $H\times \fh$, where $r$ or more singular hyperplanes meet, plus a boundary contribution:
\be
\label{final formula summary}
Z_{S^2 \times S^1} = \frac1{|W|} \sum_{\fm \in \Gamma_\fh} \bigg[ \sum_{\;x_* \in\, \fM_\text{sing}^*} \hspace{-.2em} \JKres_{x=x_*} \big( \sQ(x_*), \eta \big) \; Z_\text{int} (x;\fm) \;+\; \text{boundary } \bigg] \;.
\ee
The precise form of the contour is discussed in section \ref{sec: higher rank}. However, in all examples considered in this paper, we will be able to extrapolate the Abelian formula to the non-Abelian case without really using the complicated machinery of section \ref{sec: higher rank}.

\section{Localization on $S^2 \times S^1$ with topological twist}
\label{sec: loc}

In this section we provide a path-integral derivation of the formul\ae{} (\ref{U(1)formula}) and (\ref{final formula summary}). The crucial technique is supersymmetric localization (see \eg{} \cite{Witten:1991zz, Pestun:2007rz} and \cite{Marino:2011nm} for a modern review) which allows us to \emph{exactly} reduce the path-integral $Z_{S^2 \times S^1}(t)$ to a finite-dimensional integral over a moduli space of BPS configurations $\cM_\text{BPS}$, where the measure is provided by the one-loop determinant $Z_\text{1-loop}$ of small quadratic fluctuations around those configurations. Schematically:
$$
Z_{S^2 \times S^1}(t) \,\equiv\, \int \cD\varphi\; e^{-S[\varphi;\, t]} \;\stackrel{\text{localization}}{=}\; \int_{\cM_\text{BPS}} \cD\varphi_0\; e^{-S[\varphi_0;\, t]} \, Z_\text{1-loop}[\varphi_0; t] \;.
$$
There $\{t\}$ is a collection of parameters of the theory on $S^2 \times S^1$, on which the path-integral depends.

More in details, the computation will be similar to the one performed in \cite{Benini:2013nda, Benini:2013xpa} for the path-integral evaluation of the elliptic genus of two-dimensional $\cN=(0,2)$ theories, and in \cite{Hori:2014tda, Hwang:2014uwa} for the Witten index of $\cN=2$ quantum mechanics. Localization provides a function to be integrated on the complex $u$-plane, with various poles corresponding to the matter fields. Because of the singularities of the integrand, we will need to use a clever regulator whose existence is naturally provided  by the off-shell multiplet of zero-modes. By integrating out the gaugino zero-modes we will reduce the integral to a contour integral. We also stress that we can have generic Wilson loop insertions at points on $S^2$ and wrapping $S^1$.

Accordingly, we first construct supersymmetry, supersymmetric actions and Wilson line operators on $S^2\times S^1$, we then study the relevant moduli space of BPS configurations $\cM_\text{BPS}$, and evaluate the ``on-shell'' action and the one-loop determinant of small quadratic fluctuations around them. It turns out that $\cM_\text{BPS}$ contains fermionic zero-modes as well as singular loci with extra bosonic zero-modes. With a suitable regulator, the two problems solve each other and we are left with a contour integral: this technical part occupies the second half of this section. For the sake of clarity, we present the derivation in the case of rank-one gauge groups first, and then move to the more intricate generic case. In section \ref{sec: refinement} we present a refined version of this computation, in which the spacetime geometry is a fibration of $S^2$ over $S^1$, that can be considered as the three-dimensional $\Omega$-background (the position of Wilson line operators will then be constrained).

\subsection{Lagrangian and supersymmetry transformations}

We start by writing the metric and the background fields that we need to turn on in order to preserve some supersymmetry. We write then the supersymmetry transformations corresponding to the topologically twisted theory and the supersymmetric Lagrangians for gauge and matter fields.

\subsubsection{The supersymmetric background}

We consider three-dimensional $\cN=2$ theories on $S^2 \times S^1$, where supersymmetry is preserved by a topological twist on $S^2$. The round metric on $S^2 \times S^1$ is
\be
ds^2 = R^2 \big( d\theta^2 + \sin^2\theta\, d\varphi^2 \big) + \beta^2 dt^2 \;,
\ee
where we take $t \simeq t+1$. We take vielbein $e^1 = R\, d\theta$, $e^2 = R\sin\theta\, d\varphi$, $e^3 = \beta\, dt$. To perform the topological twist, we turn on a background for the R-symmetry proportional to the spin connection:
\be
\label{Rcharge}
V = \frac12 \cos\theta\, d\varphi = - \frac12 \omega^{12} \;,
\ee
which corresponds to a flux $\frac1{2\pi} \int_{S^2} W = -1$ for the R-symmetry curvature $W=dV$. In our notation%
\footnote{We use gamma matrices: $\gamma_1 = \smat{0 & 1 \\ 1 & 0}$, $\gamma_2 = \smat{0 & -i \\ i & 0}$, $\gamma_3 = \smat{1 & 0 \\ 0 & -1}$.}
the supersymmetry spinor $\epsilon = \smat{\epsilon_+ \\ \epsilon_-}$ has R-charge $-1$ so that the Killing spinor equation $D_\mu \epsilon = \partial_\mu \epsilon + \frac14 \omega_\mu^{ab} \gamma_{ab} \epsilon +i V_\mu \epsilon=0$ is solved by 
\be
\label{cov}
\epsilon = \Big( \begin{array}{c} \epsilon_+ \\ 0 \end{array} \Big) \qquad\qquad\text{with}\qquad\qquad \epsilon_+ = \text{constant} \;.
\ee
Because of the R-symmetry background magnetic flux, we will restrict to theories with integer R-charges.
Notice that the same setup works with a generic metric on $S^2$, and when $S^2$ is replaced by a Riemann surface $\Sigma_g$ of arbitrary genus $g$, with the same choice of $V=- \frac12 \omega^{12}$ and the same covariantly constant spinor (\ref{cov}). In general the R-symmetry field strength is related to the scalar curvature by
\be
W_{12} = \frac12 \varepsilon^{\mu\nu} W_{\mu\nu} = - \frac14 R_s \qquad\quad\text{and}\quad\qquad \frac1{2\pi} \int_{\Sigma_g} W = g - 1 \;.
\ee

If the metric on $S^2$ has a $U(1)$ isometry, we can introduce a rotation of $S^2$ along the circle, which essentially gives the $\Omega$-background on $S^2\times S^1$, 
\be
\label{deformed metric}
ds^2 = R^2 \big( d\theta^2 + \sin^2\theta (d\varphi - \varsigma\, dt)^2 \big) + \beta^2 dt^2 \;,
\ee
where the coordinates have the same periodicity as before: $t \cong t+1$, $\varphi \cong \varphi + 2\pi$. We can still perform the topological twist with $V = - \frac12 \omega^{12}$ and the covariantly constant spinor (\ref{cov}). We  call this the ``refined'' case, and we discuss it in section \ref{sec: refinement}.

\subsubsection{Supersymmetry transformations and BPS equations}
\label{sec: SUSY algebra and eqns}

We use SUSY transformations in terms of commuting spinors and anticommuting supercharges (as given in appendix B.2 of \cite{Benini:2013yva}). The supersymmetry parameters are two positive-chirality covariantly constant spinors $\epsilon$, $\tilde\epsilon$ satisfying $D_\mu \epsilon = 0$, $\gamma_3\epsilon = \epsilon$ and similarly for $\tilde\epsilon$, with the same R-charge $-1$. Notice that, in fact, $\epsilon$ and $\tilde\epsilon$ are multiples of the unique covariantly constant spinor (\ref{cov}). 
The algebra has two complex supercharges $Q$, $\wt Q$ of vanishing R-charge.

We consider the following types of multiplets: vector multiplets $\cV = (A_\mu, \sigma, \lambda, \lambda^\dag, D)$ whose components in Lorentzian signature are a gauge field, two real scalars $\sigma,D$ and a Dirac spinor, all in the adjoint representation of the gauge group; chiral multiplets $\Phi = (\phi, \psi, F)$ whose components are two complex scalars $\phi, F$ and a Dirac spinor, all in a representation $\fR$ of the gauge group; anti-chiral multiplets $\Phi^\dag = (\phi^\dag, \psi^\dag, F^\dag)$ with the same components as a chiral multiplet, all in the conjugate representation $\wb\fR$. In Euclidean signature all fields are complexified, and $^\dag$-ed fields are not adjoints but rather independent fields.

The transformations of a  vector multiplet are:
\bea
\label{gaugemultiplet}
Q A_\mu &= \frac i2 \lambda^\dag \gamma_\mu \epsilon &
Q\lambda &= + \frac12 \gamma^{\mu\nu} \epsilon F_{\mu\nu} - D\epsilon + i \gamma^\mu \epsilon \, D_\mu\sigma \\
\wt Q A_\mu &= \frac i2 \tilde\epsilon^\dag \gamma_\mu \lambda &
\wt Q \lambda^\dag &= - \frac12 \tilde\epsilon^\dag \gamma^{\mu\nu} F_{\mu\nu} + \tilde\epsilon^\dag D + i \tilde\epsilon^\dag \gamma^\mu D_\mu\sigma \\
QD &= - \frac i2 D_\mu \lambda^\dag \gamma^\mu \epsilon + \frac i2 [\lambda^\dag \epsilon, \sigma ] \qquad\quad &
Q \lambda^\dag &= 0 \qquad\qquad\qquad
Q\sigma = - \frac12 \lambda^\dag \epsilon \\
\wt Q D &= \frac i2 \tilde\epsilon^\dag \gamma^\mu D_\mu \lambda + \frac i2 [\sigma, \tilde\epsilon^\dag\lambda] &
\wt Q \lambda &= 0 \qquad\qquad\qquad
\wt Q \sigma = - \frac12 \tilde\epsilon^\dag \lambda \;.
\eea
We have turned on the  background  (\ref{Rcharge})  for the R-symmetry and  used  the R-charge assignment  $(0,0,-1,1,0)$ for $(A_\mu,\sigma,\lambda,\lambda^\dag,D)$.%
\footnote{In particular, in our notation the ``standard'' gaugino with R-charge $+1$ is $\lambda^c$, the charge-conjugate to $\lambda$.}
For the chiral multiplet the supersymmetry transformations are%
\footnote{We define charge conjugate spinors $\epsilon^c = C \epsilon^*$ and  $\epsilon^{c\dag} = \epsilon^\trans C$, where $C$ is the charge conjugation matrix such that $C\gamma^\mu C^{-1} = - \gamma^{\mu\trans}$. We choose $C = \gamma_2$ so that $C = C^{-1} = C^\dag = - C^\trans = - C^*$. Moreover $\epsilon^{cc} = - \epsilon$.}
\bea
Q\phi &= 0 \qquad\quad & Q\psi &= \big( i\gamma^\mu D_\mu\phi + i \sigma\phi \big)\epsilon & \wt Q\psi &= \tilde\epsilon^c F \\
\wt Q\phi &= - \tilde\epsilon^\dag\psi \qquad & \wt Q\psi^\dag &= \tilde\epsilon^\dag \big( -i\gamma^\mu D_\mu \phi^\dag + i \phi^\dag \sigma \big) & Q\psi^\dag &= - \epsilon^{c\dag} F^\dag \\
Q\phi^\dag &= \psi^\dag\epsilon \qquad & QF &= \epsilon^{c\dag} \big( i \gamma^\mu D_\mu \psi - i \sigma\psi - i \lambda\phi \big) & \wt QF &= 0 \\
\wt Q\phi^\dag &= 0 \qquad & \wt Q F^\dag &= \big( -i D_\mu\psi^\dag \gamma^\mu - i \psi^\dag \sigma + i \phi^\dag \lambda^\dag \big) \tilde\epsilon^c \qquad\quad & QF^\dag &= 0 \;.
\eea
The R-charges of $(\phi,\psi,F)$ are $(q, q-1, q-2)$, and those of $(\phi^\dag, \psi^\dag, F^\dag)$ are the opposite.

This transformations realize the superalgebra $\su(1|1)$, whose bosonic subalgebra $\mathfrak{u}(1)$ generates rotations of $S^1$ mixed with gauge/flavor rotations:
\be
\label{susyalgebra}
\{Q, \wt Q\} = -i \cL^A_v - \delta_\text{gauge}(\tilde \epsilon^\dag\epsilon\, \sigma) \;,\qquad\qquad Q^2 = \wt Q^2 = 0 \;,\qquad\qquad v_\mu = \tilde\epsilon^\dag \gamma_\mu \epsilon \;.
\ee
Here $\cL^A_v$ is the gauge-covariant Lie derivative (including the R-symmetry connection) along the covariantly constant (and Killing) vector field $v^\mu$.%
\footnote{The explicit expression of the Lie derivative of fields of various spins can be found in appendix B.1 of \cite{Benini:2013yva} and appendix A.2 of \cite{Closset:2014pda}.}
Using the flat basis $e^a$, in the unrefined case one finds $v = \beta^{-1} \tilde\epsilon^\dag\epsilon \, \partial_t$, while in the refined case $v = \beta^{-1} \tilde\epsilon^\dag\epsilon (\partial_t + \varsigma \partial_\varphi)$. In order to perform localization, we will use the supercharge
\be
\cQ = Q + \wt Q \;,
\ee
which behaves as an equivariant differential: $\cQ^2 = -i \cL^A_v - \delta_\text{gauge}(\tilde \epsilon^\dag\epsilon\, \sigma)$.

\subsubsection{Supersymmetric Lagrangians}
\label{sec: SUSY Lag}

We now proceed with the construction of supersymmetric Lagrangians on $S^2 \times S^1$. We consider Yang-Mills-Chern-Simons theories with matter, therefore we construct the Yang-Mills action, the various Chern-Simons terms, the matter kinetic action and superpotential interactions. Whenever the theory has some continuous flavor symmetry $G_F$, we couple it to an external vector multiplet and turn on background values for the bosonic fields therein. This corresponds to magnetic flavor fluxes on $S^2$, flat flavor connections on $S^1$, and real masses. We recall that whenever the gauge group has an Abelian factor, the flavor group includes a ``topological'' $U(1)$ subgroup (possibly enhanced to a larger subgroup in the IR quantum theory).

We work in Euclidean signature, therefore all fields get complexified. However, when performing the path-integral, we have to choose a middle-dimensional contour in field space. We choose the ``natural'' one, in which ``real'' fields are real while $^\dag$ is identified with the adjoint operation: we call such a contour the \emph{real contour}. We have chosen conventions in which all Lagrangian terms have a non-negative real bosonic part, ensuring convergence of the path-integral.

The supersymmetric Yang-Mills (YM) Lagrangian is
\be
\label{YM}
\cL_\text{YM} = \Tr\bigg[ \frac14 F_{\mu\nu} F^{\mu\nu} + \frac12 D_\mu\sigma D^\mu\sigma + \frac12 D^2 - \frac i2 \lambda^\dag \gamma^\mu D_\mu\lambda - \frac i2 \lambda^\dag [\sigma,\lambda] \bigg] \;.
\ee
One can verify that
\be
Q \wt Q \Tr \Big( \tfrac12 \lambda^\dag\lambda + 2D\sigma \Big) \,\cong\, \tilde\epsilon^\dag\epsilon\, \cL_\text{YM}
\ee
up to total derivatives, therefore the YM action is also $\cQ$-exact.

The supersymmetric Chern-Simons (CS) Lagrangian is, for each simple or Abelian factor:
\be
\label{CS}
\cL_{CS} = - \frac{ik}{4\pi} \Tr \bigg[ \epsilon^{\mu\nu\rho} \Big( A_\mu \partial_\nu A_\rho - \frac{2i}3 A_\mu A_\nu A_\rho \Big) + \lambda^\dag \lambda + 2D\sigma \bigg] \;.
\ee
In general one can have a different CS level $k$ for each simple or Abelian factor in the gauge group, however we will often be schematic with our notation and use the simple expression above. The CS action is supersymmetric but not $\cQ$-exact. If there are multiple Abelian factors in the gauge group, one can introduce mixed CS terms between them:
\be
\label{L mixed CS}
\cL_\text{mCS} = - \frac{ik_{12}}{2\pi} \bigg[ \epsilon^{\mu\nu\rho} A_\mu^{(1)} \partial_\nu A_\rho^{(2)} + \frac12 \lambda^{(1)\dag} \lambda^{(2)} + \frac12 \lambda^{(2)\dag}\lambda^{(1)} + D^{(1)}\sigma^{(2)} + D^{(2)}\sigma^{(1)} \bigg] \;.
\ee

The mixed  CS terms play an important role in turning on background fluxes or holonomies for the topological symmetries. Recall that in three dimensions, any $U(1)$  gauge symmetry gives rise to a global symmetry 
whose current  is given by $J_T^\mu = (*F)^\mu = \frac12 \epsilon^{\mu\nu\rho} F_{\nu\rho}$. The current  is automatically conserved by the Bianchi identity, $d*J_T=dF=0$, and the corresponding global symmetry $U(1)_T$ is called
topological. In order to turn on a background gauge field $A^{(T)}$ for the topological symmetry, we couple it though
\be
\label{top}
\int A^{(T)} \wedge * J_T = \int d^3x\, \sqrt g\, \epsilon^{\mu\nu\rho} A_\mu^{(T)} \partial_\nu A_\rho \;.
\ee
Here $A^{(T)}$ belongs to an external vector multiplet whose other bosonic components are $\sigma^{(T)}$ and $D^{(T)}$. In order to have a supersymmetric background, we need to set to zero the variation of the fermions in the external multiplet. From (\ref{gaugemultiplet}) we obtain that we should set $D^{(T)} = i F_{12}^{(T)}$, while $\sigma^{(T)}$ can be an arbitrary constant. The full action is  the supersymmetric completion of (\ref{top}), which is obtained from (\ref{L mixed CS}) by taking $k_{12}=1$ and regarding $(1)$ as  the background topological symmetry and $(2)$ as the gauge symmetry:
\be
\label{L topological}
\cL_\text{T} = - i \frac{A_3^{(T)}}{2\pi} \Tr F_{12} - i \frac{F_{12}^{(T)}}{2\pi} \Tr (A_3 + i\sigma) - i \frac{\sigma^{(T)}}{2\pi} \Tr D \;.
\ee
The three terms are separately supersymmetric. We see that $\sigma^{(T)}$ (a real mass for the topological symmetry) is in fact a Fayet-Iliopoulos (FI) term. 

We can also consider a mixed CS term between the R-symmetry and an Abelian flavor (or gauge) symmetry:
\be
\label{RCS}
\cL_\text{RCS} = -\frac{ik_R}{2\pi} \Big( \epsilon^{\mu\nu\rho} A_\mu \partial_\nu V_\rho + i \sigma W_{12} \Big) \;.
\ee
This term is the specialization to our background of the term in (4.19) of \cite{Closset:2012vp}.

The supersymmetric matter kinetic action is
\be
\label{matter}
\cL_\text{mat} = D_\mu \phi^\dag D^\mu\phi + \phi^\dag \big( \sigma^2 + iD - q W_{12} \big) \phi + F^\dag F + i \psi^\dag ( \gamma^\mu D_\mu -\sigma ) \psi - i \psi^\dag \lambda\phi + i \phi^\dag \lambda^\dag \psi \;,
\ee
where $q$ is the R-charge of $\phi$. One can verify that
\be
Q \wt Q \big( \psi^\dag \psi + 2i \phi^\dag \sigma \phi \big) \,\cong\, \tilde\epsilon^\dag\epsilon\, \cL_\text{mat}
\ee
up to total derivatives, therefore the matter kinetic action is also $\cQ$-exact.

Superpotential interactions are controlled by a holomorphic function $W(\Phi)$, gauge-invariant and of R-charge 2. The two Lagrangians
\be
\cL_W = i F_W \;,\qquad\qquad \cL_{\wb W} = i F_W^\dag \;,
\ee
where
\be
F_W = \parfrac{W}{\Phi_i} F_i - \frac12\, \parfrac{^2W}{\Phi_i \partial\Phi_j} \psi_j^{c\dag} \psi_i\pdag \;,\qquad\qquad F_W^\dag = \parfrac{\wb W}{\Phi_i^\dag} F_i^\dag - \frac12\, \parfrac{^2 \wb W}{\Phi_i^\dag \partial\Phi_j^\dag} \psi_j^\dag \psi_i^c
\ee
are the F-terms of the chiral multiplet $W(\Phi)$ and its antichiral partner, are separately supersymmetric. Since $\cQ \big( i \epsilon^{c\dag} \psi_W \big) \cong \tilde\epsilon^\dag\epsilon\, \cL_W$ and $\cQ\big( {-i \psi_W^\dag \tilde\epsilon^c} \big) \cong \tilde\epsilon^\dag\epsilon\, \cL_{\wb W}$ up to total derivatives, the two Lagrangians lead to $\cQ$-exact actions. Notice that although in Euclidean signature the two functions $W$ and $\wb W$ can be independent, since we consider the Wick rotation of real Lorentzian Lagrangians, we take them complex conjugate.%
\footnote{If we stay off-shell, the matter kinetic action has the positive-definite term $F^\dag F$, while the real bosonic part of $\cL_W + \cL_{\wb W}$ vanishes. On the other hand, if we integrate out $F_i,F^\dag_i$ to go on-shell, we obtain the positive-definite term $\sum_i \big| \partial_i W \big|^2$.}

The covariant derivatives in (\ref{matter}) contain the gauge fields, the background field $V$ for the R-symmetry and background fields  for the flavor symmetries of the theory.
A background vector multiplet for a flavor symmetry contains the bosonic components $F_{12}^F$, $A_3^F$, $\sigma^F$ and $D^F$ which should satisfy $D^F = i F_{12}^F$ in order to preserves supersymmetry. We see that $F_{12}$ represents a background magnetic flux for the flavor symmetry, $A_3$ is a flat flavor connection along $S^1$, and $\sigma^F$ is a real mass associated with the flavor symmetry.

Finally, we can include Wilson lines in representation $R$ defined as
\be
\label{def Wilson loop}
W = \Tr_R \Pexp \oint d\tau \big( iA_\mu \dot x^\mu - \sigma \, | \dot x| \big)
\ee
as in \cite{Kapustin:2009kz}. Here $x^\mu(\tau)$ is the worldline of the loop, $\tau$ is a parameter on it, $\dot x^\mu$ is the derivative with respect to $\tau$ and $| \dot x|$ is the length of $\dot x^\mu$. Its supersymmetry variation is
\be
Q W \;\propto\; - \frac12 \lambda^\dag \gamma_\mu \epsilon \, \dot x^\mu + \frac12 \lambda^\dag \epsilon \, |\dot x| \;.
\ee
One gets $QW=0$ (and $\wt QW = 0$) if $\dot x^1 = \dot x^2 = 0$, \ie{} if the loop is along the vector field $e_3$. In the unrefined case $e_3 = \frac1\beta \partial_t$: we can place the loop at an arbitrary point on $S^2$ and along $t$. In the refined case $e_3 = \frac1\beta(\partial_t + \varsigma \partial_\varphi)$ and $x^\mu(\tau) = (\theta_0, \varsigma\tau, \tau)$, therefore for irrational values of $\varsigma$ the loop does not close; we can either tune $\varsigma$ to rational values, or place the loop at one of the two poles of $S^2$.

\subsection{Localization on $S^2\times S^1$}
\label{sec: localization}

To compute the path-integral $Z_{S^2 \times S^1}(t)$, we use the localization method. We deform the action $S[\varphi;t] \to S[\varphi;t] + \mathsf{u} \cQ V[\varphi]$, where $\mathsf{u}$ is a positive parameter while $V$ has non-negative real bosonic part and $\cQ^2V=0$. By the standard argument, the path-integral does not depend on $\mathsf{u}$. Evaluating in the $\mathsf{u}\to +\infty$ limit, the path-integral localizes around configurations for which the real bosonic part of $\cQ V[\varphi]$ vanishes. Therefore $Z_{S^2 \times S^1}$ reduces to a semi-classical computation around those configurations.

\subsubsection{The BPS equations}

As a deformation Lagrangian we choose $\cL_\text{YM} + \cL_\text{mat}$ in (\ref{YM}) and (\ref{matter}), since each term leads to a non-negative $\cQ$-exact action. Let us start with the gauge sector. Setting to zero the real bosonic part of $\cL_\text{YM}$ along the real contour, one gets
$$
F_{\mu\nu} = D_\mu \sigma = D = 0 \;.
$$
On the other hand, without imposing any reality condition, the BPS equations $\cQ\lambda = \cQ \lambda^\dag = 0$ lead to a much larger set of complexified configurations:
$$
D = i F_{12} \;,\qquad D_1\sigma = iF_{13} \;,\qquad D_2\sigma = iF_{23} \;,\qquad D_3\sigma = 0 \;.
$$
When evaluating a standard integral in the saddle-point approximation, it is common that saddle points in the complex plane and off the original integration contour contribute to the integral; therefore we might worry that something similar happens here. In fact, with a very careful analysis, we will see that it does. It will be convenient to define $\wt \cM_\text{BPS}$ as the space of BPS configurations where the reality condition is applied to all physical fields but not to the auxiliary field $D$:
\be
\label{general BPS config}
\wt\cM_\text{BPS} = \Big\{ D = i F_{12} \;,\quad F_{13} = F_{23} = 0 \;,\quad D_\mu \sigma = 0 \Big\}  / \cG \;,
\ee
where $\cG$ is the infinite dimensional group of gauge transformations.

These equations are easily solved. Let us choose a gauge $\partial_t A_t = 0$. Let  $g \in G$ be the holonomy around $S^1$, which may depend on the position on $S^2$. The Bianchi identity implies $D_t F_{12} = 0$. Single-valuedness of $F_{12}$ along $S^1$ implies $[g,F_{12}] = 0$. This in turn implies that $F_{12}(x)$ is constant on $S^1$, $g$ is constant on $S^2$, and they commute. In particular, we can represent $F_{12}$ by a connection on $S^2$ that is constant on $S^1$ and, if $g$ is connected to the identity, we can represent $g$ by a connection on $S^1$ that is constant on $S^2$, and they commute.%
\footnote{If it is not connected, one has to introduce a discrete element.}
Integrating $D_3\sigma=0$ along $S^1$ we get $[g,\sigma]=0$ and $\sigma$ is constant on $S^1$. Finally we have to solve $D_\mu\sigma=0$ on $S^2$, which implies $[F_{12},\sigma]=0$ at all points on $S^2$, and $\sigma$ is constant on $S^2$.

Summarizing, the equations are solved by 
\be
[g,F_{12}(x)] = [g,\sigma] = [\sigma, F_{12}(x)] = 0 \;,
\ee
where $g$ and $\sigma$ are constant, while $F_{12}$ may depend on $S^2$ but is constant along $S^1$. This space $\wt \cM_\text{BPS}$ is infinite-dimensional.

If we further restrict $F_{12}$ to be constant (we will show how that comes out of the path-integral), the moduli space reduces to
\be
\label{M_BPS}
\cM_\text{BPS} = \big( H \times \fh \times \Gamma_\fh \big) / W \;,
\ee
where $H$ is a maximal torus in $G$, $\fh$ is the Cartan subalgebra, $\Gamma_\fh \subset \fh$ is the co-root lattice of $G$ that parameterizes quantized fluxes, and $W$ is the Weyl group.

Let us now move to the matter sector. The BPS equations along the real contour give
\be
D_3\phi = 0 \;,\qquad\qquad \sigma\phi = 0 \;,\qquad\qquad (D_1+iD_2)\phi = 0 \;,\qquad\qquad F=0 \;.
\ee
These equations generically imply $\phi=0$. However they admit extra nontrivial solutions when $\sigma$ and $e^{iA_t}-\unit$ have a common zero eigenvalue: the extra solutions are then zero-modes of $D_1 + i D_2$ on $S^2$ (or more generally on $\Sigma_g$).

\subsubsection{The zero-modes}
\label{sec: zero-modes}

Around each of the general complex BPS configurations (\ref{general BPS config}) there are bosonic and fermionic zero-modes of the Yang-Mills and matter actions. 

For generic configurations, only the YM action has zero-modes and they are finite in number. The bosonic modes parameterize the constant diagonal values of $\sigma$ and $A_t$, and describe the connected submanifolds
\be
\fM = H \times \fh
\ee
of the BPS manifold (to be divided by $W$ together with $\Gamma_\fh$). Since the observable gauge quantity is $e^{i\oint A_t dt}$, we define the dimensionless complex combinations
\be
u = A_t + i \beta\sigma = \beta\big( A_3 + i\sigma \big) \;,\qquad \bar u = A_t - i \beta\sigma \;,\qquad x = e^{iu}\, .
\ee
We call $u$ a complexified flat connection.
Notice that $u$ represents $r$ modes, where $r$ is the rank of the gauge group. For a $U(1)$ group,  $u \simeq u+2\pi$ lives on a cylinder while $x\in\bC^*$. For a generic group $G$, $u \simeq u+2\pi \zeta$ where $\zeta$ is an element of the co-root lattice. Configurations related by the Weyl group have to be identified. We parameterize magnetic fluxes by
\be
\frac1{2\pi} \int_{S^2} F = \fm \;,
\ee
where $\fm \in \Gamma_\fh$ satisfy $e^{2\pi i \fm} = \unit_G$, \ie, in physical terms, they are GNO quantized \cite{Goddard:1976qe}.

Besides the bosonic zero-modes, there are also fermionic zero-modes and together they form complete supermultiplets. Each bosonic zero-mode is paired with a fermionic zero-mode coming from the Cartan gaugini. The Cartan gaugini $\lambda$ are not lifted because they are charged only under the R-symmetry, and we cannot turn on a flat connection for the R-symmetry without breaking supersymmetry. In fact  the fermionic zero-modes have the same quantum numbers as $\epsilon$ and are proportional to it; we can thus define scalar fermionic zero-modes
\be
\lambda_0 = \beta\, \epsilon^\dag\lambda \;,\qquad\qquad \lambda_0^\dag = \beta\, \lambda^\dag \epsilon \;,
\ee
which are obtained from $Q \bar u = i \lambda_0^\dag$, $\wt Q \bar u = i \lambda_0$. We can close the supersymmetry algebra ``off-shell'' if we introduce an auxiliary bosonic zero-mode $D_0$:
$$
D_0 = \beta\, \epsilon^\dag\epsilon \, (D - i F_{12}) \;.
$$
This is obtained from $\wt Q \lambda_0^\dag = D_0$ or $Q \lambda_0 = - D_0$. This zero-mode corresponds to a \emph{constant} profile for $D-iF_{12}$. In the following we will choose a normalization $\beta\, \epsilon^\dag\epsilon=1$ for the zero-modes. Notice that, as usual, setting to zero the auxiliary component gives BPS configurations. The supersymmetry algebra closes and we find 
\bea
Qu&=0 \qquad& Q \bar u &= i\lambda_0^\dag \qquad& Q \lambda_0 &= - D_0 \qquad& Q\lambda_0^\dag &= 0 \qquad& QD_0 &= 0 \\
\wt Qu &= 0 \qquad& \wt Q\bar u &= i\lambda_0 \qquad& \wt Q \lambda_0 &= 0 \qquad& \wt Q \lambda_0^\dag &= D_0 \qquad& \wt Q D_0 &= 0 \;.
\eea
Notice that $Q^2 = \wt Q^2 = \{Q,\wt Q\} = 0$ on the zero-mode subspace, since the zero-modes are translationally invariant and commute with $\sigma$. We will  call ``almost BPS'' the configurations which satisfy the BPS conditions, except for a constant $D_0$---in other words $D = iF_{12} + D_0$.

All chiral multiplets give rise to bosonic and fermionic Landau levels on $S^2$, however such potential zero-modes are charged under flavor and gauge symmetries, therefore the complexified flat connections on $S^1$ generically lift them. There are, however, special hyperplanes (linear submanifolds of complex codimension $1$) on $\fM$ where the total flat connection has zero eigenvectors, and extra zero-modes appear. In particular each chiral multiplet $\Phi_i$ can give rise to a hyperplane $H_i$ such that when $u\in H_i$, $\Phi_i$ develops a bosonic zero-mode. The manifolds $H_i$ are determined by the poles of the one-loop determinant and are of the form
\be
H_i = \big\{ u \in \fM \,\big|\, e^{i\rho_i(u) + i\rho_f(v)} = 1 \big\} \;,
\ee
where $\rho_i$ is the weight of the gauge representation, $\rho_f$ of the flavor representation, and $v$ is the complexified holonomy for the flavor group. To each hyperplane we associate the covector $Q_i \equiv \rho_i \in \fh^*$ equal to the gauge weight (in the Abelian case, $Q_i$ is a list of charges).
We call
\be
\label{def M sing}
\fM_\text{sing} = \bigcup\nolimits_i H_i
\ee
the singular manifold and, as we will see, we will only be interested in $\fM \setminus \fM_\text{sing}$.

In the rank-one case, $H_i$ are just isolated points on $\fM$. For $r>1$, instead, $H_i$ are proper hyperplanes. We define $\fM_\text{sing}^* \subset \fM_\text{sing}$ the set of isolated points in $\fM$ where at least $r$ linearly independent hyperplanes meet:
\be
\label{def M sing *}
\fM_\text{sing}^* = \big\{ u_* \in \fM \,\big|\, \text{at least $r$ linearly independent $H_i$'s meet at $u_*$} \big\} \;.
\ee
Given $u_* \in \fM_\text{sing}^*$, we denote by $\sQ(u_*)$ the set of charges of the hyperplanes meeting at $u_*$:
\be
\sQ(u_*) = \big\{ Q_i \,\big|\, u_* \in H_i \big\} \;.
\ee
For a technical reason as in \cite{Benini:2013nda, Benini:2013xpa}, we will assume the following condition: \emph{for any $u_* \in \fM_\text{sing}^*$, the set of charges $\sQ(u_*)$ is contained in a half-space in $\fh^*$}. Notice that if the number of hyperplanes at $u_*$ is exactly $r$, this condition is automatically met.

\subsubsection{The classical on-shell action}
\label{sec: on-shell}

We evaluate the classical action $\cZ_\text{cl}(u,\bar u, D_0;\fm)$ on almost-BPS configurations, where $u$ is constant, $D = iF_{12} + D_0$ with constant $D_0$, but $D$ and $F_{12}$ are not necessarily constant. The moduli $u$, $\bar u$, $D_0$, $\fm$ control the expectation values of the bosonic fields in vector multiplets, which can be either dynamical (gauge group $G$) or external (flavor group $G_F$), in the Cartan subalgebra. There can also be global topological symmetries, whose moduli are denoted by $\xi = e^{iz}$ and $\ft$ with $z = A_t^{(T)} + i \beta \sigma^{(T)}$.

The classical action terms in the Abelian case are the following.
\begin{itemize}
\item The Abelian YM action (\ref{YM}) contributes
\be
\cZ_\text{cl}^\text{YM}  = e^{- \frac{1}{e^2} \int d^3x \sqrt g\, \cL_\text{YM} |_\text{on-shell}} = e^{- \frac{2\pi\beta}{e^2}  ( R^2 D_0^2 + i \fm D_0) } \;.
\ee
As $S_\text{YM}$ is $\cQ$-exact, it vanishes on actual BPS configurations where $D_0 = 0$.

\item The Abelian  Chern-Simons action (\ref{CS}) contributes
\be
\label{CS on-shell}
\cZ_\text{cl}^\text{CS}  = e^{-  \int d^3x \sqrt g\, \cL_\text{CS} |_\text{on-shell}} = x^{k\fm} \, e^{2ik \beta R^2 \sigma D_0} \;,
\ee
where $x=e^{iu}$.
Notice that if $k$ is half-integer, this is not a single-valued function of $x$. This is related to the fact that a half-integer CS level might be required to cancel a parity anomaly from the matter sector.

\item The mixed CS term between two Abelian symmetries (\ref{L mixed CS}) contributes
\be
\label{Mclass}
\cZ_\text{cl}^\text{mCS} = e^{-  \int d^3x \sqrt g\, \cL_\text{mCS} |_\text{on-shell}}  = x_1^{k_{12} \fm^{(2)}} x_2^{k_{12} \fm^{(1)}} \, e^{2ik_{12} \beta R^2 (\sigma^{(2)}D_0^{(1)} + \sigma^{(1)} D_0^{(2)} )} \;.
\ee
If one of the two Abelian symmetries is flavor, we drop the corresponding $D_0$ term.

\item The topological  term (\ref{L topological}) contributes
\be
\label{Tclass}
\cZ_\text{cl}^\text{T} = e^{-  \int d^3x \sqrt g\, \cL_\text{T} |_\text{on-shell}} = x^\ft \, \xi^\fm \, e^{2i \beta R^2 \sigma^{(T)} D_0} \;.
\ee
We recall that $\im z$ is essentially a FI term, while $\re z$ is a sort of 2d $\theta$-angle on $S^2$.

\item The mixed CS term between the R-symmetry and an Abelian gauge or flavor symmetry (\ref{RCS})  contributes
\be
\cZ_\text{cl}^\text{RCS}  = e^{-  \int d^3x \sqrt g\, \cL_\text{RCS}  |_\text{on-shell}} = x^{-k_{1R}} \;.
\ee
\end{itemize}
The previous expressions are straightforwardly generalized to the non-Abelian case by replacing  $u, D_0$ and $\fm$ with elements of the Cartan subalgebra $\fh$ and contracting all products with the Killing form. The total classical action $\cZ_\text{cl}$ is the product of the relevant terms.

Let us also evaluate the Wilson loop defined in (\ref{def Wilson loop}). Using $x^\mu(\tau) = (\theta_0, \varphi_0, \tau)$ and the fact that $A_t, \sigma$ are constant, the loop reduces to $W = \Tr_R \exp\big[ i\oint dt\, (A_t + i\beta\sigma) \big]$. Therefore
\be
W = \Tr_R e^{iu} = \Tr_R x = \sum\nolimits_{\rho \in R} x^\rho \;.
\ee

Notice that the classical action for $D_0=0$, which we denote as
\be
Z_\text{cl}(u;\fm) \equiv \cZ_\text{cl}(u,\bar u, 0;\fm)
\ee
is a holomorphic function of $u$.

\subsubsection{One-loop determinants}
\label{sec: one-loop}

Next we compute the one-loop determinants from chiral and vector multiplets, obtained by integrating out all their non-zero modes and keeping the dependence on $D_0$ which serves as a regulator of the final expression.%
\footnote{While the determinants at $D_0 = 0$ can be computed, with a cohomological method in appendix \ref{sec: locA}, for generic metric on $S^2$ and generic supersymmetric $F_{12}$, for $D_0 \neq 0$ we use the round metric on $S^2$ and take constant $F_{12}$.}
The derivation is given in appendix \ref{sec: locA}.

\paragraph{Chiral multiplet.}
Consider a chiral multiplet $\Phi$ transforming in some representation $\fR$ of the gauge and flavor symmetry group $G \times G_F$, and immersed in a \emph{constant} magnetic flux $\fm$ on $S^2$ along the Cartan subalgebra. Consider a single component $\Phi_\rho$, transforming as the weight $\rho \in \fR$ and with R-charge $q_\rho$ (to weights in different irreducible representations we can assign different R-charges). We write
\be
b \,\equiv\, \rho(\fm) - q_\rho
\ee
for the total flux seen by the scalar component of $\Phi_\rho$. The bosonic determinant is given by
\be
\label{one-loop chiral bosons}
\det \cO_\phi = \prod_{n\geq 0} \prod_{k\in\bZ} \bigg[ \frac{ (2n+1) |b| - b + 2n(n+1)}{2R^2} + \rho(\sigma)^2 + \frac{\big( 2\pi k - \rho(A_t)\big)^2}{\beta^2} + i \rho(D_0) \bigg]^{2n + |b|+1}  \;.
\ee
Notice that for non-vanishing values of $\rho(\sigma)$ or $\rho(A_t)$, the zeros of $\det \cO_\phi$ are strictly in the half-plane $\im \rho(D_0)>0$ in the complex $D_0$-plane. The fermionic determinant is given by
\begin{multline}
\label{one-loop chiral fermions}
\det \cO_\psi = \prod_{k\in\bZ} \Big[ i s \frac{2\pi k - \rho(A_t)}\beta - \rho(\sigma) \Big]^{|b+1|} \times \\
\times \prod_{n\geq 1} \bigg[ \frac{ n \big( n+|b+1| \big)}{R^2} + \rho(\sigma)^2 + \frac{ \big( 2\pi k - \rho(A_t) \big)^2}{\beta^2} \bigg]^{2n + |b+1|}
\end{multline}
where $s = \sign(b+1)$. The one-loop determinant is the ratio of the fermionic and bosonic determinants,
\be
\cZ_\text{1-loop}^\text{chiral}(u,\bar u, D_0; \fm) = \frac{\det \cO_\psi}{\det \cO_\phi}
\ee
and in general it is a function of $u$ and $\bar u$. For generic $D_0$ there are no cancelations among the massive modes. 

We will be eventually interested in the one-loop determinant for $D_0=0$. In that case all massive modes cancel out and we are left with
\be
\label{1-loop chiral before reg}
\frac{\det \cO_\psi}{\det \cO_\phi} = \prod_{k\in\bZ} \Big( \frac i\beta \Big)^{-b-1} \Big( \rho(A_t + i\beta \sigma) - 2\pi k \Big)^{-b-1} \;.
\ee
This expression requires regularization. We use $\prod_{k\in\bZ} \alpha (u - 2\pi k) = -2i \sin \frac u2$, where the prefactor $\alpha$ is irrelevant, and find
$$
\frac{\det \cO_\psi}{\det \cO_\phi} = \Big( -2i \sin \frac{\rho(u)}2 \Big)^{-\rho(\fm) + q_\rho -1} = \Big( \frac{x^{\rho/2}}{1-x^\rho} \Big)^{\rho(\fm) - q_\rho +1} \;,
$$
where $x = e^{iu}$ and $x^\rho \equiv e^{i\rho(u)}$. This determinant is a meromorphic function of $u$, and we denote it by $Z_\text{1-loop}^\text{chiral}(u;\fm) \equiv \cZ_\text{1-loop}^\text{chiral}(u,\bar u, 0; \fm)$. It has a simple Hamiltonian interpretation, and our normalization was chosen accordingly. The magnetic flux on $S^2$ generates Landau levels, which in the quantum mechanics on $S^1$ are either $|b+1|$ Fermi multiplets for $b+1<0$, or $b+1$ chiral multiplets for $b+1>0$ (a similar phenomenon has been recently discussed in \cite{Almuhairi:2011ws, Kutasov:2013ffl}). In the first case, the Fermi multiplet contains a spinor whose Hilbert space is a fermionic Fock space, and assigning charge $-\frac\rho2$ and fermion number $0$ to the vacuum, the index is $x^{-\frac\rho2} - x^\frac\rho2$. In the second case, the chiral multiplet contains a complex scalar $\phi$ and a spinor, and the Hilbert space is the product of a bosonic Fock space generated by $\phi,\phi^\dag$ and a fermionic Fock space; assigning fermion number $1$ to the vacuum, the index is $(-x^{-\frac\rho2} + x^\frac\rho2) \sum_{n\geq 0} x^{n\rho} \sum_{m \geq 0} x^{-m\rho} = (x^{-\frac\rho2} - x^\frac\rho2)^{-1}$.

Eventually, taking into account all weights $\rho$ of the representation $\fR$, we obtain:
\be
\label{one-loop 1d chiral}
\boxed{
Z_\text{1-loop}^\text{chiral} = \prod_{\rho \in \fR} \Big( \frac{x^{\rho/2}}{1-x^\rho} \Big)^{\rho(\fm) - q_\rho +1} \;.
}
\ee
Notice  that this one-loop determinant may have singularities when $x^\rho=1$, corresponding to the presence of bosonic zero-modes.
Notice also that (\ref{one-loop 1d chiral}) may not be single-valued in $x$: this is a manifestation of the parity anomaly and, when $x$ parameterizes gauge flat connections, must be canceled by a choice of half-integer CS level in (\ref{CS on-shell}).

At this point we would like to comment on a sign ambiguity, which originates from the ambiguity in the quantization of the fermionic Fock space (the Ramond vacuum). The determinant (\ref{one-loop 1d chiral}) gets contributions from fermionic chiral zero-modes on $S^2$. The fermionic operators $\psi_0, \bar\psi_0$ satisfy the anticommutation algebra $\psi_0^2 = \bar\psi_0^2 = 0$, $\{\psi_0, \bar\psi_0\} = 1$, so we interpret them as creation and annihilation operators for a fermionic Fock space $|\pm\rangle$ with $\bar\psi_0 |-\rangle = 0$, $\psi_0 |-\rangle = |+\rangle$. The two states have opposite fermion number, and flavor charges that differ by $\rho$. We could decide that $|-\rangle$ is the vacuum---a bosonic neutral state: this leads to a determinant $1-x^\rho$. However the two states have the same energy, therefore we could rather decide that $|+\rangle$ is the vacuum, leading to $-x^{-\rho} + 1$. A more democratic choice is to assign the two states flavor charges $\pm\frac\rho2$, leading to $x^{-\rho/2} - x^{\rho/2}$, which is our choice. These ambiguities correspond precisely to ambiguities in the regularization of the determinant. Even with our democratic choice, we are still left with a sign ambiguity in the assignment of the fermion number, leading to an ambiguity $(-1)^{\rho(\fm)}$ for the determinant. If $\fm$ is for a global symmetry and so it is a fixed parameter, this is just a total sign ambiguity of the index; but if $\fm$ is for a gauge symmetry and so it is summed over, it may appear that it drastically affects the partition function. However in three dimensions, if the gauge group is Abelian the ambiguity can be reabsorbed in a redefinition of the fugacity $\xi$ for the associated topological symmetry, see (\ref{Tclass}); if instead the gauge group is semi-simple the ambiguity cancels out. Because of this ambiguity, in our examples we will choose the signs at our convenience.

When we integrate out a chiral multiplet with large real mass $M$, we obtain an effective shift of the Chern-Simons levels of all groups the chiral multiplet is charged under \cite{Redlich:1983kn, Redlich:1983dv, Aharony:1997bx}. Let us check that this is reproduced by the one-loop determinant. Consider, for example, a chiral multiplet charged under many different $U(1)$s.
A real mass $M$ can be turned on by giving an expectation value to the scalar $\sigma^F$ in the flavor vector multiplet that rotates the chiral multiplet.  
The one-loop determinant of such a chiral multiplet, which we take of R-charge $1$ for simplicity,%
\footnote{For R-charge different from $1$, the fermions are charged under the R-symmetry and a mixed gauge/R CS term is generated. Such term is correctly reproduced by the one-loop determinant formula.}
becomes
\be
\label{largeM}
Z^\text{chiral}_\text{1-loop} = \bigg( \frac{\prod_a x_a^{\rho_a/2} e^{-\beta M/2}}{1-\prod_a x_a^{\rho_a} e^{-\beta M} } \bigg)^{\sum_b \rho_b \fm_b} \;\xrightarrow[M\to\pm\infty]{}\; \prod_{a,b} x_a^{\frac12 \sign(M) \, \rho_a \rho_b \fm_b} \; \big( \sign(M) \, e^{-\frac\beta2 |M|} \big)^{\sum_c \rho_c \fm_c} \;.
\ee
Comparing with (\ref{Mclass}), we recognize in the first term a shift of the $U(1)$s CS levels,
\be
\delta k_{ab} = \frac12 \rho_a \rho_b \sign(M) \;,
\ee
which precisely reproduces the known result  \cite{Aharony:1997bx}. The sign ambiguity in (\ref{largeM}) can be reabsorbed  in the fugacity $\xi$ of the topological symmetries---see (\ref{Tclass})---and the exponential in $M$ is a renormalization.

For a simple group, the same computation reproduces the shift
\be
\delta k = \frac12 T_2(\fR) \sign(M) \;,
\ee
where $T_2(\fR)$ is the quadratic index of $\fR$ defined by $\sum_{\rho\in\fR} \rho^a \rho^b = T_2(\fR)\, K^{ab}$ in terms of the Killing form $K^{ab}$ (and $h= \frac12 T_2(\text{adj})$ is the dual Coxeter number). For instance for $SU(2)$, $T_2(\text{spin }I) = 2I(I+1)(2I+1)/3$.

\paragraph{Higher genus.}
If we place the theory on a Riemann surface $\Sigma_g$ of genus $g$, instead of on the sphere $S^2$, we can still compute the one-loop determinant (see appendix \ref{sec: locA}). The only difference is the number of units of R-symmetry flux: $\frac1{2\pi} \int_{\Sigma_g} W = g-1$. By the index theorem, the number of right-moving minus left-moving modes on $\Sigma_g$ is $n_R - n_L = \rho(\fm) + (g-1)(q_\rho-1)$, therefore the one-loop determinant is
\be
Z_\text{1-loop}^\text{chiral} = \prod_{\rho\in\fR} \Big( \frac{x^{\rho/2}}{1-x^\rho} \Big)^{\rho(\fm) + (g-1)(q_\rho-1)} \;.
\ee
Notice in particular that this is independent of the flat-connection moduli on $\Sigma_g$.

\paragraph{The gauge multiplet.}
The one-loop determinant for the gauge multiplet can be computed in many ways. On the round sphere with constant magnetic flux we could compute it mode by mode as in \cite{Benini:2012ui, Doroud:2012xw}. In the general case, we could use the cohomological argument of appendix \ref{sec: locA} along the lines of \cite{Hama:2011ea}. More quickly, we notice that the modes along the Cartans are not charged under gauge or flavor symmetries, and so can be discarded. The off-diagonal modes contribute, possibly up to a flux-dependent sign, as chiral multiplets with R-charge 2 and transforming as the roots $\alpha$ of the gauge group. This can be understood from the Higgs mechanism. Suppose that a generator $\alpha$ is broken, then the gauge field will eat a chiral multiplet and become massive. The eaten chiral multiplet has no flavor charges, it transforms as $\alpha$ under the gauge group and it has R-charge zero (otherwise its VEV would break some global symmetry). Since massive fields do not contribute, we have $Z^\text{gauge}_\text{1-loop} Z^\text{chiral}_\text{1-loop} = 1$, up to a flux-dependent sign. This equation is also satisfied by two chiral multiplets of R charge two and zero which can by paired by a quadratic superpotential term and integrated out. This determines $Z^\text{gauge}_\text{1-loop}$.

Moreover, because of the bosonic zero-modes, it is natural to interpret $Z^\text{gauge}_\text{1-loop}$ as a middle-dimensional holomorphic form on $\fM$, therefore we attach the differential $d^ru$ to it, where $r$ is the rank of the gauge group. We thus have:
\be
\label{one-loop 1d vector}
\boxed{
Z_\text{1-loop}^\text{gauge} = (-1)^{2\delta(\fm)} \prod_{\alpha\in G} \bigg( \frac{x^{\alpha/2}}{1-x^\alpha} \bigg)^{\alpha(\fm)-1} \; (i\, du)^r = \prod_{\alpha \in G} (1-x^\alpha) \; (i\, du)^r \;.
}
\ee
Here $\delta = \frac12 \sum_{\alpha>0} \alpha$ is the Weyl vector,%
\footnote{Recall that $2\delta$ is always a weight, therefore $2\delta(\fm) \in \bZ$. For semi-simple groups also $\delta$ is a weight, but this is not true for Abelian factors. For instance for $U(N)$: $(-1)^{2\delta(\fm)} = (-1)^{(N-1) \sum_i \fm_i}$.}
and we have fixed the sign ambiguity for later convenience. We see that $Z_\text{1-loop}^\text{gauge}$ is just the Haar measure for the group $G$. The measure does not have any divergence: this is related to the fact that there are no flat connections on $S^2$.

\

We define the total classical and one-loop contribution as
\be
Z_\text{int}(u;\fm) =Z_\text{cl} \; Z_\text{1-loop}^\text{chiral} \; Z_\text{1-loop}^\text{gauge} \;,
\ee
which is a holomorphic $r$-form. 

\paragraph{Higher genus.} We can similarly write the gauge one-loop determinant on a Riemann surface of genus $g$:
\be
Z^\text{gauge}_\text{1-loop} = (-1)^{2\delta(\fm)} \prod_{\alpha \in G} \bigg( \frac{x^{\alpha/2}}{1-x^\alpha} \bigg)^{\alpha(\fm) + g - 1} \; (i\, du)^r = \prod_{\alpha \in G} (1-x^\alpha)^{1-g} \; (i\, du)^r \;.
\ee
This time there are singularities associated to $Z^\text{gauge}_\text{1-loop}$. This is expected: when $x^\alpha = 1$, there is enhanced non-Abelian gauge symmetry and there are extra bosonic zero-modes parameterizing the flat connections on $\Sigma_g$, which are associated to poles of the determinant.

\subsubsection{Asymptotic behavior}

In the following sections we will need the asymptotic behavior of the one-loop determinant for large values of the moduli $\sigma$, which, as we saw around (\ref{largeM}), is related to a one-loop shift of the Chern-Simons levels. In an ${\cal N}=2$ $U(1)$ theory with chiral multiplets of charges $Q_i$ and Chern-Simons coupling $k$, we can define an effective Chern-Simons coupling
\be
\label{keff}
k_\text{eff}(\sigma) = k+ \frac12 \sum_i Q_i^2 \sign( Q_i \sigma)
\ee
as a function of the vacuum expectation value of the scalar $\sigma$  \cite{Aharony:1997bx}. The shift comes from integrating out the matter  fermions which have mass $|Q_i \sigma|$. 

The correction in (\ref{keff}) is reflected in the asymptotic behavior of the one-loop determinant for a chiral field. The bare CS term contributes like in (\ref{CS on-shell}), $x^{k \fm} \, e^{2ik\beta R^2 \sigma D_0}$, while the one-loop determinant for a field of charge $Q_i$ provides for large $|\sigma|$:
\be
\label{correction}
x^{\frac12 Q_i^2 \sign(Q_i\sigma) \, \fm} \; e^{ i Q_i^2 \sign(Q_i\sigma) \, \beta R^2 \sigma D_0} \;.
\ee
The two contributions combine into
$$
x^{k_\text{eff}(\sigma) \, \fm} \, e^{2ik_\text{eff}(\sigma) \, \beta R^2 \sigma D_0}
$$
and precisely reproduce the correction in (\ref{keff}). 

To see how this works, we need to study the  asymptotic behavior of the one-loop determinant as $\sigma \to \pm\infty$ for generic values of $D_0$. For $b>1$, the determinant for a chiral field of charge $1$ is
\be
\label{totone}
\frac{\det \cO_\psi}{\det \cO_\phi} = \prod_{k\in\bZ} \frac1{\big( \frac{2\pi k - A_t}{i\beta} - \sigma \big)^{b+1}} \prod_{n\geq 0} \Bigg( \frac{ \frac{n(n+b+1)}{R^2} + \sigma^2 + \frac{(2\pi k - A_t)^2}{\beta^2} }{ \frac{n(n+b+1)}{R^2} + \sigma^2 + \frac{(2\pi k - A_t)^2}{\beta^2} + i D_0 } \Bigg)^{2n+b+1} \;,
\ee
while for generic charge we simply have to reinstate $Q_i$ in front of $A_t$, $\sigma$, $D_0$ and $\fm$.
The second product becomes $1$ when $D_0=0$, and the first product, after regularization, is the determinant we found in (\ref{one-loop 1d chiral}). 
The limit of (\ref{one-loop 1d chiral}) for large $|\sigma|$ produces the first factor in (\ref{correction}). Consider now the second factor in (\ref{totone}). Its product over $k$ is convergent and can be performed explicitly: calling $F$ its product over $k$ and $n$, we find
\be
F = \prod_{n\geq0} f(n)^{2n+b+1} \;,\quad f(n) = \frac{\cosh\big( \beta \sqrt z \big) - \cos A_t }{ \cosh\big( \beta \sqrt{ z + i D_0} \big) - \cos A_t} \;,\quad z = \frac{n(n+b+1)}{R^2} + \sigma^2 \;.
\ee
We are interested in the behavior of $F$ for $|\beta\sigma| \gg 1$. In this limit we have
\be
\log f = \beta \sqrt z - \beta \sqrt{z + i D_0 } + \cO \big( e^{-\beta\sqrt z} \big) \;.
\ee
It follows that $\log F$ is a linearly divergent sum over $n$. The divergent term can be computed with $\zeta$-function regularization: $- \sum_{n\geq 0} i\beta R D_0 = \frac i2 \beta R D_0$.  We can approximate the remaining convergent sum over $n$ with an integral:
$$
\beta \int_0^\infty \hspace{-.15cm} dn \Big[ (2n + b + 1)\big( \sqrt z - \sqrt{z+i D_0} \big) + i R D_0 \Big] = i \beta R^2 |\sigma| D_0 - \frac i2 (b+1) \beta R D_0 + \cO\big( \tfrac{\beta R^2 D_0^2}\sigma \big) \;.
$$
Reinstating the charge $Q_i$, we finally find
\be
F = \exp\bigg[ i\beta R^2 \sign(Q_i\sigma) \, Q_i^2 \sigma D_0 - \tfrac i2 \beta R b Q_i D_0 + \cO\big( \tfrac{\beta R^2 D_0^2}\sigma \big) \bigg] \;.
\ee
The first dominant term gives the second factor in (\ref{correction}). A similar computation works for $b<1$. 

For a  general theory we can have mixed Chern-Simons terms and the expression in (\ref{keff}) is replaced by
\be
\label{keff2}
k_\text{eff}^{ab}(\sigma) = k^{ab}+ \frac12 \sum_{i,c} Q_i^a Q_i^b \sign( Q_i^c \sigma_c) \;,
\ee
where the indices $a,b,c$ run over the generators of the Abelian gauge groups, $i$ runs over the different matter fields and $Q_i^a$ are the gauge charges. 
The correction (\ref{keff2}) is correctly reproduced by the asymptotic behavior of the one-loop determinant.

\subsection{The final formula: rank-one case}
\label{sec: rank one}

The last step is to integrate the classical contribution and the one-loop determinant over the moduli space of BPS configurations, taking properly into account the various zero-modes. We follow the strategy used in  \cite{Benini:2013nda, Benini:2013xpa}. There are some new features related to the non-compactness of the moduli space and the presence of magnetic fluxes. In order to clearly explain the physical ideas, we first consider the case of rank-one gauge groups.

\subsubsection{The integral and the dangerous regions}
\label{sec: dangerous regions}

We place $1/\se^2$ in front of the Yang-Mills Lagrangian $\cL_\text{YM}$ and $1/\sg^2$ in front of the matter Lagrangian $\cL_\text{mat}$, and consider localization as $\se,\sg \to 0$. For non-zero couplings $\se, \sg$, the path-integral takes the form of an integral over the supermanifold of vector multiplet bosonic and fermionic zero-modes. We can write it as
$$
Z  = \int_\fM d^2u \; F_{\se,\sg}(u, \bar u) \;,
$$
where the integration is over the bosonic zero-modes---the moduli space $\fM \cong \bC/2\pi$ of flat connections---while $F_{\se,\sg}$ is the result of the path-integral over the fermionic zero-modes and all other massive modes (we will soon see that this expression is not complete).

There are some dangerous regions in $\fM$ when we take $\se\to0$ and/or $\sg \to 0$. The dangerous regions are the points $u_* \in \fM_\text{sing}$ defined in (\ref{def M sing}) where, in the $\se\to0$ limit, extra scalar zero-modes from chiral multiplets appear. Suppose that for $u \sim u_*$ there are $M$ quasi-zero-modes $\phi_i$, whose charges $Q_i$ have---by assumption---the same sign. Then the integral over the modes looks like
$$
I = \int d^{2M}\phi\, \exp\bigg[ - \frac1{\sg^2} \sum\nolimits_i |Q_i(u - u_*)|^2 |\phi_i|^2 - \frac{\se^2}2 \Big( \zeta_\text{eff} - \sum\nolimits_i \frac{Q_i}{\sg^2} |\phi_i|^2 \Big)^2 \bigg] \;,
$$
where $\zeta_\text{eff}$ is the effective FI term at the point $u$.
Here $\sg$ can be reabsorbed in the measure for the quasi-zero-modes, therefore $\sg\to0$ does not pose any problem. On the contrary, the second term comes from the D-term potential and it ensures that the integral is convergent, even at $u=u_*$, therefore taking the limit $\se\to0$ is problematic. Let us find an upper bound on $|I|$ at small but fixed $\se$. As a function of $u$, $|I|$ is maximized at $u=u_*$. By rescaling $\phi_i \to \phi_i \big| \frac{\sg^2}{Q_i e} \big|^{1/2}$, we obtain the bound
$$
|I| \leq \frac{\sg^{2M}}{\se^M \prod_i |Q_i|} \int d^{2M}\phi\, \exp\bigg[ - \frac12 \Big( \zeta_\text{eff} \, \se \sign(Q_i) - \sum_i |\phi_i|^2 \Big)^2 \bigg] \;.
$$
In the limit that $\se$ is small, we can neglect the term in $\zeta_\text{eff}$ and the integral can be performed:
\be
|I| \lesssim \frac{C}{\se^M} \;,\qquad\qquad\qquad C = \frac{\sg^{2M}}{\prod_i |Q_i|} \,\frac{2^{\frac{M-2}2} \pi^M \Gamma(M/2)}{\Gamma(M)} \;.
\ee
So, taking the limit of $F_{\se,0}(u,\bar u)$ as $\se \to 0$ at $u=u_*$ is problematic, since we remove the quartic potential and generate illusive singularities. The resolution is the same as in \cite{Benini:2013nda}. We first remove from $\fM$ an $\varepsilon$-neighborhood $\Delta_\varepsilon$ of $\fM_\text{sing}$ and split the integral in two pieces:
\be
Z = \int_{\fM\setminus\Delta_\varepsilon} d^2u\, F_{\se,0}(u,\bar u) + \int_{\Delta_\varepsilon} d^2u\, F_{\se,0}(u,\bar u) \;.
\ee
The second integral is bounded by $\varepsilon^2 /\se^M$ up to constants, therefore in a scaling limit $\se, \varepsilon \to 0$ such that $\varepsilon^2/\se^M\to0$ as well, the second term does not contribute. We thus have
\be
\label{Z nf limit formula}
Z = \lim_{\se, \varepsilon\to0} \int_{\fM\setminus\Delta_\varepsilon} d^2u\, F_{\se,0}(u,\bar u) \;.
\ee

With respect to \cite{Benini:2013nda} we have to be more careful, though, because $\fM$ is non-compact. This is similar to the setup in \cite{Hori:2014tda, Hwang:2014uwa, Cordova:2014oxa}. Since $\fM \cong \bC/2\pi$ is a cylinder, we might have a problem when integrating over the zero-mode $\im du = \beta \, d\sigma$. Let us estimate the behavior of the integral. For $\sigma\to \pm\infty$, all chiral multiplets are massive and their effect is to shift the bare CS level $k$ as in (\ref{keff}). Let us call  $k_\pm$ the values of the effective CS for $\sigma\to \pm\infty$. The dangerous part of the integral is then, after integrating out $D$:
$$
\int d\sigma\, \exp\Big[ - \frac{\se^2}2 \big( k_\pm \sigma + \zeta \big)^2 \Big] \;,
$$
where some unimportant constants have been dropped.
When $k_\pm\neq 0$, the integral is convergent for any $\se\neq 0$ but becomes singular in the limit $\se\rightarrow 0$. The resolution is again to remove an $\varepsilon$-neighborhood of infinity in $\fM$ by considering a large number $L(\varepsilon)$ and including in $\Delta_\varepsilon$ the two regions $|\sigma| \geq L$.  Consider the integral on the region $\sigma \geq L$ (the case $\sigma \leq -L$ is equivalent). For $\se\to0$ we can neglect $\zeta$, and we are left with
$$
I \simeq \int_L^\infty d\sigma\, \exp\Big[ - \frac{\se^2 k_+^2}2 \sigma^2 \Big] = \frac1{\se |k_+|} \int^\infty_{\se |k_+| L} e^{-z^2/2} dz \;.
$$
Since $\int_x^\infty e^{-z^2/2}dz = \frac{e^{-x^2/2}}x \big( 1 + \cO(x^{-2}) \big)$, it is sufficient to take a scaling limit $\se, \frac1L\to 0$ such that $\se L$ grows as a negative power of $\se$, then the integral over $\sigma  \geq L$ does not contribute. In fact, we will take a stronger scaling limit in which $\se^2 L$ diverges as $\se \to 0$.
When $k_\pm = 0$ the integral is potentially divergent. The trick we will employ is to introduce a Lagrangian term $i\kappa_\text{reg} \sigma D$ and take the limit $\kappa_\text{reg}\to 0^\pm$. We will verify that the result is the same for the two limits, and it is finite.

In conclusion, by using the convention that we include in  the definition of $\Delta_\varepsilon$ also the two regions at infinity  and we take a suitable scaling limit,
 the path-integral is still given by (\ref{Z nf limit formula}).

\subsubsection{Configurations with flux}

There is another important difference with respect to the elliptic genus computation in \cite{Benini:2013nda} and the quantum-mechanical index in \cite{Hori:2014tda, Hwang:2014uwa, Cordova:2014oxa}. In those cases, the superalgebra  fixes $F_{\mu\nu} = D = 0$ on BPS configurations independently of the real contour chosen, while in our case the superalgebra allows generic $D(x) = i F_{12}(x)$ for complex $D(x)$. It is well-known that the saddle-point approximation to an integral along the real line can get contributions from saddle points away from the real line; therefore, let us investigate whether configurations with flux contribute.

Consider a generic real configuration $F_{12}(x)$ and $D(x)$. As long as $\se>0$, this configuration is suppressed by the classical action weight
\be
\label{suppression factor}
e^{-\frac1{\se^2} S_\text{YM}} = e^{-\frac1{2\se^2} \int d^3x \sqrt g\, \smath( F_{12}^2 + D^2 \smath)} \;.
\ee
This configuration is not BPS. However if $F_{12}(x)$ and $D(x)$ are actually \emph{constant}, then the configuration is on the complex orbit of the auxiliary zero-mode $D_0$ originating from the BPS configuration $D(x) = i F_{12}(x) = \frac{i\fm}{2R^2}$. In other words,
$$
D(x) = D_0 + \frac{i\fm}{2R^2} \,\in\,\bR \qquad\text{ for }\qquad D_0 \,\in\, \bR - \frac{i\fm}{2R^2} \;.
$$
We have computed the effective action $\cZ(u, \bar u, D_0; \fm)$ for the multiplet of zero-modes, obtained by integrating out all massive modes, around generic BPS configurations with complex $D(x)$. Such an action depends on the constant mode $D_0$, and it is valid for all $D_0 \in \bC$. We thus learn that configurations with constant $F_{12}$ are special because, starting from the BPS point and taking $D_0 \in \bR - \frac{i\fm}{2R^2}$, we reach the real contour $D(x)\in\bR$ we are integrating over, even though such real configurations are no-longer BPS. Let us analyze the contribution of these almost-BPS configurations to see whether it vanishes in the limit $\se \to 0$ or not.

The contribution of configurations with flux $F_{12} = \frac\fm{2R^2}$ can be written as
$$
Z_\fm = \cN \lim_{\se,\varepsilon\to0} \int_{\fM \setminus\Delta_\varepsilon} \hspace{-1em} d^2u \int_{\bR - \frac{i\fm}{2R^2}} \hspace{-1em} dD_0 \int d\lambda_0^\pdag \, d\lambda_0^\dag\; \cZ\big( u, \bar u, \lambda_0^\pdag, \lambda_0^\dag, D_0; \fm \big) \;,
$$
where $\cN$ is a normalization constant we will fix later.
Here $\cZ(u, \bar u, \lambda_0^\pdag, \lambda_0^\dag, D_0; \fm)$ is the effective action for all zero-modes (including the fermionic ones) in the multiplet, and it is the result of integration over all massive modes. We will analyze this function more in details in the next subsection. To compute the integral, we shift the $D_0$ integration contour along the imaginary axis until it reaches the real axis. When we do that, we can encounter poles of $\cZ$ located at the zeros of $\det\cO_\phi$ in (\ref{one-loop chiral bosons}), and we should pick the residues. However such residues are weighted by the suppression factor (\ref{suppression factor}):
$$
\re S_\text{YM} = - 2\pi \beta R^2 \im D_0 \big( \im D_0 + \tfrac{\fm}{R^2} \big) > 0 \qquad\text{ for }\qquad -\frac{\fm}{R^2}  < \im D_0  < 0 \;,
$$
using the fact that the poles are at $\re D_0 = 0$. Therefore, all these residues are suppressed in the limit $\se \to 0$ and we can neglect them.%
\footnote{As we reduce $\varepsilon$, we cross a larger number of poles. However the number of poles is polynomial, while the suppression factor is exponential.}
Once the contour has been shifted to the real axis, the result is no longer exponentially suppressed by (\ref{suppression factor}), and therefore it survives in the limit $\se\to 0$.

What about all other configurations where $F_{12}(x)$ is not constant? The corresponding BPS configurations have imaginary D-term $D(x) = i F_{12}(x)$, and the complexified orbit of the auxiliary zero-mode $D_0$ spans $D(x) = i F_{12}(x) + \bC$. If $F_{12}(x)$ is not constant, then the orbit never intersect the real contour $D \in \bR$ and the BPS configurations do not play a r\^ole in the exact saddle-point approximation to the real path-integral.

To summarize, the full path-integral reduces to a sum/integral over the bosonic moduli space $\cM_\text{BPS}$ in (\ref{M_BPS}) of BPS configurations with constant magnetic flux,
\be
\cM_\text{BPS} = \big( \fM \times \Gamma_\fh \big) / W \;,
\ee
as well as an integral over the fermionic zero-modes.

\subsubsection{Reduction to a contour integral}

Eventually, the expression for the path-integral that we need to evaluate is
\be
Z_{S^2 \times S^1} = \frac1{|W|} \sum_{\fm \in \Gamma_\fh} Z_\fm \;,
\ee
where $Z_\fm$ is the contribution from configurations with constant flux $F_{12} = \frac\fm{2R^2}$, the sum is over the co-root lattice, and $|W|$ is the order of the Weyl group. In particular
\be
\label{Z_m starting exp}
Z_\fm = \frac{i}{2\pi^2} \lim_{\se, \varepsilon\to 0} \int_{\fM\setminus\Delta_\varepsilon} \hspace{-1em} d^2u \int_{\bR + i \eta} \hspace{-1em} dD_0 \int d\lambda_0^\pdag \, d\lambda_0^\dag \; \cZ(u,\bar u, \lambda_0^\pdag, \lambda_0^\dag, D_0; \fm) \;,
\ee
the normalization has been fixed comparing with one example, and $ \cZ(u, \bar u, \lambda_0^\pdag, \lambda_0^\dag, D_0; \fm)$ is the effective action for the complete multiplet of zero-modes, obtained by integration over the massive modes around configurations with flux $\fm$. Setting $\lambda_0^\pdag = \lambda_0^\dag = 0$ we recover the  classical and one-loop expressions discussed in sections \ref{sec: on-shell} and \ref{sec: one-loop}:
$$
\cZ(u,\bar u, 0, 0, D_0;\fm) \equiv \cZ(u,\bar u, D_0;\fm) = \cZ_\text{cl} \; \cZ_\text{1-loop}^\text{gauge} \; \cZ_\text{1-loop}^\text{chiral} \;,
$$
while setting $D_0=0$ we obtain the holomorphic expression $\cZ(u,\bar u, 0;\fm) \equiv \cZ(u;\fm)$. The function $\cZ$ is holomorphic in $D_0$ around the origin as long as $u\not\in \Delta_\varepsilon$. Therefore we have the freedom to shift the real integration contour on the complex $D_0$-plane along the imaginary direction, as long as this shift is small: in (\ref{Z_m starting exp}) we have called $\eta$ such a shift.

The action $\cZ(u,\bar u, \lambda_0^\pdag, \lambda_0^\dag, D_0; \fm)$ depends on the gaugino zero-modes because of the Lagrangian couplings $\lambda\psi \phi$ to the matter fields we have integrated out. The dependence on $\lambda_0$ and $\lambda_0^\dag$ could be determined by an explicit computation as in  \cite{Benini:2013nda, Benini:2013xpa},  but we can use a shortcut exploiting supersymmetry. The integration over the fermionic zero-modes gives 
$$
\int d\lambda_0^\pdag \, d\lambda_0^\dag \; \cZ(u,\bar u, \lambda_0^\pdag, \lambda_0^\dag, D_0;\fm)  =  \parfrac{}{\lambda_0^\pdag} \parfrac{}{\lambda_0^\dag} \cZ(u,\bar u, \lambda_0^\pdag, \lambda_0^\dag, D_0;\fm) \Big|_{\lambda_0^\pdag = \lambda_0^\dag = 0} \;. 
$$
This expression can be simplified using the fact that $\cZ$ is supersymmetric. From
\be
\label{SUSY of Z}
0 = Q\cZ = \Big( i \lambda_0^\dag \parfrac{}{\bar u} - D_0 \parfrac{}{\lambda_0} \Big) \cZ \;,\qquad\qquad
0 = \wt Q\cZ = \Big( i \lambda_0 \parfrac{}{\bar u} + D_0 \parfrac{}{\lambda_0^\dag} \Big) \cZ \;,
\ee
it follows that
\be
D_0 \parfrac{}{\lambda_0^\pdag} \parfrac{}{\lambda_0^\dag} \cZ \Big|_{\lambda_0^\pdag = \lambda_0^\dag = 0} = - i \parfrac{}{\bar u} \cZ\Big|_{\lambda_0^\pdag = \lambda_0^\dag = 0} \;.
\ee
We can thus write%
\footnote{We use $d^2u = \frac i2 du \wedge d\bar u$ and $\partial ( \fM\setminus\Delta_\varepsilon)  = - \partial \Delta_\varepsilon$.}
\be
\label{fin}
 Z_\fm = \frac1{2\pi^2} \lim_{\se, \varepsilon\to 0} \int_{\fM\setminus\Delta_\varepsilon} \hspace{-1.3em} d^2u \int_{\bR + i \eta} \hspace{-.2em} \frac{dD_0}{D_0} \, \parfrac{\cZ(u,\bar u, D_0; \fm)}{\bar u}= \frac{i}{4\pi^2} \lim_{\se,\varepsilon\to0} \int_{\partial\Delta_\varepsilon} \hspace{-.8em} du \int_{\bR + i\eta} \hspace{-.2em} \frac{dD_0}{D_0} \, \cZ(u,\bar u, D_0; \fm) \;.
\ee
The same expression was found  in a similar context in \cite{Benini:2013nda, Benini:2013xpa}. The higher-rank generalization is discussed  in section \ref{sec: higher rank}.
  
The second expression in (\ref{fin}) seems to have a pole at $D_0 = 0$, however there is no pole in the first expression because $\partial_{\bar u} \cZ(u,\bar u, 0;\fm)=0$. In fact each separate connected component of $\partial\Delta_\varepsilon$ gives rise to a pole, while their sum does not. Let us consider each of them separately.
\begin{itemize}
\item Consider a component of $\partial\Delta_\varepsilon$ around a point $u_* \in \fM_\text{sing}$. Suppose that we have chosen $\eta>0$. From the unregularized expression (\ref{one-loop chiral bosons}) of the denominator $\det\cO_\phi$ of the chiral one-loop determinant, we see that the poles in the $D_0$-plane are at
$$
\rho(D_0) = i \rho(\sigma)^2 + i \frac{ \big( \rho(A_t) - 2\pi k \big)^2}{\beta^2} + i C' \;,
$$
where $C'$ is non-negative, vanishing only for $n=0$ and $b\geq 0$.

If $\rho<0$ the poles are in the negative half-plane. As $\varepsilon \to 0$, the poles for $n=0$ collapse towards $D_0 = 0$ (because $|u| \sim \varepsilon$ on the contour $\partial\Delta_\varepsilon$), however the contour $\bR + i \eta$ is safely far from them. The $D_0$-integral remains finite as $u \to 0$, and then the $u$-integral vanishes because its contour shrinks. On the contrary, if $\rho > 0$ the poles are in the upper half-plane and, as $\varepsilon \to 0$, they would cross the contour $\bR + i \eta$. To avoid that, we shift the contour to $\bR - i \eta$ and we collect minus the residue at $D_0 = 0$. As before, the integral along $\bR - i\eta$ does not yield any contribution as $\varepsilon \to 0$. Minus the residue at $D_0 = 0$, though, gives
$$
\frac1{2\pi} \lim_{\se,\varepsilon\to0} \int_{\partial\Delta_\varepsilon} \hspace{-.2cm} du\; \cZ(u, \bar u, 0; \fm) =  i \Res_{u=u_*} \cZ(u;\fm) \;,
$$
since $\cZ(u,\bar u,0;\fm)$ is holomorphic in $u$ and there is no dependence on $\se$ anymore. Suppose, instead, that we have chosen $\eta<0$. A similar argument goes through, and we obtain minus the residue at $u=u_*$ if $\rho<0$, zero if $\rho>0$.

We reach the conclusion, as in \cite{Benini:2013nda}, that for $\eta>0$ we collect the residues of $\cZ(u;\fm)$ at the points $u_* \in \fM_\text{sing}^+$ corresponding to chiral fields with positive charges, while for $\eta < 0$ we collect minus the residues at the points $u_* \in \fM_\text{sing}^-$ corresponding to chiral fields with negative charges.
This operation is called the Jeffrey-Kirwan residue \cite{JeffreyKirwan}:
\be
Z_\fm^\text{bulk} = \sum_{u_* \in \,\fM_\text{sing}} \JKres_{u = u_*}\big( \sQ(u_*), \eta \big) \; \cZ(u;\fm)  \, i\, du \;,
\ee
where $\sQ(u_*)$ is the set of charges of the fields responsible for the pole of $\cZ(u;\fm)$ at $u_*$. We can rewrite the expression in the $x$-plane:
\be
Z_\fm^\text{bulk} = \sum_{x_* \in \,\fM_\text{sing}} \JKres_{x = x_*} \big( \sQ(x_*),\eta \big) \; \cZ(x; \fm )  \frac{dx}x \;.
\ee

\item Consider the two components of $\partial\Delta_\varepsilon$ around $\im u = \pm \infty$. They give a contribution $Z_\fm^\text{bdy}$ as in (\ref{fin}), with 
\be
\cZ(u,\bar u, D_0; \fm)  \simeq  \exp\bigg[ -\frac{2\pi \beta R^2}{\se^2} D_0 \Big( D_0 +  \frac{i \fm}{R^2} \Big) + 2i \beta R^2 k_\pm \sigma D_0 + 2i \beta R^2 \sigma^T D_0 \bigg] \, \cZ(u;\fm)
\ee
for large $|\im u|$. The three terms come from (\ref{suppression factor}), (\ref{CS on-shell}) and (\ref{Tclass}), respectively.
We have  used that the asymptotic behavior of the one-loop determinants shifts the bare CS level $k$ as in (\ref{keff}), and we have denoted
\be
k_\pm = k_\text{eff}( \pm \infty )
\ee
the effective CS level for $\im u = \pm\infty$.

Consider first the case that $k_\pm \neq 0$. We have to evaluate the integral over $D_0 \in \bR+i\eta$ in the scaling limit $\se \to 0$ with $\se^2\sigma \to \infty$, therefore the terms containing $\fm$ and $\sigma^{(T)}$ are negligible. We can make use of
\be
\label{integral D0 at infinity}
\lim_{\se \sigma \to \pm\infty} \int_{\bR + i\eta} \hspace{-.2em} \frac{dD_0}{D_0} \, e^{- D_0^2 + i k_\pm \se \sigma D_0} \, \cZ(u; \fm) = \begin{cases} -2\pi i \, \cZ(u;\fm) & \text{if } \eta>0 \,,\; k_\pm \sigma < 0 \\ 0 & \text{if } \eta \, k_\pm \sigma > 0 \\ 2\pi i \, \cZ(u;\fm) & \text{if } \eta < 0 \,,\; k_\pm \sigma > 0 \;. \end{cases}
\ee
We are left with a contour integral of $\cZ(u;\fm)$ around the two infinities, which can be written more elegantly as a Jeffrey-Kirwan residue on the $x$-plane,%
\footnote{The residue at infinity is defined with a clockwise contour: $\Res\limits_{x=\infty} f(x) = \oint_{\circlearrowright,\, \infty} \frac{dx}{2\pi i}\, f(x) = - \oint_{\circlearrowleft,\, 0} \frac{dw}{2\pi i}\, \frac{f(1/w)}{w^2}$.}
\be
Z_\fm^\text{bdy} = \sum_{x_* = \, 0, \infty} \JKres_{x=x_*} \big( Q_{x_*}, \eta \big) \; \cZ(x ; \fm) \frac{dx}x \;,
\ee
if we assign charge vectors to the singularities at $x=0,\infty$ according to the effective Chern-Simons levels:
\be
Q_0 = - k_+ \;,\qquad\qquad Q_\infty = k_- \;.
\ee

If $k_+ = 0$ or $k_- = 0$ we need to regularize the integral on $\sigma$: we choose to do it by adding a Lagrangian term $- i \kappa_\text{reg} \sigma D$ and then taking the limit $\kappa_\text{reg} \to 0^\pm$. We show in section \ref{sec: boundary cancelation} that the result is independent of the sign of $\kappa_\text{reg}$, and in fact it is zero. Hence the prescription is that  we do not take any residue at infinity when $k_\pm=0$.
\end{itemize}
The full path-integral is obtained by summing $Z_\fm^\text{bulk}$ and $Z_\fm^\text{bdy}$ over all magnetic fluxes. In the rank-one case, we can elegantly write both contributions as JK residues on the complex $x$-plane. Moreover, the holomorphic 1-form $\cZ(x;\fm) \frac{dx}x$ is precisely the product of classical and one-loop contributions of section \ref{sec: on-shell} and \ref{sec: one-loop}, $\cZ(x;\fm) \frac{dx}x = Z_\text{int}(x;\fm)$, therefore the final expression is
\be
Z_{S^2 \times S^1} = \frac1{|W|} \sum_{\fm \in \Gamma_\fh} \bigg[ \sum_{\; x_* \in\, \fM_\text{sing}} \hspace{-.2em} \JKres_{x=x_*} \big( \sQ(u_*), \eta \big) \; Z_\text{int} (x; \fm) \;+\; \JKres_{x=0,\infty} ( Q_x,\eta) \; Z_\text{int}(x ; \fm) \bigg] \;.
\ee

\subsubsection{Cancelation of boundary contributions}
\label{sec: boundary cancelation}

It remains to verify that our regularization of the boundary contribution through $\kappa_\text{reg}$, when $k_\pm = 0$, leads to zero (and in particular it is independent of the sign of $\kappa_\text{reg} \to 0^\pm$).

Consider one boundary component, either $u=+\infty$ ($x=0$) or $u = -\infty$ ($x = \infty$). For one sign of $\kappa_\text{reg}$, we simply do not collect the residue for any value of $\fm$ and we obtain trivially zero. For the other sign of $\kappa_\text{reg}$, instead, we should sum the residues for all values of $\fm$. Since $k_+ = 0$ or $k_- = 0$ by assumption, the leading behavior of $Z_\text{1-loop}$ around $x=0$ or $x=\infty$ does not depend on $\fm$ (a CS term appears as a factor $x^{k\fm}$), although there can be a dependence on $\fm$ in the subleading terms in the series expansion. It follows that, depending on the value of the external fluxes, either we have a pole (of the same order) for all values of $\fm$ or for none. In the latter case we get zero. In the former case, after a suitable expansion and up to a shift in $\fm$, the residues will be sums of terms of the form $\fm^a z^\fm$. We should then evaluate the objects
$$
s_a(z) = \sum\nolimits_{\fm\in\bZ} \fm^a z^\fm \;.
$$
These sums are not convergent, but can be defined via $\zeta$-function regularization or analytic continuation. First of all
\be
s_0(z) = \sum_{\fm\in\bZ} z^\fm = \sum_{\fm \geq \fm_0} z^\fm + \sum_{\fm \geq -\fm_0 + 1} z^{-\fm} = 0 \;.
\ee
Then all $s_a(z)$ can be formally obtained by taking derivatives of $s_0(z)$, and therefore they all vanish. We conclude that the sum over $\fm$ of the residues vanish.

As a further check, we will confirm in some of our examples that, as we change $\eta$ and the JK residue picks up different contributions from the singularities in the ``bulk'' and from the boundaries, we always find convergent and well-defined expressions which eventually do not depend on $\eta$.

\subsection{The integral: higher-rank case}
\label{sec: higher rank}

The generic case of a gauge group $G$ of higher rank $r$ can be tackled with the same physical ideas, however it becomes technically more involved because of the richer topology of the moduli space $\fM$ and the singular subset $\fM_\text{sing}$. The space $\fM = H \times \fh$ is the product of $r$ complex cylinders. The singular subset $\fM_\text{sing}$ is a collection of hyperplanes $H_i$. Moreover we have to decide how to regularize the non-compact manifold $\fM$ at infinity. Eventually, $\fM \setminus \fM_\text{sing}$ has a complicated topology.

\subsubsection{Integration domain}

In section \ref{sec: zero-modes}, to each chiral multiplet $\Phi_i$ we have associated a charge covector $Q_i \equiv \rho_i \in \fh^*$ equal to the gauge weight, and a ``singular'' hyperplane $H_i = \{ u\in \fM \,|\, e^{i\rho_i(u) + i \rho_f(v)} = 1 \} \subset \fM$ (with the topology of $T^{r-1}\times \bR^{r-1}$) which is the dangerous locus where a would-be zero-mode may appear as $\se\to 0$. Since the hyperplanes are defined by an equation with real coefficients, their restriction (or imaginary projection) to $\fh$ is well-defined. To each hyperplane we associate an $\varepsilon$-neighborhood
\be
\Delta_\varepsilon(H_i) = \big\{ u \in \fM \;\big|\; |\rho_i(u) + \rho_f(v) + 2\pi k| < \varepsilon \;, \text{ for some $k \in \bZ$} \big\} \;.
\ee
We also need to introduce ``hyperplanes at infinity'' and remove their neighbourhoods.  The simplest choice would be to remove, for each $a=1,\ldots,r$, the locus $ \pm \im u_a > L$, where $L$ scales in a suitable way with $\varepsilon$. This, however, would lead to an expression difficult to evaluate. As we already saw in the rank-one case, the integral over $D_0$ near the boundary depends on the asymptotic value of the effective CS levels (\ref{keff2}), and the formula has a jump as we cross the restriction of an hyperplane on $\fh$. When an hyperplane $H_i$ intersects a boundary locus, it divides its restriction on $\fh$ into parts with different values of the effective CS levels. To avoid this complication, we cut a series of boundaries $H_\alpha^\infty$ at infinity, defined by linear equations
\be
H^\infty_\alpha = \big\{ u \in \fM \;\big|\; \gamma_\alpha(\im u) = L_\alpha \big\} \;,\qquad\qquad \gamma_\alpha \in \fh^* \;,
\ee
where $L_\alpha(\varepsilon)$ is a large cut-off. The $H_\alpha^\infty$s have the topology of $T^r \times \bR^{r-1}$, and their restriction to $\fh$ defines a convex polyhedron around infinity. We choose the polyhedron with the property that every face $H^\infty_\alpha$ that intersect one or more matter hyperplanes $H_i$, is orthogonal to all of them with respect to the Killing form $K^{ab}$ (actually, we could use any arbitrary positive-definite metric $K^{ab}$ on $\fh$):
\be
\gamma_\alpha^a \, K_{ab} \, Q_i^b = 0 \;.
\ee
We can then associate a charge vector $Q_\alpha \in \fh^*$ to each face, in analogy with what we did in the rank-one case:
\be
\label{chargeinfinity}
Q_\alpha^c = - \gamma_\alpha^a \, K_{ab} \, k_\text{eff}^{bc} \;,
\ee
which is well-defined on $H^\infty_\alpha$ and it does not jump. The convex polyhedron we have constructed can have a large number of faces, but it certainly exists. For each face $H_\alpha^\infty$, we define $\Delta_\varepsilon(H_\alpha^\infty)$ as the region of $\fM$ bounded by $H_\alpha^\infty$ and lying outside the polyhedron.  

Now the arguments of section \ref{sec: dangerous regions} go through. For large $|\sigma|$, the path-integral contains
$$
\int d^r\sigma\, \exp\Big[ - \frac{\se^2}2 \big( k^{ab}_\text{eff} \sigma_b + \zeta^a \big) K_{ac} \big( k^{cd}_\text{eff} \sigma_d + \zeta^c \big) \Big] \;.
$$
If the matrix $k_\text{eff} K^{-1} k_\text{eff}$ is positive definite, we can neglect $\zeta$ and the integral is convergent---in particular the region outside the convex polyhedron has vanishing contribution as the polyhedron is expanded. If, instead, the matrix has some zero eigenvalue, we can always introduce the regularization term $\sigma_a \kappa_\text{reg}^2 K^{ab} \sigma_b$.

For simplicity, we will use the index $i$ for all neighbourhoods, including those at infinity. For the hyperplanes at infinity we take cut-offs $L_\alpha(\varepsilon)$ which suitably scale with $\varepsilon$. We then define
\be
\Delta_\varepsilon = \bigcup\nolimits_i \Delta_\varepsilon(H_i) \;,
\ee
and consider the integral over $\fM \setminus \Delta_\varepsilon$.

\subsubsection{Stokes relations}

As in the rank-one case, the path-integral reduces to $Z_{S^2 \times S^1} = \frac1{|W|} \sum_{\fm \in \Gamma_\fh} Z_\fm$, with
\be
Z_\fm = \cN \lim_{\se, \varepsilon\to 0} \int_{\fM\setminus\Delta_\varepsilon} \hspace{-1em} d^ru \; d^r\bar u \int_{\fh + i \delta} \hspace{-.5em} d^r D\, \frac{\partial^{2r}}{\partial\lambda_1^\pdag \partial\lambda^\dag_1 \dots \partial\lambda_r \partial\lambda^\dag_r} \, \cZ(u,\bar u, \lambda, \lambda^\dag, D; \fm) \Big|_{\lambda = \lambda^\dag = 0}\;.
\ee
In order not to clutter formulas, we have dropped the subscript $0$ from the zero-modes.
Since $\cZ$ is holomorphic in $D$ around the origin, we have shifted the $D$ integration contour by a small vector $\delta \in \fh$. Following \cite{Benini:2013xpa}, we can use some ``Stokes relations'' to reduce the integral over $\fM$ to the contour integral of a meromorphic $r$-form along a specific $r$-dimensional cycle. We refer to \cite{Benini:2013xpa, Hori:2014tda} for the detailed argument, while here we just point out the peculiarities of our case.

To absorb the zero-modes, we use the following construction. First, as in (\ref{SUSY of Z}), supersymmetry guarantees that
\be
\label{SUSY of Z higher rank}
D_a \parfrac{}{\lambda^\dag_a} \cZ = -i \, \lambda_a \parfrac{}{\bar u_a} \cZ \;.
\ee
We define the following $(r-n)$-forms, for $n=0, \dots, r$:
\be
\Omega_{a_1 \dots a_n} = \frac1{(r-n)!^2} \; d\bar u_{c_1} \wedge \dots \wedge d\bar u_{c_{r-n}} \; \epsilon_{b_1 \dots b_{r-n} a_1 \dots a_n} \; \parfrac{^{2(r-n)}}{\lambda_{c_1}^\pdag \partial \lambda^\dag_{b_1} \ldots \partial\lambda_{c_{r-n}}^\pdag \partial \lambda^\dag_{b_{r-n}}} \; \cZ \Big|_{\lambda = \lambda^\dag = 0} \;.
\ee
With a little algebra, one can show that (\ref{SUSY of Z higher rank}) implies the relations
\be
\bar\partial\, \Omega_{a_1 \dots a_n} = i\, (-1)^{r-n} n \, D_{[a_1} \Omega_{a_2 \dots a_n]} = i \, (-1)^{r-n} \sum_{i=1}^n (-1)^{i-1} D_{a_i} \Omega_{a_1 \dots \widehat a_i \dots a_n} \;,
\ee
where {}$\,\widehat{}\,${} means omission and $\bar\partial \,\equiv\, d\bar u_a \, \parfrac{}{\bar u_a}$.

Then define the forms
\be
\mu_{Q_1,\dots, Q_s} = i^s \, d^ru \wedge \Omega_{a_1 \dots a_s} \wedge \frac{Q_1^{a_1} \dots Q_s^{a_s}}{Q_1(D) \dots Q_s(D)} \, d^rD \;,
\ee
where $Q_i \in \fh^*$ are $s$ covectors. Using the previous relation we get
\be
d\mu_{Q_0, \dots, Q_s} = \bar\partial \mu_{Q_0, \dots, Q_s} = \sum_{i=0}^s (-1)^{s-i} \mu_{Q_0, \dots \widehat Q_i \dots, Q_s} \;.
\ee
These forms are useful because with no vectors,
\be
\label{intform}
\mu = d^ru \wedge d^r\bar u \wedge d^r D\, \frac{\partial^{2r}}{\partial\lambda_1^\pdag \partial\lambda^\dag_1 \dots \partial\lambda_r \partial\lambda^\dag_r} \, \cZ \Big|_{\lambda = \lambda^\dag = 0}
\ee
is the integrand of the partition function $Z_\fm$, while with $r$ vectors,
\be
\label{rform}
\mu_{Q_1,\dots,Q_r} = i^r d^ru\, d^rD\, \frac{\det(Q_1 \cdots Q_r)}{Q_1(D) \dots Q_r(D)} \, \cZ\Big|_{\lambda = \lambda^\dag = 0}
\ee
where the last term is the classical and one-loop action.

\subsubsection{Reduction to a contour integral}

The boundary of the integration domain $\fM \setminus \Delta_\varepsilon$ is separated into ``tube regions'' 
\be
S_i = \partial\Delta_\varepsilon \cap \partial\Delta_\varepsilon(H_i) \;.
\ee
We also introduce
\be
S_{i_1 \dots i_s} = S_{i_1} \cap \ldots \cap S_{i_s} \;,
\ee
with the natural orientation which makes them antisymmetric in the indices. They satisfy
\be
\label{decomposition}
\partial \Delta_\varepsilon = \bigcup\nolimits_i S_i \;,\qquad\qquad \partial S_{i_1 \dots i_s} = - \bigcup\nolimits_j S_{i_1 \dots i_s j} \;,
\ee
as proven in \cite{Benini:2013xpa}. Each manifold $S_{i_1 \dots i_s}$ has real dimension $2r-s$ if not empty. Therefore the decompositions in (\ref{decomposition}) are almost disjoint: every intersection has dimension lower than the components, and the integral over the union is the sum of the integrals.

We can construct, as in \cite{Benini:2013xpa}, a cell decomposition of $\fM\setminus\Delta_\varepsilon$ such that:
\be
\label{celldomain}
\fM \setminus \Delta_\varepsilon = \bigsqcup\nolimits_i C_i \;,\qquad\qquad \partial C_{i_1 \dots i_k} = \sum\nolimits_j C_{i_1 \dots i_k j} - S_{i_1 \dots i_k} \;,
\ee
and each  $C_{i_1 \dots i_k}$ is associated to the set of charges $Q_{i_1},\ldots, Q_{i_k}$ of the hyperplanes $\{H_i\}$. Recall that we are using the index $i$ for
all hyperplanes including those at infinity. 

We can  use repeatedly the Stokes relations to reduce the integral over $\fM\setminus \Delta_\varepsilon$ to an integral over a middle-dimensional cycle in $du$.
The argument goes exactly as in  \cite{Benini:2013nda, Benini:2013xpa}.  We do not repeat all the steps of the argument, which has been spelled out in details in  
\cite{Benini:2013nda, Benini:2013xpa} and \cite{Hori:2014tda}, but we simply review the logic and mention the necessary modifications to deal with the boundary components.

The partition function is given by the integration of the form  $\mu$ (\ref{intform}) on a $3r$-cycle
given by $\Gamma \times \fM\setminus \Delta_\varepsilon$ where $\Gamma = \fh +i \delta$ is the contour for the $D$-integration, shifted from the real``axis" by a small vector $\delta\in \fh$. The integral can be manipulated by using iteratively  the Stokes relations starting from the cell decomposition  (\ref{celldomain}); for example, as in  \cite{Benini:2013nda, Benini:2013xpa}, one can derive  
\be
\label{man}
\int_{\Gamma \times \fM\setminus \Delta_\varepsilon} \mu = -\sum_i \int_{\Gamma \times S_i} \mu_{Q_i} -\sum_{i<j} \int_{\Gamma \times S_{ij}} \mu_{Q_iQ_j} + \ldots  \;.
\ee
In using the Stokes relations  we generate poles in $D$ with denominator $Q_i(D)$ and we restrict the  variable $u$ to live near the hyperplanes $H_i$.  
At each step we can deform the contour on $D$ and pick residues at the poles in analogy with what we did in the Abelian case.

Consider for example the integral  $\mu_{Q_i}$ on the contour $\Gamma\times S_i$. We need to distinguish the case where $i$ refers to a matter hyperplane and the case where
$i$ refers to a boundary at infinity. The latter case is the new ingredient compared to \cite{Benini:2013nda, Benini:2013xpa}.

Near a singularity due to a chiral field with vector charge $Q_i$, the $D$ integral is zero if $Q_i(\delta)<0$ because we can shrink the integration domain $S_i$ without encountering singularities, exactly as we did in the Abelian case.  If $Q_i(\delta)>0$, we modify the $D$ contour by changing $\delta$ until we can shrink the integration domain.  We are left with an integral over a contour $\Gamma_i$, which consists of a circle around zero in the variable  $Q_i(D)$ and is parallel to the real "axis" with an imaginary shift $i \delta_i$ in the remaining $r-1$ variables with $Q_i(D)=0$. The shift $\delta_i$ must satisfy $Q_i(\delta_i)=0$, and is different from the original $\delta$.

Near a boundary component specified by the equation $ \gamma_i(\im u) = L_i$, we can parameterize $\beta \sigma_a = L_i K_{ab} \gamma^b_i / |\gamma_i|^2 + \eta_a$, where $\eta_a$ satisfies $\gamma(\eta) = 0$ and spans the plane. The integral in $D$ contains the terms
\be
e^{ - \frac{2 \pi \beta R^2}{\se^2} K^{ab} D_a D_b \,+\, 2 i \beta R^2 \sigma_a  k^{ab}_\text{eff}(\sigma) D_b } = e^{ - \frac{2 \pi \beta R^2}{\se^2} K^{ab} D_a D_b \,-\, 2 i R^2  \frac{L_i}{|\gamma_i|^2} Q_i^a  D_a \,+\, 2 i R^2 \eta_a  k^{ab}_\text{eff}(\sigma) D_b } \;,
\ee
where the charge of the hyperplane at infinity was introduced in (\ref{chargeinfinity}). For large $L_i$, the imaginary part of the exponent is controlled by the sign of $Q_i(D)$, and we can close the contour in the variable $Q_i(D)$ by adding a semi-circle in the lower half-plane. By rescaling $D \rightarrow \se D$, the classical and one-loop contributions become independent of $D$ in the limit $\se \rightarrow 0$ and the only pole comes from $1/Q_i(D)$. If $Q_i(\delta)<0$, there are no poles in the integration contour and we obtain zero.%
\footnote{One might worry that, moving along the hyperplane at infinity, $\eta_a$ can become so large to change the sign of the imaginary part. However, in the limit $\se\rightarrow 0$, the boundary integrals are dominated by the region $D\sim 0$;  for large $\eta_a$ we are far from the matter singularities, the integrand is a regular function of $D$  and the rescaling $D\rightarrow \se D$ shows that  $\int_{\Gamma \times S_{i_1,\cdots, i_s}} \mu_{Q_{i_1},\cdots,Q_{i_s}}$ with $s<r$ vanishes with some power of $\se$. It remains to analyze the  terms with $r=s$ which are $r$-dimensional integral  in $du$ near the intersection of $r$ hyperplanes (we can always choose a boundary contour  such that no more than $r$ hyperplanes intersect in a point). If we are at the intersection of a boundary hyperplane with matter hyperplanes, the $\eta_a$ are finite and there is no problem. If we are at the intersection of two or more  boundary hyperplanes, the $\eta_a$  can be large. However, we can always arrange our cut-off at infinity in a hierarchy $L_1\gg L_2\cdots $. On the boundary hyperplane with the largest value of $L_i$ the corresponding $\eta_a$ will be necessarily smaller than $L_i$ and the argument applies.}
If $Q_i(\delta)>0$, we modify the contour by changing $\delta$ and we are left with an integral over a contour $\Gamma_i$ which circles around zero in the variable $Q_i(D)$, and is parallel to the real ``axis" but shifted by an imaginary shift $i\delta_i$, with $Q_i(\delta_i)=0$, in the remaining variables.

Each term in (\ref{man}) can be further manipulated  using the Stokes identities; for example  we can derive, as in \cite{Benini:2013nda, Benini:2013xpa},
\be
\int_{\Gamma_i \times S_i} d\mu_{Q_i} = -\sum_j \int_{\Gamma_i \times S_{ij}} d\mu_{Q_iQ_j}   -\sum_{j<k} \int_{\Gamma_i \times S_{ijk}} d\mu_{Q_iQ_jQ_k}  + \ldots \;.
\ee
At this point, using the same argument as above, $\Gamma_i$ can be  deformed to a contour $\Gamma_{ij}$ which circles around $Q_i(D)=Q_j(D)=0$. 
The process can be iterated until  we obtain a sum of terms of the form 
$$
\int_{\Gamma_{i_1,\ldots, i_r} \times S_{i_1,\ldots, i_r}} \mu_{Q_{i_1},\ldots,Q_{i_r}} \;,
$$
where $\Gamma_{i_1,\ldots, i_r}$ is a $T^r$ contour around the origin in the $D$ plane. The integral over $D$ picks up the residue of the denominator $1/Q_{i_1}(D) \cdots Q_{i_r}(D)$ and gives $\pm 1$. After the $D$ integration, the integrand becomes the classical and one-loop action $Z_\text{int}(u;\fm)$. The difficult part of the story is to keep track of all non-vanishing contributions, or their signs and of the necessary shifts in the $D$ contour. This can be done by the method explained in \cite{Benini:2013nda, Benini:2013xpa} which  introduces a reference covector $\eta\in \fh^*$. The final result is given, after summing over the fluxes, by
\be
\label{finres}
Z_{S^2\times S^1} = \frac{1}{|W|} \; \sum_{\fm\in \Gamma_\fh} \bigg[ \; \sum_{i_1,\ldots, i_r} c_{Q_{i_1},\ldots,Q_{i_r}; \eta} \; \int_{ S_{i_1,\ldots, i_r}} Z_\text{int}(u;\fm) \bigg]
\ee
where
\be
c_{Q_{i_1},\ldots,Q_{i_r}; \eta} = \begin{cases} 1 & \text{if } \eta \in \text{Cone}(Q_{i_1}, \ldots , Q_{i_r}) \\ 0 & \text{otherwise} \end{cases}
\ee
and it is independent of the choice of the reference covector $\eta$.

The integrand $Z_\text{int}(u;\fm)$ has singular points where $r$ or more linearly independent hyperplanes intersect and (\ref{finres}) reduces to a computation of residues. At  points $u_*$ where only matter hyperplanes intersect---the set of such points was called $\fM_\text{sing}^*$ in (\ref{def M sing *})---the expression in (\ref{finres}) is precisely the definition of the Jeffrey-Kirwan residue \cite{JeffreyKirwan}, as shown in  \cite{Benini:2013nda, Benini:2013xpa}. We can thus write  the partition function as
\be
\label{finalformula}
Z_{S^2\times S^1} =  \frac{1}{|W|} \; \sum_{\fm\in \Gamma_\fh} \; \sum_{u_* \in\, \fM_\text{sing}^*} \JKres_{u=u*} \big( \sQ_{u_*}, \eta \big) \; Z_\text{int}(\fm; u) \;+\; \text{boundary contribution} \;. 
\ee
In this formula, $u_*$ are intersections of the matter hyperplanes only, and $\sQ(u_*)$ is the set of charges of the hyperplanes intersecting at $u_*$. The boundary contribution refers to the intersection of hyperplanes at infinities, among themselves or with matter hyperplanes, and it should be computed using (\ref{finres}). We can simplify 
the evaluation of the boundary residues by choosing a convenient boundary polyhedron. With an appropriate sets of linear forms $\gamma_i$, we can restrict  to
the case where the intersections at infinity are transverse and no more than $r$ boundary or matter hyperplanes meet at the same point: in that case, the contour $S_{i_1,\ldots, i_r}$ is simply a $r$-dimensional torus. For $r-s$  boundaries meeting $s$ matter hyperplanes, we need to perform 
an integration over  the $r-s$ angles $\re u$ of the boundary component and an integration over an $s$-dimensional contour  which  
circles around  the $s$ matter hyperplane singularities, computing the residue at the corresponding pole. 

It would be interesting to give a proper geometrical interpretation of the boundary contribution, maybe  as some generalization of the Jeffrey-Kirwan residue \cite{JeffreyKirwan}.

\section{Examples}
\label{sec:examples}

In this section we compute the partition function for various examples of Abelian and non-Abelian theories in order to demonstrate the use of our formula. We discuss examples of Yang-Mills-Chern-Simons theories with (anti)fundamental and adjoint  matter. We will be able to recover and generalize standard results about Chern-Simons theories  and to confirm various dualities between three-dimensional theories with matter. Our results can be interpreted both as a check for our prescription as well as further evidence for three-dimensional dualities.

\subsection{Yang-Mills-Chern-Simons theories with fundamental matter}
\label{sec:QCD}

We first discuss in details the case of  $U(1)$ theories which nicely exemplify our prescription for computing the topologically twisted partition function and illustrate
many subtleties. As non-Abelian examples we consider the Aharony \cite{Aharony:1997gp} and Given-Kutasov \cite{Giveon:2008zn} dual pairs.

\subsubsection{$U(1)_{1/2}$ with one fundamental}
\label{sec:U(1)k=1/2}

We start considering a $U(1)$ theory with  Chern-Simons coupling $k=\frac12$ and one chiral multiplet $X$ of gauge charge $1$. Since this is our first example, we are pedantic and give many details. 

The chiral fields and charges are:
\be\nn
\begin{array}{c|ccc}
 & U(1)_g & U(1)_T & U(1)_R \\
\hline
X & 1 & 0 & 1 \\
\hline
T & 0 & 1 & 0 \\
\tilde T & -1 & -1 & 0
\end{array}
\ee
Here $T, \tilde T$ are the monopole operators $V_\fm$ corresponding to magnetic fluxes $\fm = 1$ and $\fm =  -1$, respectively, which play an important r\^ole in the identification of the dual theory \cite{Aharony:1997bx}. Their charge $Q$ under a generic flavor or R-symmetry is determined using the formula
\be
Q(V_\fm) = - \frac12 \sum_{\psi_i} Q(\psi_i) \, |\rho_i \fm| \;,
\ee
where the sum runs over all fermions in the theory and $\rho_i$ are their gauge charges. The same formula determines the one-loop contribution to the gauge charge of the monopole $V_\fm$ to be added to the classical contribution $k \fm$. The only flavor symmetry in the theory is the topological one, denoted as $U(1)_T$, under which only the monopoles are charged. We choose R-charge $q=1$ for the chiral multiplet, so that the fermion has R-charge zero, and no mixed gauge-R-symmetry CS term is necessary.%
\footnote{The gaugino has R-charge $1$ and is gauge neutral. Therefore, strictly speaking, the theory requires a half-integral R-R CS term. However we neglect such term because it does not introduce a dependence on the parameters, and it gives at most a constant phase.}

The matter content of the theory is not invariant under charge conjugation, signalling potential parity anomalies. These are however compensated by the half-integral Chern-Simons coupling. One can see this by noticing that the effective Chern-Simons coupling (\ref{keff0}),
\be
k_\text{eff}(\sigma) = \frac12+ \frac12\text{sign}(\sigma) \;,
\ee
is always an integer, implying the absence of parity anomalies.  

According to our rules, we construct the partition function by including the following ingredients:
\begin{itemize}
\item a measure $\frac{dx}{2\pi i \, x}$ on the $x$ plane;
\item the classical CS action contribution (\ref{classical}) which, for $k=1/2$, reads $x^{\fm/2}$;
\item the one-loop contribution (\ref{chiral1-loop}) of a chiral multiplet of gauge charge $1$ and R-charge $1$ which reads  $\big( \frac{x^{1/2}}{1-x} \big)^\fm$;
\item the contribution (\ref{topological})  $x^{\ft} \xi^{\fm}$ for the topological symmetry,  where $\xi$ and $\ft$ are the fugacity and the background flux for $U(1)_T$.
\end{itemize}
The partition function is then given by the contour integral
\be
\label{Zk=1/2}
Z = \sum_{\fm\in\bZ} \int \frac{dx}{2\pi i x} x^{\ft} (-\xi)^{\fm} x^{\fm/2} \Big( \frac{x^{1/2}}{1-x} \Big)^\fm = \sum_{\fm\in\bZ} \int \frac{dx}{2\pi i}\, (-\xi)^\fm \frac{x^{\fm + \ft -1}}{(1-x)^\fm} \;.
\ee
We included an extra $(-1)^\fm$, which can be reabsorbed in the definition of $\xi$, for later convenience. 

The contour integral should be evaluated according to formula (\ref{U(1)formula}). The integrand has singularities  at $x=1$, $x=0$ and $x=\infty$. The charge vector for $x=1$ is given by the charge of $X$,  $Q_{x=1}=1$.  The charges at ``infinity" are related to the asymptotic CS levels by (\ref{chargesboundary}): they  read  $Q_{x=0}=-k_\text{eff}(+\infty) = -1$, $Q_{x=\infty}=k_\text{eff}(-\infty) = 0$.

In order to use formula (\ref{U(1)formula}) we need to choose a number $\eta$. If we choose $\eta > 0$, the formula instructs us to take the residues at the singularities with positive charge. We thus need to pick the residue at $x=1$ which exist for $\fm \geq 1$. We get
\be
\label{precise computation}
Z = \sum_{\fm\geq1} (-\xi)^\fm \Res_{x=1} \frac{x^{\fm + \ft-1}}{ (1-x)^\fm} = \sum_{\fm\geq1} \xi^\fm \, \frac{(\ft+1)_{\fm-1}}{(\fm-1)!} = \frac\xi{(1-\xi)^{\ft+1}} \;,
\ee
where $(x)_n = \prod_{j=0}^{n-1} (x+j)$ is the Pochhammer symbol (for $n\in \bZ_{\geq 0}$). 

We could obtain the same result by resumming the integrand over $\fm\geq1$ first and then taking the residue. We are interested in the poles at $x=1$, which exist for $\fm\geq 1$, and we can take a contour with $|1-x|=\alpha$ as long as $\alpha < 1$. The series to sum is $\sum_{\fm\geq1} \big( \xi x / (x-1) \big)^\fm$, which converges uniformly along the contour for sufficiently small $\xi$. We find
\be
Z = \int_{x=1} \frac{dx}{2\pi i} x^{\ft-1} \sum_{\fm\geq1} \Big( \frac{\xi x}{x-1} \Big)^\fm = \oint_{x=\frac1{1-\xi}} \frac{dx}{2\pi i} \; \frac{x^\ft \xi}{1-\xi} \; \frac1{x - \frac1{1-\xi}} = \frac\xi{(1-\xi)^{\ft+1}} \;.
\ee
We have taken the residue at $x = (1-\xi)^{-1}$, which is the only pole inside the integration contour.

If we choose $\eta < 0$, instead, we should take minus the residues at $x=0$. The result is the same.  The residues at zero are indeed
\be
\Res_{x=0} \, (-\xi)^\fm \frac{x^{\fm+\ft-1}}{(1-x)^\fm} = \xi^\fm
\begin{cases} 0 & \fm + \ft \geq 1 \\
(-1)^\ft \frac{(\ft+1)_{-\fm-\ft}}{(-\fm-\ft)!} = (-1)^\fm \frac{(\fm)_{-\fm-\ft}}{(-\fm-\ft)!} & \fm+\ft \leq 0 \;.
\end{cases} 
\ee
We can sum them for $|\xi|>1$: $- \sum_{\fm \leq -\ft} \Res_{x=0} = \frac\xi{(1-\xi)^{\ft+1}}$.

According to our prescription, since $Q_{x=\infty}=0$, we have taken no residue at $x=\infty$. The residues there are non-vanishing both for indefinitely positive  and negative values of $\fm$ (for $\fm>1$ and $\fm<-\ft$ if $\ft\geq 0$, otherwise they are all zero). Their sum is not convergent, but it can be broken in two halves which converge in different regions of the complex plane of fugacities, and then defined by analytic continuation. The result is indeed
$$
\sum_\fm \Res_{x=\infty} = \sum_{\fm\geq 1} \Res_{x=\infty} \Big|_{|\xi|<1} + \sum_{\fm \leq -\ft} \Res_{x=\infty} \Big|_{|\xi|>1} = 0 \;,
$$
confirming our argument in section \ref{sec: boundary cancelation}.

\

It is well known that the theory above is dual to the theory of a free chiral multiplet, which can be identified with the monopole $T$ \cite{Dimofte:2011ju, Benini:2011mf}. 
Hence the dual theory is a free chiral $T$ of flavor charge 1 under $U(1)_T$ and R-charge 0. The dual theory also has half-integral CS terms $k_{TT} = -\frac12$ and $k_{RT} = -\frac12$.%
\footnote{We can start from the duality between $U(1)$ SQED with one fundamental $Q$ and one antifundamental $\tilde Q$, and the Wess-Zumino model $MT\tilde T$ \cite{Aharony:1997bx} discussed in section \ref{sec:SQED}---see in particular the table of charges in (\ref{table charges SQED}).
We turn on a real mass $m > 0$ for the axial symmetry $U(1)_A$ and for $U(1)_T$, the latter corresponding to a FI term $\zeta = m$. In the electric theory, the vacua are at the zeros of the effective D-term ``potential'' specified by $v_D'(\sigma) = k_\text{eff}(\sigma)$, $v_D(0) = \zeta$. $Q$ and $\tilde Q$ are generically massive, with the exception of $\sigma = - m$ where $Q$ is massless, and $\sigma = m$ where $\tilde Q$ is massless. Integrating them out we find $k_\text{eff} = 0$ for $|\sigma|>m$, $\frac12$ for $|\sigma|=m$, 1 for $|\sigma|<m$. It follows that the vacua are at $\sigma = -m$ (where $Q$ is massless) as well as along the flat direction $\sigma < -m$ parameterized by $T \sim e^{-\sigma}$. Integrating out $\tilde Q$, the effective theory at $\sigma = -m$ is $U(1)_\frac12$ with a fundamental $Q$. On the magnetic side, the effective theory is the free field $T$ with $k_{TT} = -\frac12$ and $k_{RT} = -\frac12$.}
We thus have:
\be
\label{dualU(1)}
Z_\text{dual} = \xi^{-\ft/2 + 1/2} \Big( \frac{\xi^{1/2}}{1-\xi} \Big)^{\ft+1} = \frac\xi{(1-\xi)^{\ft+1}} \;.
\ee
This agrees with the result (\ref{precise computation}) for the original theory.

\subsubsection{SQED with one flavor}
\label{sec:SQED}

Consider a $U(1)$ theory with two chiral multiplets of charges $\pm1$, and no Chern-Simons couplings. The theory is parity-invariant (although turning on a background for the R-symmetry breaks parity, and indeed an R-R CS term is necessary because of the gaugino). The chiral fields are:
\be
\label{table charges SQED}
\begin{array}{c|cccc}
 & U(1)_g & U(1)_T & U(1)_A & U(1)_R \\
\hline
Q & 1 & 0 & 1 & 1 \\
\tilde Q & -1 & 0 & 1 & 1 \\
\hline
M = Q\tilde Q \rule{0pt}{1.1em} & 0 & 0 & 2 & 2 \\
T & 0 & 1 & -1 & 0 \\
\tilde T & 0 & -1 & -1 & 0
\end{array}
\ee
Here $T$ and $\tilde T$ are monopole operators of magnetic charge $\pm 1$, respectively. According to our rules, the partition function is
\be
Z = \sum_{\fm\in\bZ} \int \frac{dx}{2\pi i\, x} x^\ft (-\xi)^\fm \Big( \frac{x^\frac12 y^\frac12}{1-xy} \Big)^{\fm+\fn} \Big( \frac{x^{-\frac12} y^\frac12}{1-x^{-1}y} \Big)^{-\fm+\fn} \;,
\ee
where $\xi$ and $\ft$ are the fugacity and background flux for the topological symmetry $U(1)_T$, while $y$ and $\fn$ are for the flavor symmetry $U(1)_A$.
We included an extra $(-1)^\fm$, which can be reabsorbed in the definition of $\xi$, for later convenience. 

We choose $\eta>0$, and  formula (\ref{U(1)formula}) instructs us to take the residues from the field $Q$ with positive charge, whose pole is at $x = \frac1y$. Since $k_\text{eff} = k=0$, we do not take any residue at $x=0,\infty$. There is a pole at $x = \frac1y$ only for $\fm \geq 1-\fn$. In order to evaluate all the residues, we take a contour around $\frac1y$ and sum the integrands. The series is $\sum_\fm \big( \xi (y-x)/(1-xy) \big)^\fm$, and we have uniform convergence along the contour for sufficiently small $\xi$. We sum over $\fm \geq 1-\fn$ and then take the residue at the unique pole inside the contour, namely at $x = \frac{1-\xi y}{y-\xi}$. The result is:
\be
\label{Z for SQCD(1)}
Z = - \frac{y^{3\fn-2} (-\xi)^\ft}{(1-y^2)^{2\fn-1} (1-\xi y^{-1})^{1 - \fn + \ft} (1-\xi^{-1}y^{-1})^{1-\fn- \ft}} \;.
\ee

\

The dual theory is a Wess-Zumino model with fields $M,T,\tilde T$ and  a cubic superpotential $W = M T\tilde T$ \cite{Aharony:1997bx}. The partition function is
\be
Z = \Big( \frac y{1-y^2} \Big)^{2\fn-1} \Big( \frac{\xi^\frac12 y^{-\frac12}}{ 1-\xi y^{-1} } \Big)^{\ft - \fn+1} \Big( \frac{\xi^{-\frac12} y^{-\frac12} }{1-\xi^{-1} y^{-1} } \Big)^{-\ft-\fn+1} \;.
\ee
This agrees with (\ref{Z for SQCD(1)}), up to an ambiguous sign $(-1)^{\ft +1}$.

\subsubsection{$U(N_c)$ with $N_f$ flavors and Aharony duality}

The previous example generalizes to higher gauge rank and number of flavors. Consider a $U(N_c)$ theory, with $N_f$ chiral multiplets $Q_a$ in the fundamental and $\tilde Q_b$ in the antifundamental representations, and no CS interactions. For simplicity, we only introduce backgrounds for the R-symmetry, the topological symmetry and the $U(1)_A$ subgroup of the flavor symmetry acting with the same charge on all chiral fields. We assign R-charge $1$ to the chiral fields. Hence:
\be\nn
\begin{array}{c|cccc}
 & U(N_c)_g & U(1)_T & U(1)_A & U(1)_R \\
\hline
Q_a & \rep{N_c} & 0 & 1 & 1 \\
\tilde Q_b & \rep{\overline N_c} & 0 & 1 & 1 \\
\hline
M_{ab} = Q_a \tilde Q_b & \rep{1} & 0 & 2 & 2 \rule{0pt}{1.1em} \\
T & \rep{1} & 1 & -N_f & -N_c+  1 \\
\tilde T & \rep{1} & -1 & -N_f & -N_c + 1
\end{array}
\ee
Here $T, \tilde T$ are the monopole operators $V_\fm$ corresponding to magnetic fluxes $\fm = (1,0, \dots ,0)$ and $\fm = (0, \dots 0, -1)$, respectively. Their charges under a generic flavor or R-symmetry $Q$ are determined using the formula
\be
\label{nonAbelian monopole charges}
Q(V_\fm) = - \frac12 \sum_{\psi_i} \sum_{\rho_i \in \fR_i} Q(\psi_i) \, |\rho_i(\fm)| \;,
\ee
where the sum runs over all fermions in the theory, $\fR_i$ denote their representations under the gauge group and $\rho_i$ are the corresponding weights.%
\footnote{Although the R-charge of the gaugini $\lambda$ is $-1$, in (\ref{nonAbelian monopole charges}) one should use their complex conjugate $\lambda^c$ with R-charge $R(\lambda^c)=1$. This is because the Dirac kinetic action, written in terms of $\lambda, \lambda^\dag$ in (\ref{YM}), has opposite sign with respect to the one for the matter fields $\psi, \psi^\dag$ in (\ref{matter}), and therefore the coupling to the gauge field has opposite sign as well. If we rewrite the Dirac term in (\ref{YM}) in terms of $\lambda^c, \lambda^{c\dag}$, it gets the same sign as the one in (\ref{matter}).}
The partition function of the theory is given by
\begin{multline}
\label{AharonyZ}
Z = \frac1{N_c!} \sum_{\vec\fm \,\in\, \bZ^{N_c} } \int \prod_{i=1}^{N_c}\frac{dx_i}{2\pi i \, x_i} \cdot (-1)^{N_f \sum\fm_i} \prod_{i\ne j}^{N_c} \Big( 1-\frac{x_i}{x_j} \Big) \\
\times \prod_{i=1}^{N_c} x_i^\ft \xi^{\fm_i} \Big( \frac{x_i^\frac12 y^\frac12}{1-x_i y} \Big)^{N_f(\fm_i+\fn)} \Big( \frac{x_i^{-\frac12} y^\frac12}{1-x_i^{-1}y} \Big)^{N_f(-\fm_i + \fn)} \;,
\end{multline}
where $\xi$ and $\ft$ are the fugacity and background flux for the topological symmetry while $y$ and $\fn$ are for the diagonal flavor symmetry.
We inserted a factor $(-1)^{N_f \sum \fm_i}$, which can be reabsorbed in a redefinition of $\xi$, for later convenience.

Since $k_\text{eff}=k =0$, we can ignore the residues at the boundaries. We should choose a vector $\vec\eta \in \bR^{N_c}$: we choose $\eta_i<0$, hence we have to collect minus the residues from the negatively charged fields $\tilde Q$.  They are located at $x_i = y$ and they exist only for $\fm_i \leq \fn-1$. In order to evaluate all the residues we can sum the geometric series in (\ref{AharonyZ}) first:
\be
\label{formula 01}
Z = \frac{ y^{N_c N_f \fn} \xi^{N_c \fn} }{ (-1)^{N_c (N_f \fn -1)} N_c!} \int_\circlearrowleft \prod_{i=1}^{N_c} \frac{dx_i}{2\pi i} \; \frac{x_i^{N_f \fn - N_c + \ft} }{ (1-x_iy)^{N_f(2\fn-1)} \big[ \xi (y-x_i)^{N_f} - (1-x_i y)^{N_f} \big] } \prod_{i\neq j} (x_i - x_j) \;.
\ee
In the limit $\xi \to\infty$ the series converges uniformly along a contour that encircles only the relevant poles. We have introduced a factor of $(-1)^{N_c}$ coming from our prescription for $\eta_i<0$, and the contours in the previous formula are counterclockwise. 

We define the degree-$N_f$ polynomial
\be
\cP(x) \equiv \xi(y-x)^{N_f} - (1-xy)^{N_f} = (-1)^{N_f} (\xi - y^{N_f}) \prod_{\alpha=1}^{N_f} (x-x_\alpha) \;,
\ee
where $x_\alpha$ are defined to be its roots, and we easily derive:
\be
\label{full products A}
\prod_{\alpha=1}^{N_f} x_\alpha = \frac{\xi y^{N_f} -1}{ \xi - y^{N_f}} \;,\qquad\qquad \prod_{\alpha=1}^{N_f} (1-x_\alpha y) = \frac{\xi (1-y^2)^{N_f}}{ \xi - y^{N_f}} \;.
\ee
In (\ref{formula 01}) we should pick the residues at $x_i= x_{\alpha_i}$ for all choices of $N_c$ integers $\alpha_i \in \{1,\ldots,N_f\}$ but, due to the Vandermonde factor $\prod_{i\neq j} ( x_i-x_j)$, only the residues where the integers $\alpha_i$ are all different give a non-zero contribution. We then obtain
\be
\label{partitionNcNf}
Z = \frac{(-1)^{N_c(N_f(\fn-1)-1)} \; y^{N_cN_f\fn} \; \xi^{N_c \fn} }{ (\xi - y^{N_f})^{N_c}} \sum_I^{C^{N_f}_{N_c}} \prod_{\alpha\in I} \frac{ x_\alpha^{N_f\fn - N_c + \ft} }{ (1-x_\alpha y)^{N_f(2\fn-1)} \prod_{\beta \in I^c} (x_\alpha - x_\beta) } \;,
\ee
where $I$ runs over all combinations $C^{N_f}_{N_c}$ of $N_c$ different integers in $\{ 1,\ldots,N_f\}$, while $I^c$ denotes the complementary set  $\{ 1,\cdots,N_f\}\setminus I$ belonging to $C^{N_f}_{N_f-N_c}$.

For $N_f < N_c$, the expression above obviously vanishes.%
\footnote{When $N_f \leq N_c-2$, this reflects the fact that the theory has no supersymmetric vacuum \cite{Aharony:1997gp, Aharony:1997bx}. When $N_f = N_c-1$ the interpretation is more subtle. The theory has a deformed moduli space given by $T \tilde T \det M = 1$ which is a smooth manifold, therefore the theory is IR free. However the equation forces $T, \tilde T, M \neq 0$, therefore the theory spontaneously breaks the global symmetry $U(1)_T \times U(1)_A \times U(1)_R$ to the subgroup $U(1)_{R'} = U(1)_R - U(1)_A$, which is an IR R-symmetry (it is not the superconformal one, though, which is accidental). For generic background fields, supersymmetry is broken and indeed (\ref{partitionNcNf}) is zero. However supersymmetry is preserved if we set to zero the backgrounds for broken symmetries and only retain the one for $U(1)_{R'}$, which corresponds to $\xi=y=1$, $\ft=0$, $\fn=1$. In this case (\ref{partitionNcNf}) is a formal $0/0$, which could potentially lead to a finite result. Unfortunately we do not have equivariant parameters in the UV that could ``compactify'' the moduli space, therefore we do not expect to be able to define a finite partition function.}
If $N_f = N_c$, there is only one $I$ while $I^c = \emptyset$. We immediately get
\be
\label{dualNc}
Z_{N_f = N_c} = (-1)^{N_c + \ft} \; \frac{ y^{N_c^2(3\fn - 2)} \; \xi^\ft }{ (1-y^2)^{N_c^2 (2\fn - 1)} \; (1-\xi y^{-N_c})^{N_c(1-\fn) + \ft} \; (1-\xi^{-1} y^{-N_c})^{N_c(1-\fn)-\ft} } \;.
\ee

The dual theory for $N_f=N_c$ is given by the fields $M_{ab}$, $T$ and $\tilde T$, coupled through the superpotential $W = T\tilde T \det M$ \cite{Aharony:1997bx}. The partition function of the dual theory is then
\be
Z_\text{dual}^{N_f = N_c} = \Big( \frac y{1-y^2} \Big)^{(2\fn-1)N_c^2} \Big( \frac{\xi^\frac12 y^{-\frac{N_c}2} }{1-\xi y^{-N_c}} \Big)^{N_c(1-\fn) + \ft} \Big( \frac{ \xi^{-\frac12} y^{-\frac{N_c}2} }{ 1-\xi^{-1} y^{-N_c} } \Big)^{N_c(1-\fn) - \ft} \;.
\ee
This agrees with (\ref{dualNc}), up to an ambiguous sign $(-1)^{N_c + \ft}$.

The expression (\ref{partitionNcNf}) for $N_f>N_c$ is more complicated but we can use it  to check Aharony dualities \cite{Aharony:1997gp}.
The dual theory is a $U(N_f-N_c)$ gauge theory with $N_f$ fundamentals $q_a$, $N_f$ anti-fundamentals $\tilde q_b$ and $N_f^2+2$ singlets $M_{ab}$,  $T$ and $\tilde T$, corresponding to the mesons and monopoles of the original theory, with a superpotential $W = M_{ab} q_a \tilde q_b + v_- T + v_+ \tilde T$, where $v_{\pm}$ are monopoles of the dual theory \cite{Aharony:1997gp}. We assign the charges consistently with the original theory:
\be\nn
\begin{array}{c|cccc}
 & U(N_f-N_c)_g & U(1)_T & U(1)_A & U(1)_R \\
\hline
q_a & \rep{N_f-N_c} & 0 & -1 & 0 \\
\tilde q_b & \overline{\rep{N_f - N_c}} & 0 & -1 & 0 \\
M_{ab} & 0 & 0 & 2 & 2 \\
T & 0 & 1 & -N_f & -N_c+1 \\
\tilde T & 0 & -1 & -N_f & -N_c+1\\
\hline
v_+ & 0 & 1 & N_f & N_c+1 \\
 v_-& 0 & -1 & N_f & N_c+1
\end{array}
\ee
Notice that the dual quarks have R-charge zero and axial flavor charge $-1$.

The partition function of the dual theory is obtained by multiplying the contribution of the gauge sector for the quarks $q_a,\tilde q_b$ with the contribution of the singlets  $M_{ab}$, $T$ and $\tilde T$. The first contribution is the partition function for a  $U(N_f-N_c)$ theory with quarks $q_a,\tilde q_b$ which we can read from (\ref{partitionNcNf}). According to our assignment of charges, we need to replace the background charge and fugacity for the flavor symmetry  by $y \leftrightarrow y^{-1}$ and $\fn \leftrightarrow 1-\fn$, as well as $N_c \leftrightarrow N_f - N_c$. We find
\be
\label{Z q qtilde}
Z_{q\tilde q} = \frac{y^{(N_f - N_c)N_f(\fn-1)} \xi^{(N_f - N_c)(1-\fn)} }{ (-1)^{(N_f - N_c)(N_f\fn-1)} (\xi - y^{-N_f} )^{N_f - N_c} } \sum_J^{C^{N_f}_{N_f-N_c}} \prod_{\beta \in J} \frac{\tilde x_\beta^{-N_f \fn + N_c + \ft} }{ (1-\tilde x_\beta y^{-1})^{N_f (1-2\fn)} \prod_{\alpha \in J^c} (\tilde x_\beta - \tilde x_\alpha) } \;.
\ee
The $\tilde x_\beta$ are the roots of $\tilde\cP(\tilde x) = \xi(y^{-1} - \tilde x)^{N_f} - (1-\tilde x y^{-1})^{N_f} = 0$, and in fact $\tilde x_\beta = 1/x_\beta$. We can thus rewrite $Z_{q\tilde q}$ in terms of $x_\beta$, and convert the products over $J$ into products over $J^c$ using the full products in (\ref{full products A}). We get:
\begin{multline}
Z_{q\tilde q} = \frac{(-1)^{N_f(N_c(\fn-1)+\fn) + N_c} y^{N_f(N_f-N_c)(1-\fn)} \xi^{N_f\fn + N_c(\fn-1)} }{ (1-y^2)^{N_f^2(1-2\fn)} (\xi-y^{N_f})^{N_f\fn - \ft} (\xi y^{N_f}-1)^{N_f\fn - N_c + \ft} } \\
\times \sum_J \prod_{\alpha \in J^c} \frac{ x_\alpha^{N_f\fn - N_c + \ft} }{ (1-x_\alpha y)^{N_f(2\fn-1)} \prod_{\beta\in J} (x_\alpha - x_\beta)} \;.
\end{multline}
The contribution of the gauge singlets is
\bea
Z_{MT\tilde T} &= \Big( \frac y{1-y^2} \Big)^{N_f^2(2\fn-1)} \Big( \frac{\xi^\frac12 y^{-\frac{N_f}2} }{ 1-\xi y^{-N_f} } \Big)^{\ft - N_f \fn+N_c} \Big( \frac{\xi^{-\frac12} y^{-\frac{N_f}2} }{1-\xi^{-1} y^{-N_f} } \Big)^{-\ft - N_f \fn + N_c} \\
&= \frac{ (-1)^{\ft - N_f\fn + N_c} y^{N_f( N_f(\fn-1) + N_c)} \xi^{-N_f\fn + N_c} }{ (1-y^2)^{N_f^2(2\fn-1)} (\xi - y^{N_f})^{\ft - N_f\fn + N_c} (\xi y^{N_f} -1)^{-\ft - N_f\fn + N_c} } \;.
\eea
Then the partition function of the dual theory, $Z_\text{dual} = Z_{q\tilde q} Z_{MT\tilde T}$, equals the one of the electric theory up to $(-1)^{N_c + \ft}$.

\subsubsection{$U(N_c)_k$ with $N_f$ flavors and Giveon-Kutasov duality}

Giveon-Kutasov (GK) duality \cite{Giveon:2008zn}  can be derived from Aharony duality giving a real mass to some of the flavors. Nevertheless, to test our formula, we check GK duality separately.
Consider a $U(N_c)_k$ theory with $N_f$ fundamentals $Q_a$ and antifundamentals $\tilde Q_b$. As before, for simplicity we only introduce backgrounds for the R-symmetry, the topological symmetry and the $U(1)_A$ axial subgroup of the flavor symmetry, and we assign R-charge $1$ to the chiral fields:
\be\nn
\begin{array}{c|cccc}
 & U(N_c)_g & U(1)_T & U(1)_A & U(1)_R \\
\hline
Q_a & \rep{N_c} & 0 & 1 & 1 \\
\tilde Q_b & \rep{\overline N_c} & 0 & 1 & 1 \\
\hline
M_{ab} = Q_a \tilde Q_b & 0 & 0 & 2 & 2 \phantom{\Big|} \\
\end{array}
\ee
The partition function is given by
\begin{multline}
\label{GK Z}
Z = \frac1{N_c!} \sum_{\vec\fm \,\in\, \bZ^{N_c} } \int \prod_{i=1}^{N_c}\frac{dx_i}{2\pi i x_i} \cdot (-1)^{N_f \sum\fm_i} \prod_{i\ne j}^{N_c} \Big( 1-\frac{x_i}{x_j} \Big) \\
\times \prod_{i=1}^{N_c} x_i^{k\fm_i + \ft} \xi^{\fm_i} \Big( \frac{x_i^\frac12 y^\frac12}{1-x_i y} \Big)^{N_f(\fm_i+\fn)} \Big( \frac{x_i^{-\frac12} y^\frac12}{1-x_i^{-1}y} \Big)^{N_f(-\fm_i + \fn)} \;.
\end{multline}
As before, we inserted a factor  $(-1)^{N_f \sum \fm_i}$ for later convenience.

Let us choose $\eta_i<0$. First consider the case $k>0$: we pick minus (introducing $(-1)^{N_c}$) the residues at $x_i = y$ which exist for $\fm_i \leq \fn -1$, and minus the residues at $x=0$ which exist for $\fm_i \leq \frac{N_c - N_f\fn - \ft -1}k$. It is convenient to resum over $\fm_i \leq M-1$ for some large positive integer $M$, in order to include all poles. Looking at the geometric series we see that in the $\xi\to\infty$ limit, the poles of the resummed function lie close to the poles of the separate terms. Resumming the integrand we obtain
$$
\frac{x_i^{kM+N_f\fn-N_c+\ft} (x_i-y)^{N_f(M-\fn)} \big( (-1)^{N_f} \xi \big)^M }{ (1-x_iy)^{N_f(M+\fn-1)} \big[ \xi x_i^k (y-x_i)^{N_f} - (1-x_iy)^{N_f} \big] } \;.
$$
This expression has poles at $x_i = y^{-1}$, which are not relevant for us, and at $x_i$ equal to one of the $N_f+k$ roots of the degree-$(N_f+k)$ polynomial in the denominator. Let us call $x_{\alpha}$, with $\alpha= 1,\dots, N_f+k$, its roots:
\be
\cP(x) \;\equiv\; \xi x^k (y-x)^{N_f} - (1-xy)^{N_f} = (-1)^{N_f} \xi \prod_{\alpha=1}^{N_f+k} (x-x_{\alpha}) \;.
\ee
Notice that as $\xi\to\infty$, the roots converge to $y$ and $0$, therefore those are the poles inside our contour. If we remove $(x-x_\alpha)$ from $\cP(x)$ for one $\alpha$, and then substitute $x \to x_\alpha$ in the resummed expression above, the dependence on $M$ disappears:
$$
\frac{x_\alpha^{(N_f+k)\fn - N_c + \ft} \xi^{\fn-1} (-1)^{N_f(\fn-1)} }{ (1-x_\alpha y)^{N_f(2\fn-1)} \prod_{\beta \neq \alpha} (x_\alpha - x_\beta) } \;.
$$
We can then write the partition function. Because of the Jacobian, the contributing poles have $x_i \neq x_j$ and they are simple in all variables. They are given by choices of combinations $I$ of $N_c$ integers in $\{1,\dots,N_f+k\}$. The result is
\be
\label{GK Z final}
Z_{k>0} = (-1)^{N_c \smath{(} N_f (\fn-1) + 1 \smath{)}} \; y^{N_cN_f\fn} \; \xi^{N_c(\fn-1)} \sum_I^{C^{N_f+k}_{N_c}} \prod_{\alpha \in I} \frac{ x_\alpha^{(N_f+k)\fn - N_c + \ft} }{ (1-x_\alpha y)^{N_f(2\fn-1)} \prod_{\beta \in I^c} (x_\alpha - x_\beta) } \;.
\ee

Then consider the case $k<0$: we pick minus the residues at $x_i=y$, and minus the residues at $x_i=\infty$ which exist for $\fm_i \leq \frac{-N_f\fn - N_c + \ft +1}{|k|}$. As before, we can sum over $\fm_i \leq M-1$, obtaining the same expression as before. However, in order to exhibit a polynomial in the denominator, we rewrite it as
$$
\frac{x_i^{-|k|(M-1)+N_f\fn-N_c+\ft} (x_i-y)^{N_f(M-\fn)} \big( (-1)^{N_f} \xi \big)^M }{ (1-x_iy)^{N_f(M+\fn-1)} \big[ \xi (y-x_i)^{N_f} - x_i^{|k|} (1-x_iy)^{N_f} \big] } \;.
$$
This expression does not have poles at $x_i=\infty$. This time the degree-$(N_f + |k|)$ polynomial is
\be
\wh\cP(x) \;\equiv\; \xi (y-x)^{N_f} - x^{|k|} (1-xy)^{N_f} = (-1)^{N_f-1} y^{N_f} \prod_{\alpha=1}^{N_f + |k|} (x - \hat x_\alpha) \;,
\ee
while the resummed expression, after removing $(x-\hat x_\alpha)$ and substituting $x\to\hat x_\alpha$, can be recast as
$$
\frac{\hat x_\alpha^{|k|(1-\fn) + N_f\fn - N_c + \ft} (-1)^{N_f(\fn-1) + 1} y^{-N_f} \xi^\fn }{ (1-\hat x_\alpha y)^{N_f(2\fn-1)} \prod_{\beta \neq \alpha} (\hat x_\alpha - \hat x_\beta) } \;.
$$
Eventually the partition function reads
\be
Z_{k<0} = (-1)^{N_cN_f(\fn-1)} y^{N_c N_f(\fn-1)} \; \xi^{N_c \fn} \sum_I^{C^{N_f+ |k|}_{N_c}} \prod_{\alpha \in I} \frac{ \hat x_\alpha^{|k|(1-\fn) + N_f\fn - N_c + \ft} }{ (1-\hat x_\alpha y)^{N_f(2\fn-1)} \prod_{\beta \in I^c} (\hat x_\alpha - \hat x_\beta) } \;.
\ee

The dual theory is $U(N_f + |k| - N_c)_{-k}$ with $N_f$ flavors and $N_f^2$ gauge singlets. Let us compare the two partition functions. For concreteness, let us take level $k>0$ in the electric theory and level $-k<0$ in the magnetic one (the other case can be obtained by a parity transformation). In the dual theory there is also a different charge assignment, obtained by $y \leftrightarrow y^{-1}$, $\fn \leftrightarrow 1-\fn$, $k \leftrightarrow -k$, $N_c \leftrightarrow N_f + k - N_c$. The partition function of the gauge sector of the dual theory is then
\be
Z^{q\tilde q}_{-k<0} = \frac{ y^{(N_f + k - N_c)N_f \fn} \xi^{(N_f + k - N_c)(1-\fn)} }{ (-1)^{(N_f + k - N_c)N_f \fn} } \sum_J^{C^{N_f+k}_{N_f + k - N_c}} \prod_{\beta \in J} \frac{ \hat{\tilde x}_\beta^{k(\fn-1) - \fn N_f + N_c + \ft} }{ (1-\hat{\tilde x}_\beta y^{-1})^{N_f(1-2\fn)} \prod_{\alpha\in J^c} (\hat{\tilde x}_\beta - \hat{ \tilde x}_\alpha ) } \;.
\ee
The $\hat{\tilde x}_\alpha$ are the $N_f + k$ solutions to the equation $\xi(y^{-1} - \hat{\tilde x})^{N_f} - \hat{\tilde x}^k(1- \hat{\tilde x} y^{-1})^{N_f} = 0$, and in fact $\hat{\tilde x}_\alpha = 1/x_\alpha$. To further massage the expression, we use
$$
\prod_{\beta\in J ,\, \alpha\in J^c} \big( \hat{\tilde x}_\beta - \hat{\tilde x}_\alpha \big)^{-1} = \prod_{\beta \in J} x_\beta^{N_c} \prod_{\alpha \in J^c} x_\alpha^{N_f + k - N_c} \prod_{\beta\in J ,\, \alpha\in J^c} (x_\alpha - x_\beta)^{-1} \;,
$$
as well as
$$
\prod_{\alpha=1}^{N_f + k} x_\alpha = (-1)^{k-1} \xi^{-1} \;,\qquad\qquad \prod_{\alpha=1}^{N_f + k} (1-x_\alpha y) = (1-y^2)^{N_f} \;.
$$
Eventually we find
\be
Z^{q\tilde q}_{-k<0} = \frac{ y^{(N_f+k - N_c) N_f (1-\fn)} \; \xi^{N_c(\fn-1) + \ft} }{ (-1)^{N_cN_f(\fn-1) + (k-1)\ft} \; (1-y^2)^{N_f^2(1-2\fn)} } \sum_I^{C^{N_f + k}_{N_c}} \prod_{\alpha \in I} \frac{ x_\alpha^{(N_f + k)\fn - N_c + \ft} }{ (1-x_\alpha y)^{N_f(2\fn-1)} \prod_{\beta \in I^c} (x_\alpha - x_\beta) } \;,
\ee
where $I$ are combinations of $N_c$ different integers, mapping $I = J^c$. This has to be multiplied by
$$
\Big( \frac y{1-y^2} \Big)^{N_f^2(2\fn-1)} \; y^{N_f(k-N_f)\fn - N_f(k-N_c)} \; \xi^{-\ft} \;,
$$
where the first term comes from the gauge singlets and the other ones from the global CS terms computed below. The total partition function $Z^\text{dual}_{-k<0} = Z^{q\tilde q}_{-k<0} Z_\text{sing} Z_\text{CS}$ agrees with the one of the electric theory (\ref{GK Z final}), up to an ambiguous sign $(-1)^{N_c + (k-1)\ft}$.

Notice that for $N_c = N_f + |k|$, $Z_{q\tilde q} = 1$, therefore the partition function of the electric theory equals that of the free mesons (magnetic theory).

Let us reproduce the global CS terms in the dual. To do that, we start with $U(N_c)$ with $N_f + k$ flavors ($k>0$). We divide the flavors in two groups: $Q_a, \tilde Q_a$ are $N_f$ with charge $1$ under $U(1)_A$, while $Q_P, \tilde Q_P$ are $k$ with charge $1$ under a new symmetry $U(1)_m$:
\be\nn
\begin{array}{c|cccc|c}
& U(N_c) & U(1)_T & U(1)_A & U(1)_R & U(1)_m \\
\hline
Q_a & \rep{N_c} & 0 & 1 & 1 & 0 \\
\tilde Q_a & \rep{\overline N_c} & 0 & 1 & 1 & 0 \\
Q_P & \rep{N_c} & 0 & 0 & 1 & 1 \\
\tilde Q_P & \rep{\overline N_c} & 0 & 0 & 1 & 1
\end{array}
\ee
We give positive mass associated to $U(1)_m$: then $Q_P, \tilde Q_P$ can be integrated out and we are left with $U(N_c)_k$ with $N_f$ flavors. In fact the only CS which is shifted is the gauge one: $\delta k_{gg} = k$.

By Aharony duality, the dual is $U(N_c^\prime)=U(N_f + k - N_c)$ with $N_f + k$ flavors and many gauge singlets:
\be\nn
\begin{array}{c|cccc|c}
& U(N_f+k-N_c) & U(1)_T & U(1)_A & U(1)_R & U(1)_m \\
\hline
q_a & \rep{N_c'} & 0 & -1 & 0 & 0 \\
\tilde q_a & \rep{\overline N_c'} & 0 & -1 & 0 & 0 \\
q_P & \rep{N_c'} & 0 & 0 & 0 & -1 \\
\tilde q_P & \rep{\overline N_c'} & 0 & 0 & 0 & -1 \\
M_{ab} & 0 & 0 & 2 & 2 & 0 \\
M_{aQ} & 0 & 0 & 1 & 2 & 1 \\
M_{Pb} & 0 & 0 & 1 & 2 & 1 \\
M_{PQ} & 0 & 0 & 0 & 2 & 2 \\
T & 0 & 1 & -N_f & -N_c+1 & -k \\
\tilde T & 0 & -1 & -N_f & -N_c+1 & - k
\end{array}
\ee
This time all fields with non-vanishing charge under $U(1)_m$ are massive and can be integrated out: we are left with $U(N_f + k - N_c)_{-k}$ with $N_f$ flavors and singlets $M_{ab}$. The shift in CS levels are computed as follows:
\bea
\delta k_{gg} &= -k \;,\qquad \delta k_{TT} = -1 \;\qquad \delta k_{RR} = k(N_f + k-1) - N_c^2 \\
\delta k_{AA} &= N_f (k-N_f) \;,\qquad \delta k_{AR} = N_f(k-N_c) \;\qquad \delta k_{TA} = \delta k_{TR} = 0 \;.
\eea

\subsection{Yang-Mills-Chern-Simons theories with adjoint matter}
\label{sec:CS}

We start with the case of YM-CS theories without matter. The topological twist has actually no effect on an $\cN=2$ Chern-Simons theory without matter: the only fields charged under the R-symmetry are the gaugini that are auxiliary. We can thus compare our results  with the CS literature.

Recall what happens for an $\cN=2$ YM-CS theory with simple gauge group $G$ and level $k$.  For $|k| < h$, where $h$ is the dual Coxeter number of $G$ (for $SU(N)$, $h=N$), the theory breaks supersymmetry; for $|k|=h$ the theory confines; for $k > h$ (we assume positive $k$ for definiteness), the theory is equivalent to the pure bosonic CS theory at level
\be
\bar k = k - h
\ee
because the extra scalars and fermions in the adjoint can be integrated out shifting the level \cite{Kao:1995gf, Witten:1999ds}. The partition function for the $\cN=2$ YM-CS theory on $S^2\times S^1$ is thus $1$ for $k \ge h$ (since the Chern-Simons theory on $S^2$ has a single vacuum) and the Wilson loops satisfy the Verlinde algebra at level $\bar k$ \cite{Witten:1988hf}.

We will verify these well-known facts in our formalism. To compare with the CS literature, we  multiply the partition function by a sign factor $(-1)^r$, where $r$ is the rank of the gauge group. We will also consider the case of YM-CS theories with adjoint matter which have been recently related to complex Chern-Simons theory  \cite{Gukov:2015sna}.

\subsubsection{$U(1)_k$ Chern-Simons theory}

Consider a pure $U(1)_k$ CS theory, which is the same as its $\cN=2$ version since all auxiliary fields are neutral. We introduce a background flux $\ft$  and a fugacity $\xi$ for the topological symmetry. We have then
\be
Z =  - \sum_{\fm\in\bZ} \int \frac{dx}{2\pi i x} \, x^{k\fm+\ft} \xi^\fm \;,
\ee
where we introduced a minus sign for later convenience. Since there are no poles in the bulk, we only have contributions from the boundary. We can choose $\eta \gtrless 0$, and then we should discuss the two cases $k \gtrless 0$ separately. Eventually we obtain
\be
Z =  \sign(k) \sum_{\fm\in\bZ} \xi^\fm \delta_{k\fm+\ft,0} = \begin{cases}  \sign(k) \, \xi^{-\ft/k} &\text{if } \ft=0 \pmod{k} \\ 0 &\text{otherwise.} \end{cases}
\ee
Correctly, for $\ft=0$ we find $|Z|=1$ since Chern-Simons theory on $S^2$ has a single vacuum \cite{Witten:1988hf}.

\subsubsection{$U(N)_k$ Chern-Simons theory}
\label{U(N)k}

We can similarly  consider the theory $U(N)_k$. The partition function is
\be
Z = \frac{(-1)^N}{N!} \sum_{\vec \fm \in \bZ^N} \int \prod_{i=1}^N \frac{dx_i}{2\pi i} \, x_i^{ k\fm_i-N} \prod_{i\neq j}^N (x_i - x_j)  \;.
\ee
We have   set $\ft=0$ and $\xi=1$. We assume $k>0$ and we choose $\eta_i <0$. We should then take minus the residues at $x_i=0$. We have poles for $\fm_i \leq \frac{N-1}k$. We can resum the geometric series for $\fm_i \leq M-1$ for some large positive integer $M$, obtaining $\sum_{\fm_i \leq M-1} x_i^{ k\fm_i-N} =  x_i^{ kM-N}/ ( x_i^k-1)$. There are no longer poles at $x=0$, but rather at
\be
x_\alpha = e^{\frac{2\pi i}k \alpha} \qquad\text{with } \alpha = 1,\dots, k \;.
\ee
Because of the Jacobian factor, only poles with $x_i\neq x_j$ contribute, and we end up with a sum over combinations $I$ of $N$ distinct integers in $\{1,\dots, k\}$, that we denote by $C^k_N$. Thus we have a non-vanishing result only for $k \ge N$. When we substitute, the dependence on $M$ disappears and we are left with%
\footnote{For the last equality we used that, for fixed $\alpha$: $\displaystyle \prod_{\beta\, (\neq\alpha)}^k (x_\alpha - x_\beta) = \lim_{x\to x_\alpha} \frac{x^k-1}{x-x_\alpha} = k\, x_\alpha^{k-1} = \frac k{x_\alpha}$.}
\be
\label{Z CS U(N)}
Z = \sum_I^{C^k_N} \prod_{\alpha\in I} \frac{x_\alpha^{-N}}{ \prod_{\beta\in I^c} (x_\alpha - x_\beta)} = \sum_I^{C^k_N} \prod_{\alpha\in I, \beta \not\in I} \Big( 1 - \frac{x_\alpha}{x_\beta} \Big)^{-1} =\frac1{k^N} \sum_I^{C^k_N} \prod_{\alpha,\beta\in I,\, \alpha\neq\beta} \Big( 1 - \frac{x_\alpha}{x_\beta} \Big)  \;.
\ee
One can explicitly check that this expression gives $Z = 1$, as expected.

\subsubsection{$SU(N)$ versus $U(N)$}
\label{sec: SU(N) vs U(N)}

The partition function of an $SU(N)_k$ YM-CS theory with matter neutral under the center of the group, and the partition function of the $U(N)_k$ theory with the same matter content (and no flux for the topological symmetry) are equal. To see this, we rewrite the Haar measure of $U(N)$ in terms of those of $U(1)$ and $SU(N)$ by  decomposing $x_i = z \,\hat x_i$ with $\prod_{i=1}^N \hat x_i = 1$, and use $(z, \hat x_{i=1,\ldots, N-1})$ as coordinates on the Cartan subalgebra of $U(1) \times SU(N)$. The measure factorizes:%
\footnote{Since $z^N = \prod_{i=1}^N x_i$,  $z$ is only defined up to $N$-th roots of unit, therefore the final integral should be divided by $N$. This is compensated  by  an analogous factor of $N$ in the Jacobian, $\det \partial (x_i, x_N) / \partial(\hat x_j, z) = N z^{N-1} \hat x_N$ where $i,j=1,\dots, N-1$.}
\be
\int \prod_{i=1}^N \frac{dx_i}{2\pi i x_i} \prod_{i\neq j}^N \Big( 1 - \frac{x_i}{x_j} \Big) = \int  \frac{dz}{2\pi i z} \; \prod_{i=1}^{N-1} \frac{d\hat x_i}{2\pi i \hat x_i} \; \prod_{i\neq j}^N \Big( 1 - \frac{\hat x_i}{\hat x_j} \Big) \;.
\ee
The partition function of a $U(N)_k$ theory with matter is, up to normalization, 
\be
Z_{U(N)} = \frac{1}{N!} \sum_{\vec \fm \,\in\, \bZ^N} \int \prod_{i=1}^N \frac{dx_i}{2\pi i x_i} \, \xi^{\fm_i} x_i^{k\fm_i} \; \prod_{i\neq j}^N \Big( 1 - \frac{x_i}{x_j} \Big) \,\cdot\, M(x) \;,
\ee
where $M(x)$ is the matter contribution and we introduced a fugacity $\xi$ for the topological symmetry. Under the assumption that no matter is charged under $U(1)$,  $M$ is a function of $\hat x_i$ only and we can perform the $U(1)$ integral, obtaining a Kronecker delta function:
\be
 \int \frac{dz}{2\pi i z} \, z^{k \sum_{i=1}^N \fm_i} = \delta\big( {\textstyle \sum_{i=1}^N \fm_i } \big) \;.
\ee
We thus find 
\be
Z = \frac1{N!} \; \sum_{\vec \fm \,\in\, \bZ^N ,\, \sum \fm_i = 0} \; \int \prod_{i=1}^{N-1} \frac{d\hat x_i}{2\pi i \hat x_i} \; \prod_{i\neq j}^N \Big( 1 - \frac{\hat x_i}{\hat x_j} \Big) \; \prod_{i=1}^N \hat x_i^{k\fm_i} \,\cdot\, M(\hat x) \;,
\ee
which is the partition function of an $SU(N)_k$ theory.%
\footnote{The sub-lattice of $\bZ^N$ with $\sum \fm_i=0$ is the co-root lattice of $SU(N)$ or, equivalently, the weight lattice of the GNO dual group $SU(N)/\bZ_N$, as appropriate for a theory with  gauge group $SU(N)$. Indeed, the weights in the $U(N)$ lattice $\bZ^N$ are also weights for $U(1)$ and $SU(N)$. The $U(1)$ weight is $\sum \fm_i$. We can restrict to the $SU(N)$ weight lattice by gauge fixing the translation symmetry $\fm_i \rightarrow \fm_i + 1$ ($\forall i$) of the lattice. Each weight can be brought to the form $\fm_N=0$ with this symmetry, and weights with $\fm_N=0$ correspond to the full $SU(N)$ weight lattice. Only those weights with $\sum \fm_i =0 \pmod N$ can be reduced to the alternative form $\sum \fm_i=0$. Since the $SU(N)$ center $\bZ_N$ acts with weight $\sum \fm_i \pmod{N}$, this is the $SU(N)/\bZ_N$ weight lattice.}
The dependence on $\xi$ has disappeared.

\subsubsection{$SU(N)_ k$ Chern-Simons theory and the Verlinde algebra}
\label{SU(N)k}

The partition function of an $SU(N)_k$ theory without matter is equal to the partition function of $U(N)_k$, and therefore  $Z=1$ for $k\ge N$, as expected on general grounds \cite{Witten:1988hf}. It is interesting then to study correlation functions of Wilson loops: their algebra in Chern-Simons theory is known as the Verlinde algebra. In particular, the structure constants of this algebra---computable as the three-point functions of Wilson loops on $S^2\times S^1$---encode the fusion rules of primary fields in the Kac-Moody algebra  $\wh\su(N)_{\bar{k}}$ \cite{Witten:1988hf} . 

We can extract general information about the Wilson loop algebra by using an argument similar to that in \cite{Kapustin:2013hpk}.  Consider a generic  normalized integral
of the form 
\be
\label{genSU2}
\langle f(x) \rangle = \frac1Z \; \frac{1}{N!} \; \sum_{\vec \fm \,\in\, \bZ^N ,\, \sum \fm_i = 0} \; \int \prod_{i=1}^{N-1} \frac{d\hat x_i}{2\pi i \hat x_i} \; \prod_{i\neq j}^N \Big( 1 - \frac{\hat x_i}{\hat x_j} \Big) \; \prod_{i=1}^N \hat x_i^{k\fm_i} \cdot\, f(\hat x)
\ee
where $f$ is a function on the Cartan subalgebra of $SU(N)$. Changing the summation variable $\vec \fm \rightarrow \vec \fm +\delta$, where $\delta$ is a generic co-root (an element of the weight lattice of $SU(N)/\bZ_N$), we find
\be
\label{rel}
\langle f(x) \, x^{k \delta} \rangle = \langle f(x) \rangle \;.
\ee
We can apply this relation to a Wilson loop $\chi_{\lambda}(x)$  associated with a representation  of highest weight $\lambda$ and obtain
\be
\label{Verlindealgebra}
\langle f(x) \, \chi_\lambda(x)  \rangle =  (-1)^w \, \langle f(x) \, \chi_{\lambda(\delta)}(x) \rangle \;,
\ee
where $\lambda(\delta)$ is determined by reflecting the weight $\lambda+\rho+k \delta$ into the interior of the fundamental Weyl chamber by an element $w$ of the Weyl group $W$: $w( \lambda+\rho+k \delta) = \lambda(\delta) +\rho$. Here $\rho$ is half the sum of all positive roots, also known as the Weyl vector. Whenever $\lambda+\rho+k \delta$ is on the boundary of the Weyl chamber and cannot be reflected into the interior, $\langle f  \, \chi_\lambda \rangle =0$. 

The identities (\ref{Verlindealgebra}) can be derived by using the Weyl Character formula 
\be
\chi_\lambda(x) \equiv \Tr_\lambda x = \frac{A_{\lambda+\rho}}{A_\rho} \;,
\ee
where $A_\sigma = \sum_{w\in W} (-1)^w x^{w(\sigma)}$. The denominator essentially cancels the Haar measure in the correlation functions (\ref{rel}). We focus on the terms in the numerator and, using an adapted co-root for each term, we can write
\be
\Big\langle f \, \frac{A_{\lambda+\rho}}{A_\rho} \Big\rangle = \sum_{w\in W} (-1)^w \Big\langle f \, \frac{x^{w(\lambda+\rho)}}{A_\rho} \Big\rangle =  \sum_{w\in W} (-1)^w \Big\langle f \, \frac{x^{w(\lambda+\rho) + k w(\delta)}}{A_\rho} \Big\rangle = \Big\langle f \, \frac{A_{\lambda+\rho + k\delta}}{A_\rho} \Big\rangle \;.
\ee
If $\lambda+k \delta$ is a dominant weight, the final expression gives the character $\chi_{\lambda+k \delta}$. Otherwise, either $\lambda+\rho+k \delta$ is on the boundary of a  Weyl chamber and $A_{\lambda+\rho + k\delta}$ vanishes by definition, or we can find an element $w$ of the Weyl group that maps $\lambda+\rho+k \delta$ into the interior of the fundamental Weyl chamber, $w( \lambda+\rho+k \delta) \equiv \lambda(\delta)+\rho$, and the final expression gives, up to a sign $(-1)^w$, the character
 $\chi_{\lambda(\delta)}$. The shifts by $k\delta$ extend the ordinary Weyl group to the affine version. 
 
The relations (\ref{Verlindealgebra}) induce an equivalence among representations and define the Verlinde algebra.  Using the relations (\ref{Verlindealgebra}), every representation can be identified with one of the integrable irreducible representations of the Kac-Moody algebra  $\wh\su(N)_{\bar{k}}$. The integrable representations  are those whose Dynkin labels $\lambda= [\lambda_1,\lambda_2,...,\lambda_{N-1}]$  satisfy $\sum \lambda_i \le \bar k = k - N$, or, equivalently,  whose Young diagram have the first row of length less than or equal to $\bar k$ (see for example \cite{DiFrancesco:1997nk}).%
\footnote{The $SU(N)$ roots and weights can be written using the standard basis $e_i$  of $\bR^N$ and restricting it to the plane orthogonal to $(1,1,\ldots,1)$, \ie{} $\hat e_i = e_i - \frac1N \sum_j e_j$.The simple roots are $\hat e_i-\hat e_{i+1}$ and the fundamental weights are $\omega_i = \sum_{j=1}^i \hat e_j$. The weight $\lambda= [\lambda_1,\lambda_2,...,\lambda_{N-1}]$ is then $\sum_i \lambda_i \omega_i = \sum_i m_i \hat e_i$, where $m_i = \sum_{j=i}^{N-1} \lambda_j$ are the lengths of the rows in the corresponding Young diagram.}

Using the relations (\ref{Verlindealgebra}), we can find  the correlation functions of Wilson loops corresponding to integrable representations. For example, for $\wh\su(2)_{\bar{k}}$ there are ${\bar k}+1$  integrable irreducible representations, corresponding to  the spins from zero to $\frac{\bar k}2$. It is easy to see that the one-point functions vanish except for the trivial representation, $\langle \chi_\lambda \rangle =  \delta_{\lambda 0}$, and the two-point functions are diagonal, $\langle \chi_\lambda \, \chi_\mu \rangle =  \delta_{\lambda\mu}$. The three-point functions can be extracted from (\ref{Verlindealgebra})---or explicitly computed with (\ref{genSU2})---and produce the known fusion coefficients \cite{DiFrancesco:1997nk} of the $\wh\su(2)_{\bar{k}}$ algebra:
\be
\langle \chi_\lambda \, \chi_\mu \, \chi_\nu \rangle = \begin{cases} 1 &\text{if } |\lambda - \mu| \leq \nu \leq \min\big( \lambda + \mu, 2\bar k - \lambda - \mu \big) \;,\quad \lambda + \mu + \nu = 0 \pmod{2} \;, \\ 0 & \text{otherwise.} \end{cases}
\ee
These results  generalize to $\wh\su(N)_{\bar{k}}$ and are in agreement  with the general expectations for Chern-Simons theories \cite{Witten:1988hf}. 

We finish this section with an observation about the partition function itself. It is interesting to rewrite the ``trivial" result $Z=1$ in a different form to make contact with  the representation theory of  $\wh\su(N)_{\bar{k}}$ and the Verlinde formula \cite{Verlinde:1988sn}. The last expression in (\ref{Z CS U(N)}) can be cast as    
\be
Z = \frac1{k^N} \sum_{\substack{k > m_1 > \cdots \quad \\ \quad \cdots > m_N \geq 0}} \hspace{-1em} \det \Big( 1 - \Ad \big( e^{2\pi i m_i/k} \big) \Big) = \frac1{N k^{N-1}} \sum_{\substack{k > m_1 > \cdots \quad \\ \quad \cdots > m_N=0}} \hspace{-1em} \det \Big( 1- \Ad \big( e^{2\pi i m_i/k} \big) \Big) \;,
\ee
where $\Ad(x_i)$ denotes the action of the Cartan element $\text{diag}(x_1,\ldots, x_N)$ on the adjoint representation of $SU(N)$.%
\footnote{The get the second equality, first rewrite the first expression in terms of unordered integers $m_i$, introducing a factor $N!$ in the denominator. The terms with coinciding $m_i$ vanish. The determinant is invariant under $m_i \rightarrow m_i + 1$ (simultaneous for all $i$): using this shift symmetry we can set $m_N=0$, and there are $k$ elements in each orbit. We can then use the Weyl group $S_{N-1}$ restricted to $m_N=0$ to order the remaining $m_i$: this cancels a factor $(N-1)!$.}
In turn, this can be written as
\be
\label{Verlinde12 formula Blau}
Z = \frac1{N k^{N-1}} \sum_\lambda \prod_{\alpha\in SU(N)} \det\Big( 1- e^{2 \pi i\alpha(\lambda+\rho)/k} \Big) \;,
\ee
where $\alpha$ are the roots of $SU(N)$, the sum is restricted to the weights  $\lambda= [\lambda_1,\lambda_2,...,\lambda_{N-1}]$ corresponding to the integrable representations of $\wh\su(N)_{\bar{k}}$, and the Weyl vector $\rho=[1,...,1]$. In fact, (\ref{Verlinde12 formula Blau}) is precisely the special case $g=0$ of the Verlinde formula 
$$
\frac1{(N k^{N-1})^{1-g}} \sum_\lambda \prod_\alpha \det\big( 1- e^{2 \pi i\alpha(\lambda+\rho)/k} \big)^{1-g} \;,
$$
for the partition function of the $SU(N)_{\bar k}$ CS theory on $\Sigma \times S^1$ or, equivalently, the number of conformal blocks of the $SU(N)$ WZW model at level $k$ on a Riemann surface $\Sigma_g$ of genus $g$ \cite{Verlinde:1988sn,Witten:1988hf,Blau:1993tv}.

\subsubsection{$U(N)_k$ with adjoint matter}

To further test our formula, we now compute the partition function of $U(N)_k$ YM-CS with a massive adjoint chiral multiplet. As discussed in \cite{Gukov:2015sna}, this theory is related to complex Chern-Simons  and the equivariant  Verlinde formula.

The partition function is
\be
Z = \frac{(-1)^N}{N!} \Big( \frac{y^\frac12}{1-y}\Big)^{\tilde\fn N} \sum_{\vec \fm \,\in\, \bZ^N} \int  \prod_{i=1}^N \frac{dx_i}{2\pi i x_i} \xi^{\fm_i} x_i^{k\fm_i} \prod_{i\neq j}^N \Big( 1 - \frac{x_i}{x_j} \Big) \Big( \frac{x_i^\frac12 x_j^{-\frac12} y^\frac12}{1-\frac{x_i}{x_j} y} \Big)^{\fm_i - \fm_j + \tilde \fn } \;,
\ee
where $\tilde \fn=\fn-R+1$, $R$ is the R-charge of the adjoint chiral field, and $y,\fn$ are the fugacity and background flux for the $U(1)$ flavor symmetry that assigns charge $1$ to the adjoint chiral. As we discussed in section \ref{sec: SU(N) vs U(N)}, this is the same as the partition function of $SU(N)_k$ and there is no dependence on $\xi$, however keeping $\xi$ will facilitate the computation.

We choose $\eta = (-1, \dots,-1)$ and take $k>0$. The partition function has no singularities at finite points since the intersections of the planes  $H_{ij} = \{x_i = y x_j\}$
collapse to $x_i=0$ for generic $y$. With our rules, it remains to compute an integral over an  intricate structure of intersections at the boundary. We can avoid this by resuming the poles first. We  write
\be
Z = \frac{(-1)^N y^{\tilde \fn N^2/2}}{N!\, (1-y)^{\tilde\fn N}} \sum_{\vec\fm\,\in\,\bZ^N} \int  \prod_{i=1}^N \frac{dx_i}{2\pi ix_i} \xi^{\fm_i} x_i^{k\fm_i} \prod_{j(\neq i)}^N \Big( \frac{x_i - x_j y}{x_j - x_i y} \Big)^{\fm_i} \cdot \prod_{i\neq j}^N \frac{1 - \frac{x_i}{x_j}}{\big( 1 - \frac{x_i}{x_j}y \big)^{\tilde \fn}} \;.
\ee
We then perform the sum over $\fm_i \leq M-1$, where $M$ is some large integer. Recalling that our prescription for the residues includes a further factor $(-1)^N$, we find
\be
Z = \frac{y^{\tilde \fn N^2/2}}{N!\, (1-y)^{\tilde \fn N}} \int_\circlearrowleft \prod_{i=1}^N \frac{dx_i}{2\pi i x_i} \; \frac{\big( \xi\, e^{iB_i(x)} \big)^M }{ \xi \, e^{iB_i(x)} - 1} \, \cdot \, \prod_{i\neq j}^N \frac{1 - \frac{x_i}{x_j}}{\big( 1 - \frac{x_i}{x_j}y \big)^{\tilde \fn}} \;,
\ee
where we defined the quantities
\be
e^{iB_i(x)} = x_i^k \prod_{j(\neq i)}^N \frac{ x_i - x_j y}{ x_j - x_i y} \;.
\ee
The presence of $e^{i M B_i(x)}$ with an arbitrarily large $M$ in the numerator guarantees that there are no poles at $x_i=0$ or $x_i=y x_j$.  
The only relevant poles are at $\xi e^{iB_i(x)}= 1$ for all $i$, and the dependence on $\xi$ can easily be reabsorbed in a rescaling of all $x_i$. At these poles   the dependence on $M$ in the numerator disappears. To compute the residues, we should take the Jacobian of the denominator. We then find
\be
Z = \frac{y^{\tilde\fn N^2/2}}{(1-y)^{\tilde \fn N}} \sum_I \frac1{\det \frac{\partial e^{iB_i}}{\partial x_j}} \; \prod_{i=1}^N \frac1{\xi x_i} \; \prod_{i\neq j}^N \frac{1 - \frac{x_i}{x_j}}{\big( 1 - \frac{x_i}{x_j}y \big)^{\tilde \fn}} \;,
\ee
where the sum is over all unordered collections of solutions to the ``Bethe ansatz'' equations $e^{iB_i(x)} = \xi^{-1}$. It is convenient to parameterize the solutions as $x_j = e^{i\theta_j}$, then
$$
\frac{\partial e^{iB_i}}{\partial x_j} = \frac{e^{iB_i}}{x_j} \, \parfrac{B_i}{\theta_j}
$$
and
\be
Z = \frac{y^{\tilde\fn N^2/2}}{(1-y)^{\tilde \fn N}} \; \sum_I \Big( \det \parfrac{B_i}{\theta_j} \Big)^{-1} \prod_{i\neq j}^N \frac{ 1 - e^{i(\theta_i - \theta_j)} }{ \big( 1 - y \, e^{i(\theta_i - \theta_j)} \big)^{\tilde \fn}} \;.
\ee
This formula with $\fn=0$ agrees with the results in \cite{Gukov:2015sna}. As observed there, the formula for R-charge $R=2$  ($\tilde \fn = -1$) is an equivariant Verlinde formula and is related to complex Chern-Simons theory; for $R=0$ ($\tilde \fn=1$) the formula has appeared in the mathematical literature as an index formula for the moduli stack of algebraic bundles over the sphere.

\subsubsection{The ``duality appetizer''}
\label{sec: duality appetizer}

In \cite{Jafferis:2011ns} a duality was proposed between $SU(2)_1$ with one adjoint chiral multiplet $\Phi$, and the theory of a free chiral multiplet $Y = \Tr \Phi^2$. The former theory is a special case of those considered in the previous section, however for low rank and Chern-Simons level we can write down the partition function explicitly and we can test the duality.

In the electric theory there is a flavor $U(1)_F$ symmetry that rotates the adjoint chiral with charge $1$: we denote by $y,\fn$ its fugacity and background flux, respectively. To cancel a parity anomaly, we introduce a CS term $k_{FF} = \frac12$. We assign R-charge $1$ to $\Phi$. The partition function of the electric theory is then
\be
\label{appetizer}
Z = \frac12 \sum_{\fm\,\in\,\bZ} \int \frac{dx}{2\pi i x}\, \frac{(1-x^2)^2}{x^2} \, x^{2\fm} \, y^{\fn/2} \, \Big( \frac{xy^\frac12}{1-x^2y} \Big)^{2\fm+\fn} \Big( \frac{y^\frac12}{1-y} \Big)^\fn \Big( \frac{ x^{-1}y^\frac12}{1-x^{-2}y} \Big)^{-2\fm + \fn} \;.
\ee

We choose $\eta <0$, therefore we should collect minus the residues at $x=0,\pm\sqrt y$. The poles at $x=\pm\sqrt y$ exist for $\fm \leq (\fn-1)/2$, whilst the pole at $x=0$ exists for $\fm \leq 1-\fn$. We can sum the series for $\fm \leq M-1$, where $M$ is some large integer, in a regime $y \ll 1$. The series is uniformly convergent along a contour right outside the unit circle, and such a contour includes the poles at $x=0, \pm\sqrt y$ and no others. The series yields the expression
$$
Z = \frac12 \int_{\circlearrowleft} \frac{dx}{2\pi i} \, \frac{ x^{6M-3} y^{2\fn} (1-x^2) \big( 1 - \frac y{x^2} \big)^{2M-n} }{ (1-y)^\fn (1-x^2 y)^{2M +\fn -2} \big( 1 + x^2 (1-2y-y^2) + x^4 \big) } \;.
$$
Now the contour includes four poles, produced by the last term in the denominator, located at $x = \big( \pm (y+1) \pm \sqrt{ y^2 + 2y-3} \big)/2$. The residues can be explicitly computed and summed: the dependence on $M$ drops out, and we obtain
\be
\label{appetizer result}
Z = y \, \Big( \frac y{1-y^2} \Big)^{2\fn-1} \;.
\ee

The dual theory is given by a free chiral multiplet $Y = \Tr\Phi^2$, with flavor charge 2 and R-charge 2 (plus  a topological sector). As we infer from the partition function, there must also be an R-flavor Chern-Simons term $k_{RF} = -1$. Then (\ref{appetizer result}) precisely agrees with the partition function of the dual theory.

\section{Refinement by angular momentum}
\label{sec: refinement}

If we choose a metric on $S^2$ invariant under a $U(1)$ isometry, we can refine the partition function by adding a fugacity $\zeta = e^{i\varsigma/2}$ for the angular momentum on $S^2$. From the path-integral point of view, this is achieved by deforming the metric on $S^2 \times S^1$ as%
\footnote{More general supersymmetric backgrounds have been constructed in \cite{Klare:2012gn, Closset:2012ru}, and the dependence of the partition function on the background has been studied in \cite{Closset:2013vra}.}
\be
ds^2 = R^2 \big( d\theta^2 + f(\theta)^2 (d\varphi - \varsigma\, dt)^2 \big) + \beta^2 dt^2 \;,
\ee
where $ds_{S^2}^2 = R^2 \big( d\theta^2 + f(\theta)^2 d\varphi^2\big)$ is a generic metric on $S^2$ with $U(1)$ isometry along $\varphi$. The special case of a round metric was considered in (\ref{deformed metric}), and in fact, since the theory is quasi-topological and independent of the metric on $S^2$, we can do all computations using the round metric. Reducing down to $S^1$, the deformed metric yields a quantum mechanical index with a fugacity $\zeta$ for the angular momentum $L_\varphi$ of rotations along $\varphi$, in other words $Z_{S^2 \times S^1}$ computes the index
$$
I = \Tr_\cH (-1)^F e^{-\beta H} e^{iJ_f A_f} \zeta^{2L_\varphi} \;.
$$
Such a refined index is easily computed by noticing that the $\big| \rho(\fm) - q_\rho + 1 \big|$ Landau zero-modes on $S^2$ form a representation of the $SU(2)$ group of rotations. The refined one-loop determinant for a chiral multiplet in representation $\fR$ is then:
\be
\label{refoneloop}
Z^\text{chiral}_\text{1-loop} = \prod_{\rho \in \fR} \; \prod_{j = - \frac{|B|-1}2}^{\frac{|B|-1}2} \bigg( \frac{x^{\rho/2} \zeta^j }{ 1-x^\rho \zeta^{2j} } \bigg)^{\sign B} \;,\qquad\qquad B = \rho(\fm) - q_\rho + 1 \;.
\ee
The factor $\zeta^j$ in the numerator cancels out and could be omitted.
The determinant can be conveniently rewritten in terms of the $q$-Pochhammer symbol $(x;q)_n$ as
\be
\boxed{
Z^\text{chiral}_\text{1-loop} = \prod_{\rho \in \fR} \frac{x^{\rho B/2}}{(x^\rho \zeta^{1-B}; \zeta^2)_B} \;.
}
\ee
Recall that, for integer values of $n$, the $q$-Pochhammer is defined as
\be
(x;q)_n = \begin{cases} \prod\nolimits_{j=0}^{n-1} (1-xq^j) & \text{for } n \geq 0 \\ \prod\nolimits_{j=-n}^{-1} (1-xq^j)^{-1} & \text{for } n \leq 0 \;, \end{cases}
\ee
and it satisfies
\be
(x;q)_n = \frac1{(xq^n; q)_{-n}} = (xq^{n-1}; q^{-1})_n \;,\qquad\qquad (x;q)_{m+n} = (x;q)_m (xq^m; q)_n \;.
\ee

We can quickly derive the one-loop determinant for gauge fields using the fact that, up to a flux-dependent sign, it equals that of a chiral multiplet with R-charge 2. We get:
\be
\boxed{
Z^\text{gauge}_\text{1-loop} = \prod_{\alpha>0} \frac{ (1-x^\alpha \zeta^{\alpha(\fm)})( 1- x^{-\alpha} \zeta^{\alpha(\fm)})}{ \zeta^{\alpha(\fm)} } \; (i\, du)^r  = \zeta^{-\sum_{\alpha>0} |\alpha(\fm)|} \prod_\alpha (1-x^\alpha \zeta^{|\alpha(\fm)|}) \; (i\, du)^r  \;.
}
\ee
The classical CS actions are the same as before. A formal derivation of the above statements  is given in appendix \ref{sec:Arefinement}.

A little bit of care has to be given to the Wilson loops. As we found at the end of section \ref{sec: SUSY Lag}, to be supersymmetric a loop has to lay along the vector field $e_3 = \frac1\beta(\partial_t + \varsigma \partial_\varphi)$, \ie{} its embedding function must be $x^\mu(\tau) = (\theta_0, \varsigma\tau, \tau)$ for some $\theta_0$, and we have to make sure that the loop closes. One possibility is to place the loops at the poles, with $\theta_0 = 0,\pi$. In this case, the classical action term---to be inserted in the localization formula---is
\be
\label{Wilson loop poles}
\text{Poles:} \hspace{2cm} W = \Tr_R \, e^{iu \,\pm\, i \varsigma \frac\fm2} = \sum_{\rho\in R} x^\rho \, \zeta^{\pm \rho(\fm)} \;. \hspace{2cm}
\ee
The signs $\pm$ refer to the North and South pole of the sphere, respectively.
Notice that this yields a deformation of the Verlinde algebra when we place multiple loops on top of each other, because subtle contact terms intervene.

Another possibility is to choose a ``rational'' value $\varsigma = 2\pi \frac pq$, where $p,q\in\bZ$ are coprime. Then the loops can be place at any $\theta_0$: they wrap $q$ times $S^1$ and $p$ times the $\varphi$ direction. The classical action term, to insert in the localization formula, is
\be
\hspace{.3cm} \text{Generic points:} \hspace{2cm}
W = \Tr_R \, e^{iqu \,+\, i \pi p \fm} = \sum_{\rho\in R} (-1)^{p \rho(\fm)} \, x^{q\rho} \hspace{2cm} \;.
\ee

Notice that the refinement by angular momentum only exists on $S^2$ and not at higher genus (although a certain refinement, the elliptic genus, exists on $T^2$ as well).

\subsection{$U(1)_{1/2}$ with one chiral}

We would like to perform some simple checks of the refined formula. As a first example, consider again the $U(1)_{1/2}$ theory with a single chiral multiplet of charge $1$. We follow the same conventions as before (and include the sign factor$(-1)^\fm$). The partition function is then
\be
Z = \sum_{\fm\in\bZ} \int \frac{dx}{2\pi i} \; \frac{(-\xi x)^\fm x^{\ft -1}}{ (x \zeta^{1-\fm}; \zeta^2)_\fm } \;.
\ee

It is convenient to choose $\eta<0$ so that we have to pick minus the residue at $x=0$ for all terms with $\fm\leq -\ft$. We use the following two formul\ae:
\be
\frac1{(x;q)_n} = \sum_{j=0}^\infty \frac{(q^n;q)_j}{(q;q)_j}\, x^j \;,\qquad\qquad
(x;q)_n = (-x)^n q^{\frac{n(n-1)}2} (x^{-1} q^{1-n}; q)_n \;.
\ee
The first one is the $q$-binomial theorem and it is valid for $|x|<1$, $|q|<1$ and $n\in\bZ$; for $n \leq 0$ only the terms with $j\leq n$ are non-vanishing and the right-hand-side is a finite polynomial. Expanding the integrand we find
\bea
Z &= - \sum_{\fm\leq-\ft} \;\Res_{x=0}\; (-\xi)^\fm \sum_{j=0}^\infty \frac{(\zeta^{2\fm};\zeta^2)_j}{(\zeta^2;\zeta^2)_j}\, \zeta^{(1-\fm)j} x^{\fm + \ft + j -1} \\
&= - \sum_{\fm\leq -\ft} (-\xi)^\fm \, \frac{(\zeta^{2\fm};\zeta^2)_{-\fm-\ft}}{ (\zeta^2;\zeta^2)_{-\fm-\ft} } \, \zeta^{(\fm-1)(\fm+\ft)} = (-1)^{\ft+1} \sum_{n\geq0} \xi^{-\ft-n} \frac{(\zeta^{2(\ft+1)};\zeta^2)_n}{ (\zeta^2;\zeta^2)_n} \zeta^{-n\ft} \\
&= - \frac{(-\xi)^{-\ft}}{ (\xi^{-1} \zeta^{-\ft}; \zeta^2)_{\ft+1}} = \frac{\xi}{(\xi \zeta^{-\ft}; \zeta^2)_{\ft+1}} \;.
\eea
This is exactly the refinement of the partition function (\ref{dualU(1)}) of the dual theory, which, as discussed in section \ref{sec:U(1)k=1/2}, is the theory of a free chiral field.

\subsection{Pure Chern-Simons theories}

As a further check, we can evaluate the refined partition function of the $\cN=2$ $U(N)_k$ CS theory with $k = \bar k + N > N$, to verify that it is independent of $\zeta$. In fact Chern-Simons theory is topological and should not depend on a continuous deformation of the metric. The refined partition function  reads
\be
Z = \frac{(-1)^N}{N!} \sum_{\vec \fm \,\in\, \bZ^N} \int \prod_{i=1}^N \frac{dx_i}{2\pi i x_i} \, x_i^{k\fm_i} \; \prod_{i > j}^N \frac{  (1 - x_i \zeta^{\fm_i}/x_j \zeta^{\fm_j}) (1 - x_j \zeta^{\fm_i}/x_i \zeta^{\fm_j})  }{\zeta^{\fm_i}/\zeta^{\fm_j} }  \;.
\ee
Choosing $\eta_i<0$, we should collect minus the residues at $x_i=0$. Since the path-integral of CS theory is saturated by flat connections, we could expect that only the sector $\fm_i=0$ contributes to the path-integral. If this is the case, the refined partition function coincides with the unrefined one---and both give $Z= 1$---being the average of the identity over the $U(N)$ Haar measure.

It is easy to see that only $\fm_i=0$ contributes. Let us re-write the partition function as
 \be
Z = \frac{1}{N!} \sum_{\vec \fm \,\in\, \bZ^N} \int_\circlearrowleft \prod_{i=1}^N \frac{dx_i}{2\pi i}  \, x_i^{k\fm_i -N} \; \prod_{i > j}^N \big( x_j \zeta^{\fm_j} - x_i \zeta^{\fm_i} \big) \Big( \frac{x_i}{\zeta^{\fm_i}} - \frac{x_j}{\zeta^{\fm_j}} \Big) \;,
\ee
and use $\prod_{i>j}^N (z_i-z_j) = \sum_{\sigma\in S_N} \epsilon(\sigma) \, z_1^{\sigma(1)-1} \cdots z_N^{\sigma(N)-1}$. We find
\be
Z=  \frac{1}{N!} \sum_{\vec\fm \,\in\, \bZ^N}  \sum_{\sigma,w\in S_N} \epsilon(\sigma) \, \epsilon(w) \int_\circlearrowleft \prod_{i=1}^N \frac{dx_i}{2\pi i}\,  x_i^{k\fm_i -N +\sigma(i) +w(i)-2}\,  \zeta^{\fm_i \left( \sigma(i)-w(i) \rule{0pt}{.6em} \right)} (-1)^{N(N-1)/2} \;.
\ee
The integral is non vanishing only for $k\fm_i  =N -\sigma(i) -w(i)+1$. Since $\sigma(i)$ and $w(i)$ run over $\{1,\ldots, N\}$, then $| N -\sigma(i) -w(i)+1| \leq N-1$, and since
$k > N$, then the previous equation can only be satisfied with all $\fm_i=0$. The solutions have $\fm_i =0$, arbitrary permutation $\sigma$, and permutation $w(i) = N +1 - \sigma(i)$. Then $ \epsilon(\sigma) \, \epsilon(w)=(-1)^{N(N-1)/2}$ and $Z = \frac1{N!} \sum_{\sigma\in S_N} 1 = 1$.

The supersymmetric Wilson loops, on the other hand, are along orbits that depend on the choice of $\varsigma$. As we discussed before, in order to have room for non-coincident loops, we should choose $\varsigma = 2\pi p/q$ with $p,q\in\bZ$ coprime. Then the loops wrap $q$ times around $S^1$ and $p$ times around $S^2$: they are knotted in a topologically nontrival way, rather than continuously deformed as $\varsigma$ is varied. This explains why correlation functions of Wilson loops depend on $\varsigma$.

\subsection{The vortex partition function}

Finally, we would like to make contact with the other partition functions, defined on different manifolds or supersymmetric backgrounds. As first observed in \cite{Pasquetti:2011fj} and later elaborated upon in \cite{Beem:2012mb, Cecotti:2013mba}, both the $S^3$ partition function of \cite{Kapustin:2009kz}, the $S^2 \times S^1$ supersymmetric index of \cite{Kim:2009wb, Imamura:2011su} and the lens space index of \cite{Benini:2011nc} ``factorize'' into building elements which can be interpreted as the partition function on $\bR^2_\Omega \times S^1$ in an $\Omega$-background. For $U(N_c)_k$ theories with (anti)fundamentals, the basic elements are K-theoretic vortex partition functions.%
\footnote{More precisely, this is true for the so-called ``maximally chiral theories'' where $|k| \leq |N_f - N_a|/2$ \cite{Benini:2013yva}.}
A mechanism behind this factorization, called ``Higgs branch localization'', has been first discovered in two dimensions \cite{Benini:2012ui, Doroud:2012xw} and later generalized to three \cite{Fujitsuka:2013fga, Benini:2013yva} and four \cite{Yoshida:2014qwa, Peelaers:2014ima} dimensions. Here we would like to show that even the twisted $S^2 \times S^1$ partition function factorizes, in terms of the very same $\bR^2_\Omega \times S^1$ elements. We show it for simple theories that lead to the vortex partition function, where the computation can be carried out explicitly.

In section \ref{sec: other dim} we will discuss generalizations of our formula to two and four dimensions. Although we only show the factorization of the twisted $S^2 \times S^1$ partition function here, the same statement is true for the twisted $S^2$ and $S^2 \times T^2$ partition functions as well. In particular in four dimensions one discovers the elliptic vortex partition function \cite{Yoshida:2014qwa, Peelaers:2014ima}, which is a building block of the 4d superconformal index of \cite{Romelsberger:2007ec}.

Consider $U(1)$ SQED with a single flavor pair, $N_f = 1$, \ie{} with one fundamental and one antifundamental. In view of a generalization to $N_f>1$, it is convenient to introduce separate parameters $(y,\fn)$ for the global symmetry that rotates the fundamental, and $(\tilde y,\tilde\fn)$ for the one that rotates the antifundamental, even though the two are the same flavor symmetry up to a gauge rotation. Assigning R-charge $0$ to the chiral fields, the partition function is
\be
Z = \sum_{\fm \in \bZ} \int \frac{dx}{2\pi i x} \, \xi^\fm x^\ft \, \frac{(y^{-\frac12} x^\frac12 )^{\fm - \fn +1} }{ (y^{-1} x \zeta^{-\fm + \fn}; \zeta^2)_{\fm - \fn+1} } \, \frac{(\tilde y^\frac12 x^{-\frac12} )^{-\fm + \tilde \fn+1} }{ (\tilde y x^{-1} \zeta^{\fm - \tilde\fn}; \zeta^2)_{-\fm + \tilde\fn+1} } \;.
\ee
We choose $\eta>0$, therefore we pick the residues from the fundamental at $x = y \zeta^{\fm - \fn - 2k}$, for $k=0, \dots, \fm - \fn$; such poles exist only for $\fm-\fn \geq 0$. We thus obtain a double sum $\sum_{\fm \geq \fn} \sum_{k=0}^{\fm-\fn}$, which is more conveniently written as $\sum_{m_1,m_2 \geq 0}$ with
\be
m_1 = \fm - \fn - k \;,\qquad\qquad m_2 = k \;.
\ee
In this notation the poles are at $x=y\zeta^{m_1 - m_2}$. Taking the residues we get
\be
Z = - \sum_{m_1,m_2\geq0} \frac{ \xi^{m_1 + m_2+\fn} \, y^\ft \, (y/\tilde y)^\frac{ m_1 + m_2 + \fn - \tilde \fn -1}2 \, \zeta^{m_1^2 - m_2^2 + (m_1 - m_2)\smath( \frac{\fn - \tilde \fn}2 + \ft \smath)} }{ (\zeta^2;\zeta^2)_{m_1} (\zeta^{-2}; \zeta^{-2})_{m_2} \big( \frac{\tilde y}y \zeta^{2m_2 + \fn - \tilde \fn}; \zeta^2 \big)_{-m_1 - m_2 - \fn +\tilde\fn + 1} } \;.
\ee
The last $q$-Pochhammer can be factorized into three factors:
$$
\frac1{\big( \frac{\tilde y}y \zeta^{2m_2+\fn-\tilde\fn} ;\zeta^2 )_{-m1-m2-\fn+\tilde\fn+1}} = \frac{ \big( \tfrac{\tilde y}y \zeta^{\tilde \fn-\fn} ; \zeta^{-2} \big)_{m_1} \big( \tfrac{\tilde y}y \zeta^{\fn - \tilde\fn}; \zeta^2 \big)_{m_2} }{ \big( \frac{\tilde y}y \zeta^{\fn - \tilde\fn}; \zeta^2 \big)_{\tilde\fn - \fn +1} } \;.
$$
We can thus write the partition function as the product of three factors:
\be
Z = Z_\text{1-loop} \; Z_\text{vortex}(\zeta) \; Z_\text{vortex}(\zeta^{-1}) \;.
\ee
The one-loop factor is
\be
Z_\text{1-loop} = - \xi^\fn y^\ft \, \frac{ (\tilde y^\frac12 y^{-\frac12} )^{\tilde\fn - \fn + 1} }{ ( \tilde y y^{-1} \zeta^{\fn - \tilde\fn}; \zeta^2)_{\tilde \fn -\fn+1} } \;.
\ee
We recognise that this is the classical action times the one-loop determinant of the antifundamental (which is not Higgsed when the FI term is positive), evaluated on the background $x=y$, $\fm=\fn$ where the fundamental is massless and gets Higgsed (with positive FI). The vortex contributions are
\bea
Z_\text{vortex}(\zeta) &= \sum_{m=0}^\infty \xi^m \Big( \frac y{\tilde y} \Big)^{\frac m2} \zeta^{m^2 + m \smath( \frac{\fn - \tilde\fn}2 + \ft \smath)} \, \frac{ \big( \frac{\tilde y}y \zeta^{\tilde\fn - \fn}; \zeta^{-2} \big)_m }{ (\zeta^2; \zeta^2)_m } \\
&= \sum_{m=0}^\infty ( \xi \, \zeta^\ft )^m \prod_{j=0}^{m-1} \frac{ \sinh \, \log \, \Big[ \big( \frac y{\tilde y} \big)^\frac12 \zeta^{ \frac{\fn - \tilde\fn}2 + j } \Big] }{ \sinh \, \log \, \zeta^{j-m} } \;.
\eea
The last expression is the classic form of the K-theoretic vortex partition function (see for instance (2.78) in \cite{Benini:2013yva}). This computation can be generalized to theories with a CS term at level $k$, as well as $N_f > 1$ and gauge group $U(N)$ with $N>1$.

\section{Other dimensions}
\label{sec: other dim}

Our three-dimensional formalism can be easily generalized to theories in two dimensions on $S^2$, and in four dimensions on $S^2 \times T^2$. In both cases supersymmetry is preserved with a topological A-twist on $S^2$, and a refinement by the angular momentum is possible.
The two-dimensional formula is obtained from the one on $S^2 \times S^1$ by shrinking the radius $\beta$ of the circle to zero. Besides, we can insert twisted chiral operators at arbitrary points of $S^2$ (they are, in a sense, the reduction of Wilson loops on $S^1$). The four-dimensional formula, instead, is obtained by considering elliptic generalizations of our expressions. The two-dimensional result can be applied to the study of amplitudes of topologically twisted gauged linear sigma models. Our formula differs from the classic result of \cite{Witten:1993yc} because we do not integrate $D$ out until the end, obtaining in this way a ``Coulomb branch localization'' as opposed to the ``Higgs branch localization'' to vortices in \cite{Witten:1993yc}---using the language of \cite{Benini:2012ui}. Here we consider the two simple cases of sigma models with target the projective space and the quintic, leaving a more general analysis for further work.

One could also use our formul\ae{} to test non-perturbative dualities in two (for instance the dualities discovered in \cite{Hori:2006dk, Benini:2012ui, Benini:2014mia, Gomis:2014eya}) and four dimensions. In this paper, we only consider a very simple example of Seiberg duality \cite{Seiberg:1994pq} in 4d $\cN=1$ SQCD.

\subsection{Two-dimensional $\cN=(2,2)$ theories on $S^2$}

We consider two-dimensional $\cN=(2,2)$ theories with a vector-like R-symmetry, defined on $S^2$ with a topological A-twist. The theories do not have to be conformal, but it must be possible to choose integer R-charges. Unfortunately this excludes most Landau-Ginzburg (LG) models.%
\footnote{Not all. For instance, the quintic GLSM in the LG phase flows to a LG model quotiented by $\bZ_5$. Since the UV GLSM is consistent with A-twist, so must be its IR phases. Indeed, an integer R-symmetry exists.}
Therefore we will be mainly concerned with GLSMs. Very practically, the two-dimensional formula can be derived from the three-dimensional one by taking the limit in which the circle shrinks: $\beta \to 0$. This corresponds to an expansion around $u=0$, or $x=1$. The moduli space of bosonic zero-modes reduces to $\fM = \fh \times \fh = \fh_\bC$ and is parameterized by the complex scalar $\sigma$ in the vector multiplet. Let us summarize the results. Details, as well as a more systematic derivation along the lines of section \ref{sec: loc}, are given in appendix \ref{sec:A2d}. 

The one-loop determinant for a chiral field is given by the $x\to 1$ limit of the three-dimensional one-loop determinant in (\ref{chiral1-loop}), namely
\be
\boxed{
Z_\text{1-loop}^\text{chiral} = \prod_{\rho\in\fR} \Big[ \frac1{\rho(\sigma)} \Big]^{\rho(\fm) - q_\rho  + 1} \;.
}
\ee
The one-loop determinant for the vector multiplet is
\be
\boxed{
Z_\text{1-loop}^\text{gauge} = (-1)^{\sum_{\alpha>0} \alpha(\fm)} \prod_{\alpha \in G} \alpha(\sigma) \,\cdot\, (d\sigma)^r\;.
}
\ee
We do not drop the sign factor in front of the determinant, because in two dimensions there are no sign ambiguities related to the regularization (see appendix \ref{sec:A2d}).

In two dimensions we can define a twisted superpotential $\wt W(\sigma)$, which is a holomorphic function and it leads to the following bosonic action:
\be
\cL_{\widetilde W} + \overline{\cL_{\widetilde W}} \Big|_\text{bos} = 2i \re \widetilde W'(\sigma) \cdot D - 2i \im \widetilde W'(\sigma) \cdot F_{12} \;.
\ee
Of particular importance is a linear twisted superpotential, yielding a complexified FI term:
\be
\widetilde W_{\text{FI},\theta} = -\frac1{4\pi} (\zeta + i\theta) \Tr \sigma \qquad\Rightarrow\qquad \cL_{\text{FI}, \theta} = - i \frac\zeta{2\pi} \Tr D + i \frac\theta{2\pi} \Tr F_{12} \;.
\ee
Evaluated on-shell on almost-BPS configurations, the action gives
\be
e^{-S_{\widetilde W}} = e^{4\pi \widetilde W'(\sigma) \cdot \fm \,-\, 8\pi i R^2 \re \widetilde W'(\sigma) \cdot D_0} \;.
\ee
When specialized to the complexified FI term it becomes
\be
e^{- S_{\text{FI},\theta}} = e^{- (\zeta + i \theta) \Tr \fm + 2i R^2 \zeta \Tr D_0} = q^{\Tr \fm} \; e^{2i R^2 \zeta \Tr D_0} \;.
\ee
Here $q \equiv e^{-\zeta - i\theta}$, according to standard notation.

Any holomorphic function $f(\sigma)$ can be inserted in the path-integral because it is supersymmetric:
\be
Q \, f(\sigma) = \tilde Q\, f(\sigma) = 0 \;.
\ee

The computation of the partition function proceeds as in the three-dimensional case. In particular, the one-loop determinant has poles along singular hyperplanes $H_i$, with associated charge covectors $Q_i$. One has to choose a covector $\eta \in \fh^*$. The path-integral reduces to a sum over the magnetic fluxes $\fm$ on $S^2$, of the JK residues evaluated at the points where $r$ linearly independent hyperplanes meet ($r = \rank G$). In other words:
\be
Z_{S^2} = \frac1{|W|} \sum_{\fm \in \Gamma_\fh} \Bigg[ \; \sum_{\sigma_* \in\, \fM_\text{sing}^*} \JKres_{\sigma=\sigma_*} \big( \sQ(\sigma_*), \eta \big) \; Z_\text{int}(\sigma;\fm) \;+\; \text{boundary} \Bigg] \;.
\ee
The only novelty is how to treat the boundary terms in the region at infinity of $\fh_\bC$. The asymptotic behavior of the one-loop determinant for large $|\sigma|$ is given by 
$$
 \exp\Big( 2i R^2 \log \big| \rho(\sigma) \big| \, \rho(D_0) \Big) \;.
$$
This corresponds to the effective twisted superpotential $\widetilde W_\text{eff} = - \frac1{4\pi} \rho(\sigma) \big( \log \rho(\sigma) - 1 \big)$. Comparing with (\ref{integral D0 at infinity}), we see that for $\eta>0$ we pick the residue at infinity iff $\sum \rho < 0$, while for $\eta < 0$ we pick minus the residue at infinity iff $\sum \rho > 0$. In other words, the r\^ole played by the effective CS level in 3d is played by the FI $\beta$-function in 2d.%
\footnote{This is correct if the running FI is the leading behavior, which is the case if we started with a renormalizable Lagrangian whose bare twisted superpotential has only FI term. If, on the contrary, we started with a bare twisted superpotential with higher order terms, then the analysis of the boundary contribution has to be repeated.}

\paragraph{Higher genus.} The expressions at higher genus are derived in a similarly easy way. We find:
\be
Z_\text{1-loop}^\text{chiral} = \prod_{\rho\in\fR} \Big[ \frac1{\rho(\sigma)} \Big]^{\rho(\fm) + (g-1)(q-1)} \;,\qquad Z_\text{1-loop}^\text{gauge} = (-1)^{\sum_{\alpha>0} \alpha(\fm)} \prod_{\alpha\in G} \alpha(\sigma)^{1-g} \, (d\sigma)^r\;.
\ee

\subsubsection{Refinement by angular momentum: the $\Omega$-background}

As in three-dimensions, if we choose a metric on $S^2$ which has $U(1)$ invariance, we can refine the partition function with a weight for the angular momentum. The proper supergravity background necessary for a full-fledged computation was constructed and analyzed in \cite{Closset:2014pda}. This is essentially the $\Omega$-background on $S^2$. On the other hand, we can more modestly obtain the relevant formula for the partition function with insertions by taking the $\beta \to 0$ limit from three dimensions. After a redefinition of the variables and a rescaling that matches the unrefined case, we obtain
\be
\boxed{
Z_\text{1-loop}^\text{chiral} = \prod_{\rho \in \fR} \prod_{j= - \frac{|B|-1}2}^{\frac{|B|-1}2} \bigg( \frac1{\rho(\sigma) + j \varsigma} \bigg)^{\sign B} \;,} \qquad\qquad B = \rho(\fm) - q + 1 \;.
\ee
Here $\varsigma$ is the parameter associated to the refinement, which can be identified with the $\epsilon$ parameter of the $\Omega$-background. This expression could be written in a more elegant way using the Pochhammer symbol:
\be
Z_\text{1-loop}^\text{chiral} = \frac{\varsigma^{-B}}{ \big( \frac{\rho(\sigma)}\varsigma + \frac{1-B}2 \big)_B} \;.
\ee
However such an expression is not valid for $\varsigma= 0$ even though the limit is not singular. For the vector multiplet we obtain
\be
\boxed{
Z_\text{1-loop}^\text{gauge} = (-1)^{\sum_{\alpha>0} \alpha(\fm)} \prod_{\alpha \in G} \Big( \alpha(\sigma) + | \alpha(\fm) | \,  \frac\varsigma2 \Big) \, (d\sigma)^r = \prod_{\alpha>0} \Big[ \alpha(\sigma)^2 - \frac{\alpha(\fm)^2 \varsigma^2}4 \Big] \, (d\sigma)^r \;.
}
\ee

Insertions of gauge-invariant twisted chiral operators constructed out of $\sigma$ can only be done at the two poles. To compute a correlator involving $f(\sigma)$ at the North or South pole, we include
$$
f \Big( \sigma \,\pm\, \varsigma \frac\fm 2\Big)
$$
in the localization formula, where the signs $\pm$ refer to the North and South pole, respectively. This follows from the reduction of the Wilson loop in (\ref{Wilson loop poles}). As we show in the next example, operator insertions at coincident points introduce specific contact terms, leading to a sort of deformation of the chiral ring. We leave the question of what this deformation is to future work.

\subsubsection{The projective spaces $\bP^{N-1}$}

In this section we compute the topological amplitudes for the projective space $\bP^{N-1}$. The model has $U(1)$ gauge group and $N$ chiral multiplets of charge $1$, to which we assign R-charge $0$. The topological amplitude with $n$ insertions of the basic (1,1) class $\sigma$ is
\be
\langle \sigma_1 \cdots \sigma_n \rangle = \sum_{\fm\in\bZ} \int \frac{dx}{2\pi i}\, \frac1{x^{(\fm+1)N}} q^\fm x^n = \begin{cases} q^\frac{n-N+1}N &\text{if } n = N-1 \pmod{N} \\ 0 &\text{otherwise.} \end{cases}
\ee
We have chosen $\eta >0$ and we have taken the residue at the singularity $x=0$; had we chosen $\eta<0$, we would have taken minus the residue at infinity. Let us understand why this reproduces the quantum cohomology of $\bP^{N-1}$. First, $\sigma$ represents the K\"ahler class $\omega$ or alternatively the hyperplane divisor class, therefore $\langle \sigma^{N-1} \rangle = 1$ because the intersection of $N-1$ hyperplanes is a single point. Then the quantum cohomology is $\sigma^N = q$, and in fact $\langle \sigma^{\fm N + N - 1} \rangle = q^\fm$. All other correlators vanish because of axial R-symmetry charge conservation \cite{Witten:1991zz}.

Next consider the case with masses $\mu_j$ for the $N$ chiral fields (the sum of the masses can be reabsorbed by a shift of $\sigma$). The amplitude for an insertion $f(\sigma)$, which in general is the product of insertions at different points, is
\be
\langle f(\sigma) \rangle = \sum_{\fm\in\bZ} \int \frac{dx}{2\pi i} \; q^\fm \, f(x) \; \prod_j \frac1{(x-\mu_j)^{\fm + 1}} \;.
\ee
By taking minus the residue at infinity (and noticing that only the sector $\fm=0$ could contribute) one easily finds
\be
\langle \sigma^n \rangle = \begin{cases} 0 &\text{for } n = 0, \dots, N-2 \\ 1 &\text{for } n = N-1\;. \end{cases}
\ee
This is the classical result. By a shift $\fm \to \fm +1$ it is easy to prove
\be
\Big\langle f(\sigma) \prod_{j=1}^N (\sigma - \mu_j) \Big\rangle = q \, \langle f(\sigma) \rangle \;,
\ee
which is the equivariant quantum cohomology of $\bP^{N-1}$.

\paragraph{Refined case.} Let us compute the amplitudes in the case with refinement, but no masses. We can consider the insertion of $\sigma^n$ at the North pole. We compute
\be
\langle \sigma^n \big|_\text{N} \rangle = \sum_{\fm\in\bZ} \int \frac{dx}{2\pi i} \; q^\fm \Big( x + \varsigma \frac\fm2 \Big)^n \prod_{j = - \frac{ |B|-1}2}^{\frac{|B|-1}2} \bigg( \frac1{x+j\varsigma} \bigg)^{N \sign B} \;,\qquad\qquad B = \fm+1 \;.
\ee
First of all notice that for $\fm \leq -1$ the integrand is a polynomial, therefore there are no residues neither at finite points nor at infinity. We reduce to
\be
\langle \sigma^n \big|_\text{N} \rangle = \sum_{\fm\geq0} \int \frac{dx}{2\pi i} \; q^\fm \Big( x + \varsigma \frac\fm2 \Big)^n \prod_{j=-\fm/2}^{\fm/2} \frac1{(x+j\varsigma)^N} \;.
\ee
By evaluating around infinity, we conclude that $\langle \sigma^n \big|_\text{N} \rangle=0$ for $n = 0, \dots, N-2$ and $\langle \sigma^{N-1}\rangle = 1$, which is the classical result. Higher amplitudes are constrained by an equation obtained by shifting $\fm \to \fm-1$ and $x \to x - \frac\varsigma2$:
\bea
\label{diff equation zeta P^N}
q \, \big\langle f( \sigma) \big|_\text{N} \big\rangle &= \sum_{\fm \geq 1} \int \frac{dx}{2\pi i}\, q^\fm f \Big( x - \varsigma + \varsigma \frac\fm2 \Big) \prod_{j = - \frac{\fm-1}2}^{\frac{\fm-1}2} \frac1{\big( x + (j-\frac12)\varsigma \big)^N} \\
&= \sum_{\fm \geq 1} \int \frac{dx}{2\pi i}\, q^\fm \big( x + \tfrac\fm2 \varsigma \big)^N f\Big( x - \varsigma + \varsigma \frac\fm2 \Big) \prod_{j = - \frac\fm2}^{\frac\fm2} \frac1{( x + j\varsigma )^N} = \Big\langle \sigma^N f(\sigma - \varsigma) \big|_\text{N} \Big\rangle \;.
\eea
Here $f(\sigma)$ is a holomorphic correlator (a polynomial). In the last equality we used that for $\fm=0$ there are no residues, so that term can harmlessly be added. In the last expression, expanding the polynomial $f$ we obtain relations between different amplitudes. Shifting $x \to x + \frac\varsigma2$, instead, we obtain
\be
\label{underformed CR P^N}
q  \, \big\langle f(\sigma) \big|_\text{N} \big\rangle = \Big\langle f(\sigma) \big|_\text{N} \sigma^N \big|_\text{S} \Big\rangle \;.
\ee
In this case, the correlator on the right contains an insertion of $\sigma^N$ at the South pole.

The equation (\ref{underformed CR P^N}) is just the flat-space chiral ring relation $\sigma^N = q$ applied at the South pole---this is independent of the insertion of $f(\sigma)$ at the North pole, as it should be in the ring. On the contrary equation (\ref{diff equation zeta P^N}), where all insertions are at the North pole, represents an interesting deformation of the chiral ring.

Equation (\ref{diff equation zeta P^N}) can be used to show that $\langle \sigma^n \rangle = 0$ for $n = N, \dots, 2N-2$ and $\langle \sigma^{2N-1} \rangle = q$, which again is the same as the classical result. For $n \geq 2N$, though, the amplitudes are deformed by $\varsigma$, for instance
$$
\langle \sigma^{2N} \rangle = \langle \sigma^{2N} \rangle_{\varsigma=0} + N \varsigma q \;.
$$
All other amplitudes for larger values of $n$ can be recursively determined in a similar way.

Since the topological amplitudes should be fixed by the twisted chiral ring relations on flat space, how do we explain the deformation by $\varsigma$? The reason is that we are forced to place local operators at coincident points, and the deformations that depend on $\varsigma$ should be attributed to contact terms. For instance, the operators in the twisted chiral ring of $\bP^{N-1}$ are $\sigma^j$ with $j=0, \dots, N-1$. Therefore for $n\geq N$, in (\ref{diff equation zeta P^N})  we are considering a two or higher point function, with at least two coincident operators at the North pole. It would be interesting to understand if the deformation of the amplitudes produced by our computation has a nice mathematical structure.

\subsubsection{The quintic}

This is a simple example of a compact Calabi-Yau manifold of complex dimension $3$. The model has 5 chiral multiplets $X_i$ of charge 1 and R-charge 0, and one multiplet $P$ of charge $-5$ and R-charge 2, as well as a superpotential $W = P f_5(X_i)$ where $f_5$ is a homogeneous polynomial of degree $5$. The topological amplitude with $n$ insertions is
\be
\langle \sigma_1 \cdots \sigma_n \rangle = \sum_{\fm\in\bZ} \int \frac{dx}{2\pi i}\, \frac1{x^{5\fm+5} (-5x)^{-5\fm -1}} \, q^\fm x^n \;.
\ee
The first factor of $x$ in the denominator comes from the $X_i$, the second from $P$. In principle there might be a problem because both the positively and negatively charged fields give a divergence at the some point $x=0$, which is not acceptable. However, we choose $\eta>0$ and therefore we take the residues from the $X_i$, which give a pole for $\fm \geq \frac{n-4}5$. Taking into account that $\fm \in\bZ$, we see that within this range we never get a pole from $P$. 

We can resum in $\fm$ and obtain the following expression:
\be
\langle \sigma_1 \cdots \sigma_n \rangle = \sum_\fm \int \frac{dx}{2\pi i}\, \frac{(-5)^{5\fm+1} q^\fm}{x^{4-n}} = \begin{cases} -\frac5{1+5^5 q} &\text{if } n=3 \\ 0 &\text{otherwise.} \end{cases}
\ee
The expression is non-vanishing only for $n=3$, and in that case we summed over $\fm \geq 0$. This expression correctly represents the quantum cohomology of the quintic CY$_3$, and it exactly matches with (4.24) in \cite{Morrison:1994fr}, up to the minus sign which is a matter of convention.

\subsection{Four-dimensional $\cN=1$ theories on $S^2 \times T^2$}

The last topic we cover is four-dimensional $\cN=1$ theories placed on $S^2 \times T^2$, with a topological twist on $S^2$. To preserve supersymmetry, the theories must possess a non-anomalous R-symmetry with integer R-charges.
The one-loop determinant for a chiral multiplet on $S^2 \times T^2$  has already been considered in \cite{Closset:2013sxa}. We will present the complete formula for gauge theories. The result of localization is simply the elliptic generalization of our formula.%
\footnote{The formula can be derived with the same steps as in section \ref{sec: loc}, using the supersymmetry transformations and actions in \cite{Closset:2013sxa, Nishioka:2014zpa}. We do not repeat those steps here.}
The one-loop determinant is a function of $q = e^{2\pi i \tau}$ and $x = e^{iu}$, where $\tau$ is the modular parameter of $T^2$ and $u$ parameterizes the Wilson lines on the two direction of the torus.

Let us consider the case refined by the angular momentum on $S^2$; the formula with no refinement is obtained simply setting $\zeta = 1$. Borrowing the expressions from the elliptic genus of \cite{Benini:2013nda,Benini:2013xpa}, the one-loop determinant for the chiral multiplet is
\be
\boxed{
Z_\text{1-loop}^\text{chiral} = \prod_{\rho \in \fR} \prod_{j=- \frac{|B|-1}2}^{\frac{|B|-1}2} \bigg( \frac{i\eta(q)}{\theta_1(x^\rho \zeta^{2j};q)} \bigg)^{\sign(B)} \;, } \qquad\qquad B = \rho(\fm) - q + 1 \;.
\ee
The two elliptic functions are $\eta(q) = q^{1/24} \prod_{n=1}^\infty (1-q^n)$ and
\be
\theta_1(x;q) = -i q^\frac18 x^\frac12 \prod_{k=1}^\infty (1-q^k)(1-xq^k)(1-x^{-1}q^{k-1}) = -i \sum_{n\in\bZ} (-1)^n e^{iu( n + \frac12)} \, e^{\pi i \tau (n+\frac12)^2} \;.
\ee
The three-dimensional determinant is reproduced in the $q \to 0$ limit (\ie{} $\tau \to i\infty$), up to a zero-point energy:
$$
\lim_{q\to0} Z_\text{1-loop}^\text{4d} = Z_\text{1-loop}^\text{3d} \prod_{\rho \in \fR} q^{-B/12}  \;.
$$
The one-loop for off-diagonal vector multiplets equals the one-loop for chiral multiplets with R-charge 2:
\bea
\boxed{ Z_\text{1-loop}^\text{gauge, off} =  (-1)^{\sum_{\alpha>0} \alpha(\fm)} \prod_{\alpha\in G} \frac{\theta_1\big(x^\alpha \zeta^{|\alpha(\fm)|};q \big) }{i\eta(q)} \;. }
\eea
This  time there is also a contribution from the vector multiplets along the Cartan generators, as evinced from \cite{Benini:2013nda,Benini:2013xpa}:
\be
\boxed{Z_\text{1-loop}^\text{gauge, Cartan} = \eta(q)^{2r} \, (i\, du)^r \;. }
\ee

For theories with semi-simple gauge group, all action terms are $\cQ$-exact and the meromorphic form $Z_\text{int}(u;\fm)$ only gets contributions from the one-loop determinants (supersymmetry transformations and actions can be found in \cite{Nishioka:2014zpa}). Instead, for Abelian factors one can also consider a Fayet-Iliopoulos term leading to a classical action contribution:
\be
\cL_\text{FI} = - i \frac\zeta{2\pi} D \qquad\Rightarrow\qquad Z_\text{class} = e^{-S_\text{FI} \big|_\text{on-shell}} = e^{- \vol(T^2) \, \zeta \, \fm } \;.
\ee

The partition function is obtained by integrating the one-loop determinant and the classical action on the space of flat connections on $T^2$ and summing over the fluxes on $S^2$. More precisely, we should consider the space of flat connections that commute with the constant flux on $S^2$, \ie{} we should look for commuting triplets
\be
[g_1, g_2] = [g_1,\fm] = [g_2,\fm] = 0 \;,\qquad\qquad g_1, g_2 \in G \;,\qquad \fm \in \fg \;.
\ee
Since the space of flat connections on $T^2$ is compact, there are no subtleties with infinity. Therefore the integration contour is simply the one provided by the Jeffrey-Kirwan residue.

\subsubsection{Example: $SU(2)$ SQCD}

As an example, we would like to consider SQCD theories in four dimensions and check that those related by Seiberg duality \cite{Seiberg:1994pq} yield the same partition function. To keep the example simple, we look at $SU(2)$ SQCD with $N_f=3$ flavors, whose dual is a Wess-Zumino model of chiral multiplets with a superpotential.

In fact, $SU(2)$ SQCD is more easily described within the $USp$ family, whose Seiberg dualities have been studied in \cite{Intriligator:1995ne}. $USp(2N_c)$ SQCD has $2N_f$ chiral multiplets $Q_i$ in the fundamental, the global symmetry is $SU(2N_f) \times U(1)_R$, and the flavors are in the fundamental of $SU(2N_f)$ and have R-charge $R = (N_f-N_c-1)/N_f$. The gauge invariants are the mesons $M_{ij} = Q_i \cdot Q_j$, where the contraction over gauge indices is done with the invariant tensor of $USp(2N_c)$, and there is antisymmetry in $ij$.

We consider $USp(2)$ SQCD with $2N_f=6$ quarks. Their R-charges are $R = \frac13$, which is not acceptable in our setup on $S^2 \times T^2$ because the R-charges must be integers. However, we can mix $U(1)_R$ with a Cartan generator of the flavor symmetry $SU(6)$, namely $\text{diag}\big( \frac23, \frac23, -\frac13, -\frac13, -\frac13, -\frac13 \big)$, to form a new R-symmetry $U(1)_R' = \text{diag}(1,1,0,0,0,0)$, non-anomalous and integer. For simplicity, we look at the unrefined case $\zeta=1$ and we do not consider the most general background for the flavor symmetry, rather we turn on a minimal background to obtain a finite partition function: we take the Cartan of $SU(6)$ given by $U(1)_A = \text{diag}(-2,-2,1,1,1,1)$. In order to write down the partition function $Z_{S^2\times T^2}$ in a compact form, we introduce the notation
\be
f_\chi(g,a,r) = \bigg( \frac{i\eta(q)}{ \theta_1(x^g y^a;q)} \bigg)^{g\fm + a\fn - r +1}
\ee
for the one-loop determinant of a chiral multiplet with gauge charge $g$ (under the Cartan of $SU(2)$), flavor charge $a$ and R-charge $r$: this is an implicit function of $(x,\fm)$ and $(y,\fn)$, and, as usual, $x=e^{iu}$ and $y=e^{iv}$. Then $Z_{S^2 \times T^2}$ is
\be
\label{Z 4d SU(2)}
Z_{SU(2)}^{N_f=3} = \frac12 \sum_{\fm\in\bZ} \frac1{2\pi i} \int i\eta(q)^2 du\; \frac{\theta_1(x^2;q)}{i\eta(q)} \; \frac{\theta_1(x^{-2};q)}{i\eta(q)} \; \prod_{g=\pm1} f_\chi(g,-2,1)^2 \, f_\chi(g,1,0)^4 \;.
\ee
The magnetic theory, which in this case is more properly interpreted as the effective IR description, is a Wess-Zumino model of fundamental fields $M_{ij}$ (antisymmetric in $ij$), subject to the cubic superpotential
\be
W = \Lambda^{-3} \,\text{Pf}\, M \;.
\ee
The partition function for these 15 mesons is simply
\be
\label{Z 4d dual}
Z_\text{dual} = f_\chi(0,-4,2) \, f_\chi(0,-1,1)^8 \, f_\chi(0,2,0)^6 \;.
\ee
We would like to check that the two partition functions coincide.

We can check that (\ref{Z 4d SU(2)}) and (\ref{Z 4d dual}) coincide at the lowest order in a $q$ expansion. After expanding the integrand of (\ref{Z 4d SU(2)}) at the lowest order in $q$, we choose $\eta>0$ and consider the poles at $x=y^2$ (existing for $\fm \geq 2\fn+1$) and at $x=y^{-1}$ (existing for $\fm \geq -\fn$). Then we follow the same strategy as in section \ref{sec: duality appetizer}. We first sum over $\fm \geq -M$, for some large positive integer $M$, to be sure to pick up all residues. This produces an expression in which the relevant poles are at the two roots of the polynomial equation
$$
2y^2(1+x^2) - x (1+2y -2y^2+2y^3+y^4) = 0 \;.
$$
The residues can be computed explicitly and summed: the dependence on $M$ disappears, and we are left with
\be
Z_{SU(2)}^{N_f=3} = - q^{-5/12} y^4 (1+y)^{-8\fn} (1-y^2)^{-5} (1+y^2)^{4\fn+1} + \cO(q^{7/12}) \;.
\ee
This is precisely the expansion of $Z_\text{dual}$ at the lowest order in $q$.

\section{Conclusions}

In this paper we have provided a general formula for the partition function of three-dimensional $\cN=2$ gauge theories placed on $S^2\times S^1$, with a topological twist along $S^2$. The result depends on a collection of background magnetic fluxes and fugacities for the flavor symmetries. There are many generalizations of our result to other dimensions and different manifolds: some of these generalizations have been discussed in this paper, others are left for future work. In particular, our result can be easily extended to the case where $S^2$ is replaced by a Riemann surface $\Sigma$ of higher genus. The new ingredient is the presence of extra zero-modes corresponding to the Wilson lines along $\Sigma$.  The case $\Sigma=T^2$ would compute the Witten index of three-dimensional theories.

It would be interesting to see whether our reformulation of the A-twist for supersymmetric two-dimensional sigma models in terms of Coulomb branch localization can provide new geometrical insight in the study of Gromov-Witten invariants and mirror symmetry. In particular, while the physics of Abelian GLSMs, and the mathematics of the corresponding low energy NLSMs (complete intersections in toric varieties), are very well understood,%
\footnote{But see \cite{Hellerman:2006zs, Caldararu:2007tc} for models where non-perturbative effects play a crucial r\^ole.}
this is not the case for non-Abelian models \cite{Hori:2006dk, Jockers:2012zr}. The partition function on the sphere with no twist have also been computed via localization in \cite{Benini:2012ui, Doroud:2012xw} and has led to a re-interpretation of the K\"ahler geometry of the Calabi-Yau moduli space \cite{Jockers:2012dk, Gomis:2012wy}. The topologically twisted partition function, instead, directly computes amplitudes.

We have not spent many words on four-dimensional $\cN=1$ theories, although this is clearly an interesting setting. A line of investigation we would like to suggest is about correspondences {\it \`a la} AGT \cite{Alday:2009aq, Gadde:2009kb}. They put in correspondence observables in certain 4d $\cN=2$ \cite{Gaiotto:2009we, Gaiotto:2009hg, Benini:2009gi} and $\cN=1$ \cite{Benini:2009mz, Bah:2012dg} supersymmetric gauge theories with 2d conformal or topological theories, by different compactifications of the mysterious 6d $\cN=(2,0)$ theory. With a new observable at our disposal---the index with twist---it is natural to ask what is the 2d theory that computes it.

Finally, we should mention that this paper grew originally from the attempt to understand in microscopical terms the entropy of AdS$_4$ static black holes. Regular static black holes have been recently found in four-dimensional ${\cal N}=2$ gauged supergravity \cite{Cacciatori:2009iz, Dall'Agata:2010gj, Hristov:2010ri}. They have magnetic charges and  an AdS$_2\times \Sigma$ horizon and  can be  interpreted as a dual description of a three-dimensional CFT placed on $\Sigma\times S^1$ with a topological twist along $\Sigma$ and various background magnetic fluxes for the global symmetries \cite{Hristov:2013spa}. They thus perfectly fit in the framework of our paper. We hope to report on the subject soon.

\section*{Acknowledgements}

We would like to thank Cyril Closset, Stefano Cremonesi, Kentaro Hori, Kimyeong Lee, Sungjay Lee and Dave Morrison for useful discussions, and especially Kiril Hristov and Alessandro Tomasiello for collaboration at an early stage of this work. 
FB is supported in part by the Royal Society as a Royal Society University Research Fellowship holder. This work is part of the Delta ITP consortium, a program of the Netherlands Organisation for Scientific Research (NWO) that is funded by the Dutch Ministry of Education, Culture and Science (OCW). AZ is supported by the INFN and the MIUR-FIRB grant RBFR10QS5J ``String Theory and Fundamental Interactions''.  AZ thanks the Galileo Galilei Institute for Theoretical Physics (Florence)  for the hospitality  during the completion of this work.

\appendix

\section{One-loop determinants}
\label{sec: locA}

In this appendix we give the details of the computation of one-loop determinants on $S^2\times S^1$: we perform an explicit computation with the round metric, we provide an alternative cohomological argument valid when $D_0=0$ with any metric, and we extend the computation to the refined case.

\subsection{Reduction on $S^2$}
\label{sec: reduction S^2}

We consider the round sphere with constant magnetic field:
\be
ds^2 = R^2 (d\theta^2 + \sin^2\theta\, d\varphi^2) \;,\qquad\qquad F = \frac b2 \sin\theta\, d\theta\wedge d\varphi \;.
\ee
The flux is quantized, $b \in \bZ$, the spin connection is $\omega^{12} = - \cos\theta\, d\varphi$ and we choose a gauge potential $A = \frac b2 \omega^{12}$. The covariant derivative for a particle of charge $1$ and spin $s_z \in \bZ/2$ is
\be
D_\mu = \partial_\mu + i s_z \omega^{12}_\mu - \frac{ib}2 \omega^{12}_\mu \;.
\ee
We define the effective spin
\be
s \equiv s_z - \frac b2 \;.
\ee
The particle transforms as a section of a line bundle with $c_1 = b-2s_z = -2s$, therefore we can regard it as a scalar particle in a magnetic flux $-2s$, or as a neutral particle of effective spin $s$.

The gauge-covariant Laplace operator is%
\footnote{The total connection (spin plus gauge) is $\cA = s\cos\theta\, d\varphi$, which is in a singular gauge because $\cA$ is singular at the poles. Moreover, for $s \in \bZ + \frac12$ a gauge transformation is required as we go around the poles, as it becomes clear computing the Wilson line very close to the poles which should vanish. Thus, for $s \in \bZ + \frac12$ the wavefunctions get a minus sign as $\varphi \to \varphi + 2\pi$. They become single-valued using standard smooth gauges on two patches. \label{foo gauge}}
\be
D^2 = \frac1{R^2} \Big[ \frac1{\sin\theta} \partial_\theta (\sin\theta\, \partial_\theta) + \frac1{\sin^2\theta} \Big( \partial_\varphi - is \cos\theta\Big)^2 \Big] \;.
\ee
Its eigenfunctions are the well-known monopole harmonics \cite{Wu:1976ge}. Let us quickly review their construction. We  introduce the following operators:
\be
\tilde J_\pm = i\, e^{\pm i \varphi} \Big( \partial_\theta \pm \frac i{\tan\theta} \partial_\varphi \pm \frac s{\sin\theta} \Big) \;,\qquad\qquad \tilde J_3 = -i\partial_\varphi \;.
\ee
They are constructed in such a way to satisfy the same commutation relations as the standard angular momentum, $[\tilde J_3, \tilde J_\pm] = \pm \tilde J_\pm$ and $[\tilde J_+, \tilde J_-] = 2\tilde J_3$, and to be related to the Laplace operator in a simple way, $- R^2 D^2 = \frac12 \{\tilde J_+, \tilde J_-\} + \tilde J_3^2 - s^2$, which implies $[D^2, \tilde J_3] = [D^2, \tilde J_\pm] = 0$. We can then diagonalize $D^2$ and $\tilde J_3$ simultaneously, using $\tilde J_\pm$ as ladder operators for $\tilde J_3$. The states in a representation satisfy $-j \leq j_3 \leq j$, in terms of a quantum number $j$ so defined. We use the notation $Y^s_{j,j_3}$, then the spectrum is
\be
-R^2 D^2 Y^s_{j,j_3} = \big( j(j+1) - s^2 \big) \, Y^s_{j,j_3} \;.
\ee
Positivity of $-R^2D^2$ implies $j \geq |s|$. This becomes clear working out the highest weight eigenfunctions: $Y^s_{j,j} = e^{ij\varphi} \big( \tan \tfrac\theta2 \big)^{-s} (\sin\theta)^j$ annihilated by $\tilde J_+$. They are well-defined for $j\geq |s|$ and $j- s\in \bN$ (see footnote \ref{foo gauge}). The other wavefunctions can be obtained by acting with $\tilde J_-$.

We also have bundle-changing operators $D_\pm$:
\be
D_\pm^{(s)} = \partial_\theta \mp \frac i{\sin\theta} \partial_\varphi \mp s \frac{\cos\theta}{\sin\theta} \;,
\ee
which map $s \to s \pm 1$ keeping $j,j_3$ fixed.%
\footnote{When acting with the operators $D_\pm$, one has to be careful to keep track of the value of $s$, as they change it. Concretely, when acting on a wavefunction of spin $s$, $[D_+,D_-] = D_+^{(s-1)} D_-^{(s)} - D_-^{(s+1)} D_+^{(s)}$.}
One verifies that $[D_\pm, \tilde J_3] = [D_\pm, \tilde J_+] = [D_\pm, \tilde J_-] = 0$, moreover $[D_+, D_-] = -2s$ and $-R^2 D^2 = - \frac12 \{D_+, D_- \} = - D_+ D_- -s = - D_- D_+ + s$. It follows that $[-R^2D^2,D_\pm] = (\mp 2s - 1)D_\pm$, confirming the map $s \to s \pm 1$. A state annihilated by $D_-$ (it has minimal $s$) has $-R^2 D^2 = -s$, therefore $j = -s$; a state annihilated by $D_+$ (maximal $s$) has $-R^2 D^2 = s$, therefore $j = s$.
The eigenfunctions annihilated by $D_-$ are $Y^{-j}_{j,j_3} = e^{ij_3\varphi} \big( \tan\tfrac\theta2 \big)^{j_3} (\sin\theta)^j$, then $Y^s_{j,j_3}$ with $-j \leq s \leq j$ can be obtained by acting with $D_+$. Finally, starting with $Y^{-j}_{j,j_3}$ and acting with $D_+$, or starting with $Y^j_{j,j_3}$ and acting with $D_-$, we get the relations
\be
D_+ D_- Y^s_{j,j_3} = - \big( j + s \big)\big( j - s + 1 \big) Y^s_{j,j_3} \;,\qquad
D_- D_+ Y^s_{j,j_3} = - \big( j - s \big)\big( j + s + 1 \big) Y^s_{j,j_3} \;.
\ee

\subsection{The one-loop determinant on $S^2\times S^1$}

We can use the spectral analysis on $S^2$ to compute the one-loop determinant on $S^2 \times S^1$ for a chiral multiplet $\Phi_\rho$, transforming as the weight $\rho$ of a representation $\fR$, and with R-charge $q$. The field is immersed in a magnetic field $\fm \in \fg$, there is a flat connection $A_t$ along $S^1$, and the D-term has expectation value $D = \frac{i\fm}{2R^2} + D_0$. Notice that in this computation we do not distinguish between gauge and flavor symmetries. We define the integer
\be
b \equiv \rho(\fm) - q \;.
\ee
We now compute the determinants for the scalar and the Dirac field in the chiral multiplet, while the full one-loop determinant is assembled in the main text.

\paragraph{The scalar $\phi$.} This field has R-charge $q$ and spin $s_z=0$, therefore, recalling that there are $-1$ units of R-symmetry flux on $S^2$ besides the magnetic flux, the effective spin is $s = - \frac b2$. The action follows from the quadratic expansion of (\ref{matter}) around the background, that we write as $\phi^\dag \cO_\phi \phi$. The eigenfunctions are $Y^s_{j,j_3} e^{2\pi i k t}$ with $j = |s|+n$ and $n \in \bZ_{\geq0}$, $k \in \bZ$. We immediately get the determinant
\be
\det \cO_\phi = \prod_{n\geq 0} \prod_{k\in\bZ} \bigg[ \frac{(2n+1)|b| - b + 2n(n+1)}{2R^2} + \rho(\sigma)^2 + \frac{\big( 2\pi k - \rho(A_t)\big)^2}{\beta^2} + i \rho(D_0) \bigg]^{2n + |b|+1} \;.
\ee

\paragraph{The Dirac spinor $\psi$.} The operator from the quadratic expansion of (\ref{matter}) is
\be
\cO_\psi = \gamma^\mu D_\mu - \rho(\sigma) = \mat{ D_3 - \rho(\sigma) & \frac1R D_+ \\ \frac1R D_- & -D_3 - \rho(\sigma)} \;.
\ee
The spinor $\psi$ has the same gauge/flavor charge $\rho$ as the scalar $\phi$, but its R-charge is $q-1$. Therefore the flux experienced by $\psi$ is $\rho(\fm) - q + 1 = b+1$. A Dirac spinor on $S^2$ has generically two components with $s_z = \pm \frac12$:
$$
Y^{-b/2}_{j,j_3} \text{ for } j \geq \Big| \frac b2 \Big| \;,\qquad\qquad Y^{-b/2 - 1}_{j,j_3} \text{ for } j \geq \Big| \frac b2 + 1 \Big| \;.
$$
If both components exist, the matrix is
$$
\mat{ \frac1\beta \big( 2\pi i k - i \rho(A_t) \big) - \rho(\sigma) & \frac1R D_+^{(-b/2-1)} \\ \frac1R D_-^{(-b/2)} & - \frac1\beta \big( 2\pi i k - i\rho(A_t) \big) - \rho(\sigma)} \mat{ Y^{-b/2}_{j,j_3} \\ Y^{-b/2 - 1}_{j,j_3}} \;,
$$
and its determinant is
\be
\det = \frac1{R^2} \Big( j-\frac b2 \Big)\Big( j + \frac b2 + 1\Big) + \rho(\sigma)^2 + \frac{\big( 2\pi k - \rho(A_t) \big)^2}{\beta^2} \;.
\ee
Now we should distinguish a few cases. For $b\geq 0$, at $j=\frac b2$ only the right-moving (RM) mode exists, while for $j = \frac b2 + n$ and $n\geq 1$ both modes exist. We then obtain for $\det \cO_\psi$:
$$
\prod_{k\in\bZ} \Big[ i \frac{2\pi k - \rho(A_t)}\beta - \rho(\sigma) \Big]^{b+1} \prod_{n\geq 1} \bigg[ \frac1{R^2} \, n (n+b+1) + \rho(\sigma)^2 + \frac{\big( 2\pi k - \rho(A_t) \big)^2}{\beta^2} \bigg]^{2n + b + 1} \;.
$$
For $b\leq -2$, at $j = - \frac b2 - 1$ only the left-moving (LM) mode exists, while for $j = - \frac b2 - 1 + n$ and $n\geq 1$ both modes exist. We then obtain for $\det \cO_\psi$:
$$
\prod_{k\in\bZ} \Big[ -i \frac{2\pi k - \rho(A_t)}\beta - \rho(\sigma) \Big]^{-b-1} \prod_{n\geq 1} \bigg[ \frac1{R^2} \, n (n-b-1) + \rho(\sigma)^2 + \frac{\big( 2\pi k - \rho(A_t) \big)^2}{\beta^2} \bigg]^{2n-b-1} \;.
$$
Finally, for $b = -1$ there are no special cases with chiral modes, because even with the smallest possible angular momentum, $j = \frac12$, both modes exist. Then we set $j = \frac12 + n$ and we obtain
$$
\prod_{k\in\bZ} \prod_{n\geq 0} \bigg[ \frac1{R^2}\, (n+1)^2 + \rho(\sigma)^2 + \frac{\big( 2\pi k - \rho(A_t) \big)^2}{\beta^2} \bigg]^{2n+2} \;.
$$
The three cases can be summarized by the following formula for the fermionic determinant:
\begin{multline}
\det \cO_\psi = \prod_{k\in\bZ} \Big[ i s \frac{2\pi k - \rho(A_t)}\beta - \rho(\sigma) \Big]^{|b+1|} \times \\
\times \prod_{n\geq 1} \bigg[ \frac{n \big( n+|b+1| \big)}{R^2} + \rho(\sigma)^2 + \frac{ \big( 2\pi k - \rho(A_t) \big)^2}{\beta^2} \bigg]^{2n + |b+1|}
\end{multline}
where $s = \sign(b+1)$.

\subsection{The cohomological argument}
\label{sec: cohomologic}

For $D_0 = 0$ (\ie{} $D = iF_{12}$) we can reproduce the one-loop determinant with an alternative cohomological argument, similar to that in \cite{Hama:2011ea}. We consider a generic metric on $S^2$ and a generic profile $F_{12}(x)$ for the magnetic field. We will need the identity
\be
\Dslash^2 = D_\mu D^\mu - \frac14 R_s - \frac i2 F_{\mu\nu} \gamma^{\mu\nu} \;,
\ee
and we will use the supersymmetry spinor $\epsilon$ in (\ref{cov}).

Now consider the two operators from the quadratic expansion of the matter action (\ref{matter}):
\be
\cO_\phi = - D_\mu D^\mu + \rho(\sigma)^2 + i \rho(D) - q W_{12} \;,\qquad\qquad \cO_\psi = \gamma^\mu D_\mu - \rho(\sigma) \;.
\ee
Suppose we have a fermionic mode $\Psi$ with $\cO_\psi \Psi = \lambda \Psi$: if $\epsilon^\dag \Psi \neq 0$ (iff $\gamma_3 \Psi \neq - \Psi$), then
\be
\cO_\psi \Psi = \lambda \Psi \qquad\Rightarrow\qquad \cO_\phi \epsilon^\dag \Psi = - \lambda \big( \lambda + 2 \rho(\sigma) \big) \epsilon^\dag \Psi \;.
\ee
On the other hand, suppose we have a scalar mode $\Phi$ with $\cO_\phi \Phi = - \lambda \big( \lambda + 2\rho(\sigma) \big) \Phi$. Then we can construct the two fermionic modes
\be
\Psi_1 = \Phi \epsilon \;,\qquad\qquad \Psi_2 = D_\mu \Phi \gamma^\mu \epsilon = \Dslash \Psi_1 \;.
\ee
The action of $\cO_\psi$ on the two-dimensional vector space is
\be
\mat{ \cO_\psi \Psi_1 \\ \cO_\psi \Psi_2} = \mat{ -\rho(\sigma) & 1 \\ \lambda \big( \lambda + 2\rho(\sigma) \big) + \rho(\sigma)^2 \;\; & -\rho(\sigma) } \mat{ \Psi_1 \\ \Psi_2} \;,
\ee
and the eigenvalues are $\lambda$ and $-\big(\lambda + 2\rho(\sigma)\big)$. We conclude that the modes in all these eigenspaces do not contribute to the one-loop determinant because their eigenvalues cancel out.

The only modes that contribute to the determinant are the unpaired ones. Let us find them. If $\epsilon^\dag \Psi = 0$, we do not have a partner scalar mode. This only happens if $\gamma_3 \Psi = - \Psi$. Then the defining equation splits into
\be
\gamma^a D_a \Psi = 0 \qquad\text{summed over } a = 1,2 \;,\qquad\qquad - D_3 \Psi - \rho(\sigma) \Psi = \lambda \Psi \;.
\ee
This implies that $\Psi$ is a LM chiral zero-mode of the twisted Dirac operator on $S^2$ and it depends on $t$ as $e^{2\pi i k t}$ for $k\in \bZ$. In the unrefined case $D_3 = \frac{\partial_t - i \rho(A_t)}\beta$, therefore
\be
\label{lambda 0}
\lambda \equiv \lambda_0 = -i \frac{2\pi k - \rho(A_t)}\beta - \rho(\sigma) \;.
\ee
If on $S^2$ (with the chosen metric and gauge flux) there are $n_L$ LM chiral zero-modes, the contribution to the index is $\lambda_0^{n_L}$. The refined case is discussed in section \ref{sec: refinement} and appendix \ref{sec:Arefinement}.

If the two modes $\Psi_1$ and $\Psi_2$ are actually parallel (including the case that $\Psi_2 = 0$), the scalar mode is paired to one fermionic partner only. Let us write $\Dslash \Psi_1 = \alpha \Psi_1$ for some $\alpha$. Since $\gamma_3\Psi_1 = \Psi_1$, the equation splits into
\be
\gamma^a D_a \Psi_1 = 0 \qquad\text{summed over } a=1,2 \;,\qquad\qquad D_3 \Psi_1 = \alpha\Psi_1 \;.
\ee
This implies that $\Psi_1$ is a RM chiral zero-mode on $S^2$. Then $\cO_\psi \Psi_1 = \lambda \Psi$ with $\lambda = \alpha - \rho(\sigma)$. In the unrefined case we find
$$
\lambda = i \frac{2\pi k - \rho(A_t)}\beta - \rho(\sigma) \;.
$$
However recall that we also have the scalar, therefore the contribution to the determinant is
$$
\frac\lambda{- \lambda \big( \lambda + 2\rho(\sigma)\big)} = \Big( -i \frac{2\pi k - \rho(A_t)}\beta - \rho(\sigma) \Big)^{-1} = \lambda_0^{-1}
$$
from each of these modes. If on $S^2$ there are $n_R$ RM chiral zero-modes, the contribution to the index is $\lambda_0^{-n_R}$. By the index theorem we have $n_R - n_L = \rho(\fm) - q +1$, therefore we are led to the same determinant (\ref{1-loop chiral before reg}) as before.

\subsection{Refined one-loop determinant}
\label{sec:Arefinement}
We now give some details about the computation of the refined one-loop determinant for a chiral multiplet. With round metric, the vielbein and its inverse are
\be
e\ud{a}{\mu} = \mat{ R & 0 & 0 \\ 0 & R\sin\theta & -R\varsigma \sin\theta \\ 0 & 0 & \beta} \;,\qquad\qquad e\ud{\mu}{a} = \mat{ \frac1R & 0 & 0 \\ 0 & \frac1{R\sin\theta} & \frac\varsigma\beta \\ 0 & 0 & \frac1\beta} \;.
\ee
A background with $F_{12} = \frac\fm{2R^2}$, $F_{13} = F_{23} = 0$ has an $F_{\theta t}$ component, and we can choose the connection to be
\be
A = - \frac\fm2 \cos\theta\, (d\varphi - \varsigma\, dt) + \tilde A_t dt = - \frac\fm{2R} \cot\theta\, e^2 + \frac{\tilde A_t}\beta\, e^3 \;,
\ee
where $\tilde A_t$ is the constant zero-mode part that commutes with $\fm$. This connection, though, is singular at the poles and therefore one should be very careful in using it to compute the Wilson loop. It is clearer to consider two patches, North and South, and two smooth connections:
\bea
A_N &= - \frac \fm2 (\cos\theta-1) (d\varphi - \varsigma\, dt) + \big( \tilde A_t + \tfrac\fm2 \varsigma \big) dt \\
A_S &= - \frac \fm2 (\cos\theta+1) (d\varphi - \varsigma\, dt) + \big( \tilde A_t - \tfrac\fm2 \varsigma \big) dt \;.
\eea
They give the following values for the angular and temporal Wilson loops:
\bea
\label{value of t phi loops}
W_\varphi &\equiv e^{i \oint_\varphi A_{N,S}} = \exp\Big[ i\pi \fm(1 - \cos\theta) \Big] \\
W_t &\equiv e^{i \oint_t A_{N,S}} = \exp\Big[ i \tilde A_t + i \varsigma \frac\fm2 \cos\theta \Big] \;.
\eea
The bosonic zero-mode is defined as
\be
u = \tilde A_t + i \beta \sigma = \beta( A_3 + i \sigma) \;.
\ee

To evaluate the classical CS actions, we should first extend $A$ to a connection on a four-manifold whose boundary is $S^2\times S^1$. We choose $S^2 \times D_2$, with $r$ the radius of $D_2$, and extend
\bea
\hat A &= - \frac\fm2 \cos\theta\, (d\varphi - \varsigma\, r^2 dt) + \tilde A_t\, r^2dt \\
\hat F &= \frac\fm2 \sin\theta\, d\theta \wedge (d\varphi - \varsigma\, r^2 dt) + \Big( \tilde A_t + \frac\fm2 \varsigma\cos\theta\Big) dr^2 \wedge dt \;.
\eea
The extension satisfies $\hat F \big|_{S^2 \times S^1} = F$, $\int_{D_2} \hat F = \int_{S^1} A$. Then $\int_{S^2 \times D_2} \hat F \wedge \hat F = 4\pi \fm \tilde A_t$ independently of $\varsigma$, therefore the on-shell CS actions are not affected by $\varsigma$.

A supersymmetric Wilson loop must be along the vector field $e_3 = \frac1\beta(\partial_t + \varsigma \partial_\varphi)$, as found after (\ref{def Wilson loop}), \ie{} it must lay along the embedding $x^\mu(\tau) = (\theta_0, \varsigma\tau, \tau)$ with parameter $\tau$. The Wilson loop equals $W = \Tr_R \Pexp i \oint d\tau\, \beta (A_3 + i\sigma)$. If we place the loop at the North ($\theta_0=0$) or South ($\theta_0 = \pi$) pole of the sphere, the direction $\varphi$ is immaterial, the loop only winds once around $t$, and there is no constraint on $\varsigma$. We obtain from (\ref{value of t phi loops}):
\be
W = \Tr_R e^{iu \,\pm\, i \varsigma \frac\fm2} = \sum\nolimits_{\rho \in R} \zeta^{\pm \rho(\fm)} \, x^\rho \;,
\ee
where the signs $\pm$ refer to the North and South pole, respectively. On the other hand, for generic values of $\theta_0 \neq 0,\pi$, the loop closes only if $\varsigma = 2\pi \frac pq$ with $p,q \in \bZ$ coprime: in this case the loop winds $p$ times around $\varphi$ and $q$ times around $t$. Combining the two integrals in (\ref{value of t phi loops}), we find
\be
W = \Tr_R e^{iqu \,+\, i\pi p \fm} =  \sum\nolimits_{\rho\in R} (-1)^{p\rho(\fm)} \, x^{q\rho} \;.
\ee

To compute the one-loop determinant, we follow the cohomological argument and count the fermionic zero-modes. For a given weight $\rho$, the flux seen by the fermions is $B = \rho(\fm) - q_\rho + 1$. If $B>0$, there are $B$ RM zero-modes on $S^2$ which are annihilated by $D_-$: they are the modes $Y^\frac{1-B}2_{j,j_3}$ with $j=\frac{B-1}2$ and $j_3 = - j, -j+1, \dots, j$, whose dependence on $\varphi$ is $e^{ij_3\varphi}$. On $S^2\times S^1$, these modes $\Psi_1$ have a dependence $e^{2\pi i kt}$. Each mode contributes a factor $-\frac1{\alpha + \rho(\sigma)}$, where $D_3 \Psi_1 = \alpha \Psi_1$. We thus find
$$
\prod_{j_3 = - \frac{B-1}2}^{\frac{B-1}2} \prod_{k\in\bZ} \Big[ \frac i\beta \big( \beta \rho(A_3 + i\sigma) - \varsigma j_3 - 2\pi k \big) \Big]^{-1} = \prod_{j_3 = - \frac{B-1}2}^{\frac{B-1}2} \frac{ x^{\rho/2} \zeta^{j_3} }{ 1 - x^\rho \zeta^{2j_3} } \;,
$$
as $\zeta = e^{i\varsigma/2}$.
If $B<0$, there are $|B|$ LM zero-modes on $S^2$ annihilated by $D_+$, \ie{} the modes $Y^\frac{|B|-1}2_{j,j_3}$ with $j = \frac{|B|-1}2$. Each mode $\Psi$ contributes a factor $\lambda$, where $-D_3\Psi - \rho(\sigma)\Psi = \lambda \Psi$. We thus find
$$
\prod_{j_3 = - \frac{|B|-1}2}^{\frac{|B|-1}2} \prod_{k\in\bZ} \frac i\beta \big( \beta \rho(A_3 + i\sigma) - \varsigma j_3 - 2\pi k \big) = \prod_{j_3 = - \frac{|B|-1}2}^{\frac{|B|-1}2} \frac{ 1 - x^\rho \zeta^{2j_3} }{ x^{\rho/2} \zeta^{j_3} } \;.
$$
This reproduces the one-loop determinant given in (\ref{refoneloop}).

\section{The two-dimensional partition function}
\label{sec:A2d}

In this appendix we give some details on localization in the two-dimensional case. The background is a generic metric on $S^2$ with volume $4\pi R^2$. The spin connection $\omega^{12}$ satisfies $d\omega^{12} = \frac{R_s}2 \dvol$ and $\frac1{2\pi} \int d\omega^{12} = 2$. We take a background vector $V = - \frac12 \omega^{12}$ coupled to the R-symmetry. Then the SUSY parameters satisfy $D_\mu\epsilon = 0$, $\gamma_3\epsilon = \epsilon$ and similarly for $\tilde\epsilon$ .

The SUSY transformations are easily derived from the three-dimensional case by mapping $A_3 \to \sigma_1$, $\sigma \to \sigma_2$ and then $\sigma_1 + i \sigma_2 \to -i\sigma$, $\sigma_1 - i \sigma_2 \to i\bar\sigma$. Using $F_{\mu 3} \to D_\mu\sigma_1$, $D_3 \to - i [\sigma_1, \,\cdot\,]$ and $[\sigma_1,\sigma_2] \to \frac i2[\sigma,\bar\sigma]$, we get the following. For the vector multiplet:
\bea
Q A_\mu &= \frac i2 \lambda^\dag \gamma_\mu \epsilon \qquad\qquad
QD = - \frac i2 D_\mu\lambda^\dag \gamma^\mu\epsilon + \frac i2 [\sigma, \lambda^\dag \epsilon] \qquad &
Q \lambda^\dag &= 0 \\
\wt Q A_\mu &= \frac i2 \tilde\epsilon^\dag \gamma_\mu \lambda \qquad\qquad
\wt Q D = \frac i2 \tilde\epsilon^\dag \gamma^\mu D_\mu \lambda - \frac i2 [ \sigma, \tilde\epsilon^\dag \lambda] \qquad &
\wt Q\lambda &= 0 \\
Q\lambda &= \Big( i F_{12} - D + \frac i2[\sigma,\bar\sigma] - i \gamma^\mu D_\mu \sigma \Big) \epsilon \qquad &
\hspace{-1.1cm} Q\sigma = 0 \qquad\qquad Q \bar\sigma &= \lambda^\dag \epsilon \\
 \wt Q \lambda^\dag &= \tilde\epsilon^\dag \Big( -i F_{12} + D + \frac i2 [\sigma,\bar\sigma] -i \gamma^\mu D_\mu \sigma \Big) \qquad &
\hspace{-1.1cm} \wt Q \sigma = 0 \qquad\qquad \wt Q \bar\sigma &= \tilde\epsilon^\dag \lambda \;.
\eea
The BPS equations for complexified fields are $D = i F_{12}$, $[\sigma,\bar\sigma]=0$ and $D_\mu\sigma = 0$. Restricting to configurations where the gauge field is real, $\bar\sigma$ is the complex conjugate to $\sigma$ but $D$ remains complex, we get
\be
D = i F_{12} \;,\qquad\qquad D_\mu \sigma = D_\mu \bar\sigma = 0 \;,\qquad\qquad [\sigma,\bar\sigma] = 0 \;.
\ee
Up to gauge transformations, the moduli space of bosonic zero-modes is $\fM = \fh \times \fh = \fh_\bC$ (to be divided by the Weyl group $W$ together with the magnetic fluxes $\fm \in \Gamma_\fh$) parameterized by a diagonal complex $\sigma$ which becomes our integration variable.

The transformations of the chiral multiplet are
\bea
Q\phi &= 0 \qquad & Q\psi &= \big( i\gamma^\mu D_\mu\phi -i \sigma\phi \big)\epsilon & Q\psi^\dag &= - \epsilon^{c\dag} F^\dag \\
\wt Q\phi &= - \tilde\epsilon^\dag\psi \qquad & \wt Q\psi^\dag &= \tilde\epsilon^\dag \big( -i\gamma^\mu D_\mu \phi^\dag -i \phi^\dag \sigma \big) & \wt Q\psi &= \tilde\epsilon^c F \\
Q\phi^\dag &= \psi^\dag\epsilon \qquad & QF &= \epsilon^{c\dag} \big( i \gamma^\mu D_\mu \psi +i \sigma\psi - i \lambda\phi \big) & QF^\dag &= 0 \\
\wt Q\phi^\dag &= 0 \qquad & \wt Q F^\dag &= \big( -i D_\mu\psi^\dag \gamma^\mu +i \psi^\dag \sigma + i \phi^\dag \lambda^\dag \big) \tilde\epsilon^c \qquad & \wt QF &= 0 \;.
\eea
The BPS equations for complexified fields are $(D_1 + i D_2)\phi = (D_1 - i D_2)\phi^\dag = 0$, $\sigma \phi = \phi^\dag \sigma = 0$, $F = F^\dag = 0$. Going to the real contour, they are complex conjugate pairs and reduce to
\be
(D_1 + i D_2)\phi = 0 \;,\qquad\qquad \sigma\phi = 0 \;,\qquad\qquad F = 0 \;.
\ee
The points (hyperplanes) of $\fM$ where $\sigma\phi = 0$ for some chiral multiplet $\phi$ form the subset $\fM_\text{sing}$: at those points, the BPS equations would allow for extra zero-modes.

The action terms that we consider are the standard ones, as in \cite{Benini:2012ui, Doroud:2012xw}: YM action, matter action and superpotential, which are $\cQ$-exact. More interesting is a twisted superpotential, which is not $\cQ$-exact. First, any holomorphic function $f(\sigma)$ can be inserted in the path-integral because it is supersymmetric:
\be
Q \, f(\sigma) = \tilde Q\, f(\sigma) = 0 \;,
\ee
in particular this allows us to make local insertions at arbitrary points on $S^2$. A twisted superpotential action must be real in Lorentzian signature, but we cannot insert $f^* (\bar \sigma)$ because this is not supersymmetric, even after integration. The twisted superpotential Lagrangian is in fact
\bea
\cL_{\widetilde W} &= i \widetilde W'(\sigma) \cdot (D + i F_{12}) - \frac i2 \widetilde W''(\sigma) \cdot \lambda^\dag (1-\gamma_3) \lambda \\
\overline{\cL_{\widetilde W}} &= i \widetilde W^{*\prime}(\bar\sigma) \cdot (D - i F_{12}) - \frac i2 \widetilde W^{*\prime\prime}(\bar\sigma) \cdot \lambda^\dag (1+\gamma_3)\lambda \;,
\eea
where $\widetilde W(\sigma)$ and $\widetilde W^*(\bar\sigma)$ are gauge-invariant holomorphic functions of their arguments. The two terms are separately supersymmetric, therefore the two functions $\widetilde W$ and $\widetilde W^*$ could be independent, however in order for the action to be real in Lorentzian signature, they should be complex conjugate. In the non-Abelian case it should be read as
$$
\cL_{\widetilde W} = i \parfrac{\widetilde W}{\sigma_A} (D + i F_{12})_A - \frac i2 \parfrac{^2\widetilde W}{\sigma_A \partial\sigma_B} \lambda^\dag_A (1-\gamma_3) \lambda_B
$$
where $A,B$ are indices of the adjoint representation, and similarly for $\overline{\cL_{\widetilde W}}$. Both terms are annihilated by $Q, \wt Q$.%

The bosonic part of the twisted superpotential Lagrangian is
\be
\cL_{\widetilde W} + \overline{\cL_{\widetilde W}} \Big|_\text{bos} = 2i \re \widetilde W'(\sigma) \cdot D - 2i \im \widetilde W'(\sigma) \cdot F_{12} \;.
\ee
Of particular importance is a linear twisted superpotential, leading to a complexified FI term:
\be
\widetilde W_{\text{FI},\theta} = -\frac1{4\pi} (\zeta + i\theta) \Tr \sigma \qquad\Rightarrow\qquad \cL_{\text{FI}, \theta} = - i \frac\zeta{2\pi} \Tr D + i \frac\theta{2\pi} \Tr F_{12} \;.
\ee
Evaluated on-shell on almost BPS configurations, the action gives
\be
e^{-S_{\widetilde W}} = e^{4\pi \widetilde W'(\sigma) \cdot \fm - 8\pi i R^2 \re \widetilde W'(\sigma) \cdot D_0} \;.
\ee
When specialized to the complexified FI term it becomes
\be
e^{- S_{\text{FI},\theta}} = e^{- (\zeta + i \theta) \Tr \fm + 2i R^2 \zeta \Tr D_0} = q^{\Tr \fm} \; e^{2i R^2 \zeta \Tr D_0} \;.
\ee
Here $q \equiv e^{-\zeta - i\theta}$, according to standard notation.

The computation of the one-loop determinants is essentially the same as in the three-dimensional case, but without the sum over the KK modes on $S^1$ that we labeled by $k$. In particular, there is no longer any regularization to do and no sign ambiguity in the final answer (besides, there are no topological symmetries nor the path-integral has a Hamiltonian interpretation as a trace, although a sign ambiguity could still be reabsorbed in the $\theta$-angle). To give some details, consider the round $S^2$ and the chiral multiplet. The scalar operator is
\be
\cO_\phi = - D_\mu D^\mu + \rho(\sigma)\rho(\bar\sigma) + i \rho(D_0) - q W_{12} \;,
\ee
and its spectrum is
\be
\det \cO_\phi = \prod_{n\geq 0} \bigg[ \frac{(2n+1) |b| - b + 2n(n+1)}{R^2} + |\rho(\sigma)|^2 + i \rho(D_0) \bigg]^{2n + |b|+1} \;,
\ee
where $b = \rho(\fm) - q_\rho$. The Dirac operator is
\be
\gamma^\mu D_\mu - i \gamma_3 \rho(\sigma_1) - \rho(\sigma_2) = \mat{ \rho(\bar\sigma) & \frac1R D_+ \\ \frac1R D_- & \rho(\sigma) } \;,
\ee
and its spectrum is
\be
\det \cO_\psi = \Big[ \rho\big( S(\sigma) \big) \Big]^{|b+1|} \prod_{n\geq 1} \bigg[ \frac1{R^2} n \big( n + |b+1| \big) + |\rho( \sigma)|^2 \bigg]^{2n + |b+1|}
\ee
where $S$ is the identity if $b \leq -2$, and complex conjugation if $b\geq 0$. The one-loop determinant is the ratio:
\be
Z_\text{1-loop}^\text{chiral} = \prod_{\rho\in\fR} \Big[ \frac1{\rho(\sigma)} \Big]^{\rho(\fm) - q_\rho + 1} \;.
\ee
Notice that this is in fact just the $x\to 1$ limit of the three-dimensional one-loop determinant in (\ref{one-loop 1d chiral}).
The one-loop determinant for the vector multiplet is
\be
Z_\text{1-loop}^\text{gauge} = (-1)^{\sum_{\alpha>0} \alpha(\fm)} \prod_{\alpha \in G} \alpha(\sigma) \;.
\ee
This time, we do not drop the sign factor in front because there are no ambiguities. This sign was noted in \cite{Hori:2013ika} for the background on $S^2$ constructed in \cite{Benini:2012ui, Doroud:2012xw}.%
\footnote{Although this sign seems not to be present in \cite{Benini:2012ui}, in fact there is a small mistake in the last appendix of that paper, and correcting it the sign is present in \cite{Benini:2012ui} too.}

One performs the localization, which proceeds as in the three-dimensional case. The only novelty is how to treat the region at infinity of $\fh_\bC$, \ie{} the complex $\sigma$-plane. For that, we need the asymptotic dependence of the one-loop determinant on $D_0$. Let us perform the analysis for $b\geq 0$. We want to compute
\be
F = \prod_{n\geq 0} \bigg( \frac{ \frac{n(n+b+1)}{R^2} + |\rho(\sigma)|^2 }{ \frac{n (n+b+1)}{R^2} + |\rho(\sigma)|^2 + i \rho(D_0) } \bigg)^{2n+b+1}
\ee
in the limit $|\sigma| \to \infty$. We have
\be
\log F = - i R^2 \rho(D_0) \sum_{n\geq 0} \frac{2n+b+1}{n(n+b+1) + a} + \cO(a^{-2}) \;,\qquad\qquad a = R^2 |\rho(\sigma)|^2 \;.
\ee
This expression diverges as $\sum \frac1n$, which cannot be regularized by $\zeta$ function (in fact it leads to dimensional transmutation). We regularize by subtracting $\frac2{n+1}$, then the sum can be performed:
\bea\nn
\log F\Big|_\text{reg} &= -i R^2 \rho(D_0) \bigg( -2\gamma - \sum_\pm \psi \Big( \frac{ 1+b \pm \sqrt{ -4a + (b+1)^2}}2 \Big) \bigg) \\
&= -i R^2 \rho(D_0) \Big( -2\gamma - \log a + \cO(a^{-1}) \Big)
\eea
where $\gamma$ is Euler's constant and $\psi(z) = \Gamma'(z)/\Gamma(z)$. We thus find the leading behavior
\be
F \simeq \exp\Big( 2i R^2 \log \big| \rho(\sigma) \big| \, \rho(D_0) \Big) \;.
\ee
This corresponds to the effective twisted superpotential $\widetilde W_\text{eff} = - \frac1{4\pi} \rho(\sigma) \big( \log \rho(\sigma) - 1 \big)$. Comparing with (\ref{integral D0 at infinity}), we see that for $\eta>0$ we pick the residue at infinity iff $\sum \rho > 0$, while for $\eta < 0$ we pick minus the residue at infinity iff $\sum \rho<0$.

\bibliographystyle{JHEP}
\bibliography{S2A}

\end{document}